\documentclass[12pt]{article}
\pdfoutput=1

\pdfoutput=1
\usepackage{geometry,amsmath,amsfonts}
\usepackage{slashed}
\usepackage{epsfig}
\usepackage{latexsym}
\usepackage{graphicx}
\usepackage{amssymb}
\usepackage[caption=false]{subfig}
\usepackage{color}
\usepackage{multirow}
\usepackage{color}
\usepackage{rotating}
\usepackage{ifthen}
\usepackage{epsfig}
\usepackage{lineno}
\usepackage{cite}

\begin{document}

\vspace*{8mm}

\begin{center}

{\large\bf  The Higgs-portal for Dark Matter: }\vspace*{2mm}

{\large\bf effective field theories versus concrete realizations }

\vspace*{9mm}

{\sc Giorgio Arcadi$^{1}$},  {\sc Abdelhak~Djouadi$^{2,3,4}$} and {\sc Marumi Kado$^{5,6}$}

\vspace*{9mm}

{\small
$^1$ Dipartimento di Matematica e Fisica and INFN, Universit\`a di Roma
Tre,\\ Via della Vasca Navale 84, 00146, Roma, Italy.\\ \vspace{0.2cm}

$^2$ CAFPE and Departamento de F\'isica Te\'orica y del Cosmos,\\ Universidad de Granada, E--18071 Granada, Spain.\\ \vspace{0.2cm}

$^3$ NICPB, R{\"a}vala pst. 10, 10143 Tallinn, Estonia.\\ \vspace{0.2cm}

$^4$ Universit\'e Savoie--Mont Blanc,  CNRS, LAPTh, F-74000 Annecy, France.\\ \vspace{0.2cm}

$^5$ Department of Physics and INFN, ``Sapienza" Universt\`a di Roma,\\ Pizzale Aldo Moro 5, I--00185 Roma, Italy.\\ \vspace{0.2cm}

$^6$ LAL, Universit\'e Paris-Sud, CNRS/IN2P3, Université Paris-Saclay, Orsay, France.\\ \vspace{0.2cm}

}
\end{center}

\vspace*{4mm}

\begin{abstract} 
Higgs-portal effective field theories are widely used as benchmarks in order to interpret collider and astroparticle searches for dark matter (DM) particles.
To assess the validity of these effective models, it is important to confront them to concrete realizations that are complete in the ultraviolet regime. In this paper, we compare effective Higgs-portal models with scalar, fermionic and vector DM with a series of increasingly complex realistic models, taking into account all existing constraints from collider and astroparticle physics. These complete realizations include the inert doublet with scalar DM, the singlet-doublet model for fermionic DM and models based on spontaneously broken dark U(1), SU(2) and SU(3) gauge symmetries for vector boson DM. We show that in large regions of the parameter space of these models, the effective Higgs-portal  approach provides a consistent limit and thus, can be safely adopted, in particular for the interpretation of searches for invisible Higgs boson decays at the LHC. The phenomenological implications of assuming or not that the DM states generate the correct cosmological relic density are also discussed.
\end{abstract}

\newpage

\section{Introduction}

Despite the overwhelming astrophysical and cosmological evidence, the nature of Dark Matter (DM) in the universe remains 
an open question, among the most prominent in modern physics. The solution of this longstanding puzzle, by means of the existence of a new weakly interacting massive particle (WIMP) \cite{Drees:1998ra,Bertone:2004pz,Arcadi:2017kky} that is stable at cosmological scales, is of paramount importance for particle physics. WIMPs are actively searched for in astrophysical experiments: in direct detection where the existence of the DM particles would be revealed by the low energy recoil of nuclei from their elastic scattering and in indirect detection when looking at signals of the annihilation of DM particles in regions of space with a high DM density. Searches at high--energy particle colliders are also gathering increasing interest and, at the CERN LHC, a prominent role is played by the searches of ``invisible" decays of the Standard Model (SM) Higgs boson into DM particles.\smallskip

Effective portal models have been introduced in order to perform simplified phenomenological studies, to easily interpret the outcome of the experimental searches and to illustrate the complementarity between different search strategies. In these effective models, the DM candidate is assumed to belong to a so-called dark or hidden sector, interacting with the SM states that form the visible sector through portals or suitable mediator fields. The most economical among them are the Higgs-portal models consisting in the extension of the SM with a single new particle, the DM state, which interacts in pairs only with the Higgs sector of the theory which is assumed to be minimal and hence involves only the unique Higgs boson observed at the LHC  \cite{Aad:2012tfa,Chatrchyan:2012xdj}; see Ref.~\cite{Arcadi:2019lka} for a recent review.  The new particle can have different spin assignments. It can be a scalar, a vector or a fermionic state (higher spins have been also considered, see e.g. Ref.~\cite{Criado:2020jkp,Falkowski:2020fsu}). All DM related observables can then be determined by just two basic parameters, namely, the mass of the DM particle and its coupling with the Higgs boson. This makes the comparison between the various searches and the experimental constraints of different nature, collider, direct and indirect detection straightforward.\smallskip 

The ATLAS and CMS collaborations at the LHC already provided  strong upper bounds on the branching fraction of the Higgs boson to invisible decays~\cite{Aaboud:2019rtt,Aaboud:2018sfi,Sirunyan:2018owy,Khachatryan:2016whc}. In the context of the Higgs-portal effective approach, they gave powerful examples of the complementarity among DM searches comparing, for instance in Refs.~\cite{Aaboud:2019rtt,Aaboud:2018sfi,Sirunyan:2018owy,Khachatryan:2016whc}, the upper bound on the DM scattering cross section on nucleons, as measured in direct detection experiments such as XENON1T \cite{Aprile:2018dbl}, with a corresponding constraint obtained by requiring that the coupling of the Higgs boson to DM particles does not generate an  ``invisible" Higgs branching fraction that exceeds the observed experimental limit. A major outcome of this comparison is that searches of invisible Higgs decays can be regarded as a fundamental probe of relatively light DM states, light enough to be produced in the decay of the Higgs boson,  as the sensitivity of direct DM searches is still limited in this mass range.\smallskip

These simple, elegant and rather model-independent models are in practice  effective field theories (EFT). Their theoretical consistency at the mass and energy scales probed by the various experimental searches therefore needs to be ascertained. It is also important to confront these simple models with more complicated but complete scenarios and frameworks in the ultraviolet (UV) regime. Finally, it is also useful to see whether these DM models are compatible with the generation of the correct cosmological relic density as precisely measured by Planck experiment \cite{Aghanim:2018eyx}.\smallskip

Realizations of these  UV complete models require the introduction of additional fields which, when corresponding to light degrees of freedom, can spoil these EFT interpretations. This is particularly the case of spin--1, and to a lesser extent spin--$\frac12$, DM simple effective models which are non-renormalizable and could lead to severe problems such as unitarity violation at the energy scales that are supposed to be experimentally explored.  A brief discussion of these issues has been recently made in Ref.~\cite{Arcadi:2020jqf}, focusing exclusively on the case of a vector DM candidate interacting with the SM Higgs sector. The EFT approach has been confronted  with its simplest UV-completion, namely the one in which the DM state is identified with the stable gauge boson of a dark U(1) gauge symmetry group, spontaneously broken by an extra complex scalar field which develops a vacuum expectation value \cite{Hambye:2008bq,Lebedev:2011iq,Baek:2012se,Farzan:2012hh,Arcadi:2016qoz,Glaus:2019itb}. A new scalar field, that bears mixing with the SM Higgs field, is introduced and the two Higgs mass eigenstates then represent the portal between the DM and the SM particles. We have shown that despite of perturbative unitarity constrains on the mass hierarchy between the two Higgs eigenstates, the EFT limit can be approximately recovered in some areas of the parameter space of the U(1) model.  Hence, one can still use the complementarity between constraints obtained from invisible Higgs decays at colliders and direct DM detection. Nevertheless, the previous statement is valid only if the U(1) dark gauge boson is not required to generate  the observed  cosmological relic density.\smallskip

A more extensive and detailed study of these aspects, considering all spin hypotheses, is presented herein. In the same spirit as in Ref.~\cite{Arcadi:2020jqf}, the predictions in the three realizations of the Higgs-portal effective field theories are compared with those of renormalizable ultraviolet complete models of increasing complexity. First, in the vector DM case, we go beyond the simplest U(1) UV-realization and discuss the possibility of a DM state that belongs to a hidden sector with larger gauge symmetries, namely SU(2) and SU(3). While the phenomenology of the former group differs only slightly from the one of the U(1) case, the dark SU(3) model can lead to two different and interesting possibilities for the DM particle, with a rather rich phenomenology.  We show that not only the EFT limit can be consistently recovered in such UV--complete scenarios but also, and in contrast to the U(1) case, the DM cosmological relic density can be generated under some conditions.\smallskip

The analysis is extended to the spin--$\frac12$  case where the SM Higgs coupling to DM particles is non-renormalizable in the EFT approach. We discuss two possible UV--complete realizations: a first one  is when the Higgs sector is extended by a singlet Higgs field with a vacuum expectation value that generates the mass of the isosinglet DM state, leading to a renormalizable Higgs--DM interaction; a second realization is the so-called singlet--doublet model in which the Higgs sector is kept minimal as in the SM, but the DM sector is enlarged, with the DM particle appearing along with additional charged and neutral leptons with which it mixes.\smallskip 

Finally, we also consider for completeness concrete UV--complete realizations in the spin--$0$ DM case, although the simple EFT approach can be described using renormalizable interactions. We perform updated analyses first in the model in which the Higgs sector is extended by the introduction of a singlet Higgs field and, second, in the so-called inert Higgs doublet model that does not develop a vacuum expectation value and in which the DM scalar comes with another neutral and two charged scalar companions.\smallskip

The phenomenology of all these scenarios is investigated, including the constraints that come from the most recent results in searches of DM signatures and the projected sensitivities of forthcoming experiments. On the collider front, the various  direct and  indirect searches of invisible decays of the Higgs boson at the LHC are considered and the impact of the refined searches to be conducted at the high-luminosity upgrade of the LHC \cite{Cepeda:2019klc} as well as at future high--energy proton colliders with energies up to 100 TeV \cite{deBlas:2019rxi,Contino:2016spe,Tang:2015qga} and high-luminosity $e^+e^-$ colliders \cite{deBlas:2019rxi,Gomez-Ceballos:2013zzn,CEPCStudyGroup:2018ghi} will be summarized. On the astroparticle side, the leading constraints from the XENON1T direct detection experiment \cite{Aprile:2018dbl}, including the dedicated search for light DM states \cite{Aprile:2019xxb} are accounted for and projected sensitivities of the future XENONnT \cite{Aprile:2020vtw} and DARWIN \cite{Aalbers:2016jon} experiments (similar results are also expected from the DarkSide~\cite{Aalseth:2017fik}, LUX--ZEPLIN (LZ)  \cite{Szydagis:2016few} and PANDA-X~\cite{Zhang:2018xdp} experiments) are also included. Whenever relevant, bounds from indirect searches for gamma-ray signals of DM annihilation as well as from the imprint of the latter on the cosmic microwave background (CMB) are also imposed. The correct DM cosmological relic density, as measured by the Planck satellite \cite{Aghanim:2018eyx}, is also required to be achieved through the conventional thermal WIMP paradigm \cite{Drees:1998ra,Bertone:2004pz,Arcadi:2017kky}.\smallskip

The rest paper is structured as follows. In the next section, a brief review of the main phenomenological aspects of the EFT Higgs-portal is presented, followed by a general and  critical discussion of the interpretation of the invisible Higgs searches in terms of the DM-nuclei elastic cross section. In section 3, a first simple renormalizable realization of the Higgs portal is discussed, in which the DM isosinglets are not directly coupled to the SM Higgs but through its  mixing with an additional real Higgs singlet. In the spin--one DM case, as already shown in Ref.~\cite{Arcadi:2020jqf}, this setup offers a viable renormalizable solution that is valid up to very high scales, provided that the new scalar degree of freedom is sufficiently massive. The discussion is then extended to the spin--0 and spin--$\frac12$ possibilities. In section 4, we further extend the UV--complete models  and investigate the cases in which the DM states belong to more complicated dark sectors with additional particle content. More specifically, we consider the inert doublet model for scalar DM, the singlet-doublet model for fermionic DM and extended SU(2) and SU(3) dark gauge symmetry scenarios in the case of vector DM. Finally, a short conclusion is given in section 5. 

\section{The effective Higgs-portal}

The effective field theory approach in the Higgs-portal scenario for dark matter has been formulated and extensively used for spin--0, spin--$\frac12$ as well as for spin--1 DM particles. Assuming CP-conserving interactions, the three different DM spins hypotheses are described by the following effective Lagrangians that involve, besides the SM Higgs doublet field $\phi$, the scalar $s$, the fermion $\chi$ and the vector $V$ DM fields 
\cite{Kim:2006af,Kanemura:2010sh,Djouadi:2011aa,Djouadi:2012zc,Mambrini:2011ik,LopezHonorez:2012kv,Goodman:2010ku,Fox:2011pm,Buckley:2014fba,Abdallah:2015ter,Baglio:2015wcg,Alanne:2017oqj}: 
\begin{eqnarray} 
\label{Lag:DM}  
\!&&\Delta {\cal L}_s = -\frac12 m_s^2 s^2 -\frac14 \lambda_s s^4 - \frac14 \lambda_{Hss}  \phi^\dagger \phi s^2 \;, \nonumber \\ 
\!&&\Delta {\cal L}_V = \frac12 m_V^2 V_\mu V^\mu\! +\! \frac14 \lambda_{V}  (V_\mu V^\mu)^2\! +\! \frac14 \lambda_{HVV}  \phi^\dagger \phi
V_\mu V^\mu , \nonumber \\  
\!  &&\Delta {\cal L}_\chi = - \frac12 m_\chi \bar \chi \chi - \frac14 {\lambda_{H\chi\chi}\over \Lambda} \phi^\dagger \phi \bar \chi \chi \;. 
\end{eqnarray} 
In the expressions above, the self-interaction term $s^4$ in the scalar and the $(V_\mu V^\mu)^2$ term in the vector cases are not essential in the EFT context and can be ignored. In the fermionic case, the DM can be either of Dirac or Majorana types but in the EFT approach the phenomenology is hardly affected by the different nature (in contrast to some specific UV--completions \cite{Arcadi:2019lka}). The presence of operators involving only an even number of DM fields, which is required to ensure that the DM states are stable, is ensured by assuming a discrete $\mathbb{Z}_2$ symmetry.\smallskip

After spontaneous electroweak symmetry breaking, the original Higgs field $\phi$ can be decomposed in the unitary gauge as $\phi^T =(0, v + H)/ \sqrt{2}$, with $v \simeq 246$ GeV being its vacuum expectation value (vev). By substituting the latter expression in the Lagrangians of eq.~(\ref{Lag:DM}), interaction vertices with size $\lambda_{H XX}v$ between the physical Higgs boson $H$ and a pair of DM states $X$, are generated, together with additional contributions to the masses of the DM candidates which then read:
\begin{eqnarray}
M_s^2 &=& m_s^2 + \frac{1}{4}\lambda_{Hss} v^2 \; , \nonumber \\ 
M_V^2 &=& m_V^2 + \frac{1}{4}\lambda_{HVV} v^2 \; , \nonumber \\
M_\chi &=& m_\chi + \frac{1}{4}{\lambda_{H\chi \chi}\over \Lambda} v^2 \;.
\end{eqnarray}
The Higgs-portal models as formulated above are extremely simple and predictive,
featuring only the DM masses $M_X$ and couplings with the $H$ state $\lambda_{HXX}$ as free parameters. They have therefore been extensively studied and became widely used benchmarks for the experimental collaborations~\cite{Aaboud:2019rtt,Aaboud:2018sfi,Sirunyan:2018owy,Khachatryan:2016whc,Aad:2015pla,Chatrchyan:2014tja}. We briefly summarize below, the main phenomenological aspects of such models; for a more detailed discussion we refer, for example, to the recent review given in Ref.~\cite{Arcadi:2019lka}.\smallskip

A major requirement for a viable DM candidate is to generate the correct cosmological relic density as determined with high precision by the Planck experiment~\cite{Aghanim:2018eyx}:
\begin{eqnarray}     
\Omega_{\rm DM} h^2 = 0.1188 \pm 0.0010 \, , \label{eq:omegah}   
\end{eqnarray}      
with $h$ the reduced Hubble constant. Assuming the WIMP paradigm, this requirement will translate into a requirement on the thermally averaged DM annihilation cross section into SM states, $\langle \sigma (XX \to {\rm SM}) v_r \rangle  \propto 1/\Omega_{\rm DM}h^2$ where $v_r$ is the DM relative velocity, and hence on the model parameters $M_X$ and $\lambda_{HXX}$. As the DM relic density is measured with high precision, the viable parameter space will be represented by very narrow bands. We note that in the numerical analyses that we perform in this study,  we take into account the full set of relevant annihilation channels and their diagrams, including possible higher order effects; we use for this purpose the numerical program micrOMEGAs \cite{Belanger:2013oya,Belanger:2014vza}.\smallskip

The DM states that couple only  to the Higgs boson feature spin independent (SI) interactions with detectors such as XENON1T which currently provides the strongest direct detection experimental constraints. These are expressed as upper limits on the elastic scattering cross section over nucleons, typically protons, as a function of the DM mass. These constraints can be simply expressed in the  Higgs-portal  effective scenarios as  
\begin{eqnarray}
&& \sigma^{\rm SI}_{sN} = \frac{\lambda_{Hss}^2}{16 \pi M_H^4} \frac{m_N^4  f_N^2}{ (M_s + m_N)^2} \;,  \nonumber\\
&& \sigma^{\rm SI}_{VN} = \frac{\lambda_{HVV}^2}{16 \pi M_H^4} \frac{m_N^4  f_N^2}{ (M_V + m_N)^2} \;, \nonumber \\
&& \sigma^{\rm SI}_{\chi N} = \frac{\lambda_{H\chi \chi}^2}{4 \pi \Lambda^2 M_H^4} \frac{m_N^4 M_\chi^2  f_N^2}{ (M_\chi + m_N)^2} \;,
\end{eqnarray}
where $m_N$ is the nucleon mass and $f_N \sim 0.3$ parameterizes the Higgs-nucleon interactions. Scalar and vector DM interacting through the Higgs-portal can also provide indirect detection signals. The corresponding constraints are, however, always subdominant in the context of the EFT approach so that we will postpone this aspect to a later discussion.\smallskip

Finally, the SM-like Higgs boson can decay into two DM particles if the latter are light enough, namely when $M_X \, \textless \, \frac12 M_H$.  The corresponding partial decay widths for, respectively, scalar, fermion and vector DM states, can again be simply written as \cite{Djouadi:1994mr}
\begin{eqnarray}
 &&\Gamma_{{\rm inv}} ( H\rightarrow ss) = \frac{\lambda_{Hss}^2 v^2 \beta_s}{64 \pi  M_H}  \;, \nonumber\\
&& \Gamma_{\rm inv} (H \rightarrow V V) = \frac{\lambda^2_{HVV} v^2 M_H^3 \beta_V }{512 \pi   M_{V}^4} \left( 1-4 \frac{M_V^2}{M_H^2}+12\frac{M_V^4}{M_H^4}
\right), \nonumber\\
&&  \Gamma_{\rm inv} (H \rightarrow \chi \chi ) = {\lambda_{H \chi\chi}^2 v^2 M_H \beta_\chi^{3}\over 32 \pi \Lambda^2 }  \;, 
\label{GammaInv}
\end{eqnarray}
where $\beta_X=\sqrt{1-4M_X^2/M_H^2}$ is the DM velocity. The invisible Higgs decay branching ratios are then given by 
\begin{eqnarray}
{\rm BR}(H\to {\rm inv}) =  \Gamma_{\rm inv} (H \rightarrow XX)/\Gamma_H^{\rm tot} 
~~~{\rm with}~~\Gamma_H^{\rm tot} = \Gamma_{\rm inv} \! + \! \Gamma_H^{\rm SM} \, , 
\label{eq:BRinv-def}
\end{eqnarray}
where $\Gamma_H^{\rm SM}=4.07$ MeV is the total width of the 125 GeV Higgs boson as determined in the SM. To precisely evaluate BR($H\to {\rm inv})$, we have adapted the program HDECAY \cite{Djouadi:1997yw,Djouadi:2001fa,Djouadi:2006bz,Djouadi:2018xqq}, which calculates all Higgs partial decay widths and branching ratios including higher order effects,  to incorporate these invisible channels.\smallskip 

Combination of direct and indirect measurements at the LHC, allowed ATLAS and CMS collaborations to set a $95\,\%$ CL upper bound on the invisible Higgs branching fraction
\begin{eqnarray} 
{\rm BR}(H \to {\rm inv}) < 11\% ~~ @~{\rm LHC~Run2}  \, . 
\end{eqnarray} 
This limit can be ultimately improved to BR$(H \to {\rm inv}) < 2.5\%$ at HL-LHC \cite{Cepeda:2019klc}. Prospective studies at future accelerators, including the ILC, CLIC, CEPC, FCC-ee and FCC-hh have shown that sensitivities to invisible branching fractions below the percent level can be reached~\cite{deBlas:2019rxi}.\smallskip

The invisible Higgs partial width $\Gamma(H\to XX)$, and the spin--independent  $X$--proton elastic cross section $\sigma^{\rm SI}_{ X p}$ are both proportional to the squared coupling $\lambda_{HXX}^2$. Consequently, the only unknown parameter in the ratio $r_X=\sigma^{\rm SI}_{X p}/\Gamma (H \to XX)$ is the DM mass $M_X$. We can thus relate the invisible branching fraction of the Higgs to the DM scattering cross section through a very simple expression of the form
\begin{equation}
    \sigma_{Xp}^{\rm SI}=r_X \Gamma(H\rightarrow XX)=r_X {\rm BR}(H \rightarrow \,\mbox{inv})\Gamma_H^{\rm tot}
\end{equation}
where BR$(H \rightarrow\,\mbox{inv})=\Gamma(H\rightarrow XX)/\Gamma_H^{\rm tot}$ with $\Gamma_H^{\rm tot}$  given in eq.~(\ref{eq:BRinv-def}). This highlights the complementarity in the EFT approach that we already mentioned, between the DM constraints obtained at colliders such as the LHC and direct detection experiments such as XENON.\smallskip

\begin{figure}[!h]
\centering
\subfloat{\includegraphics[width=0.43\linewidth]{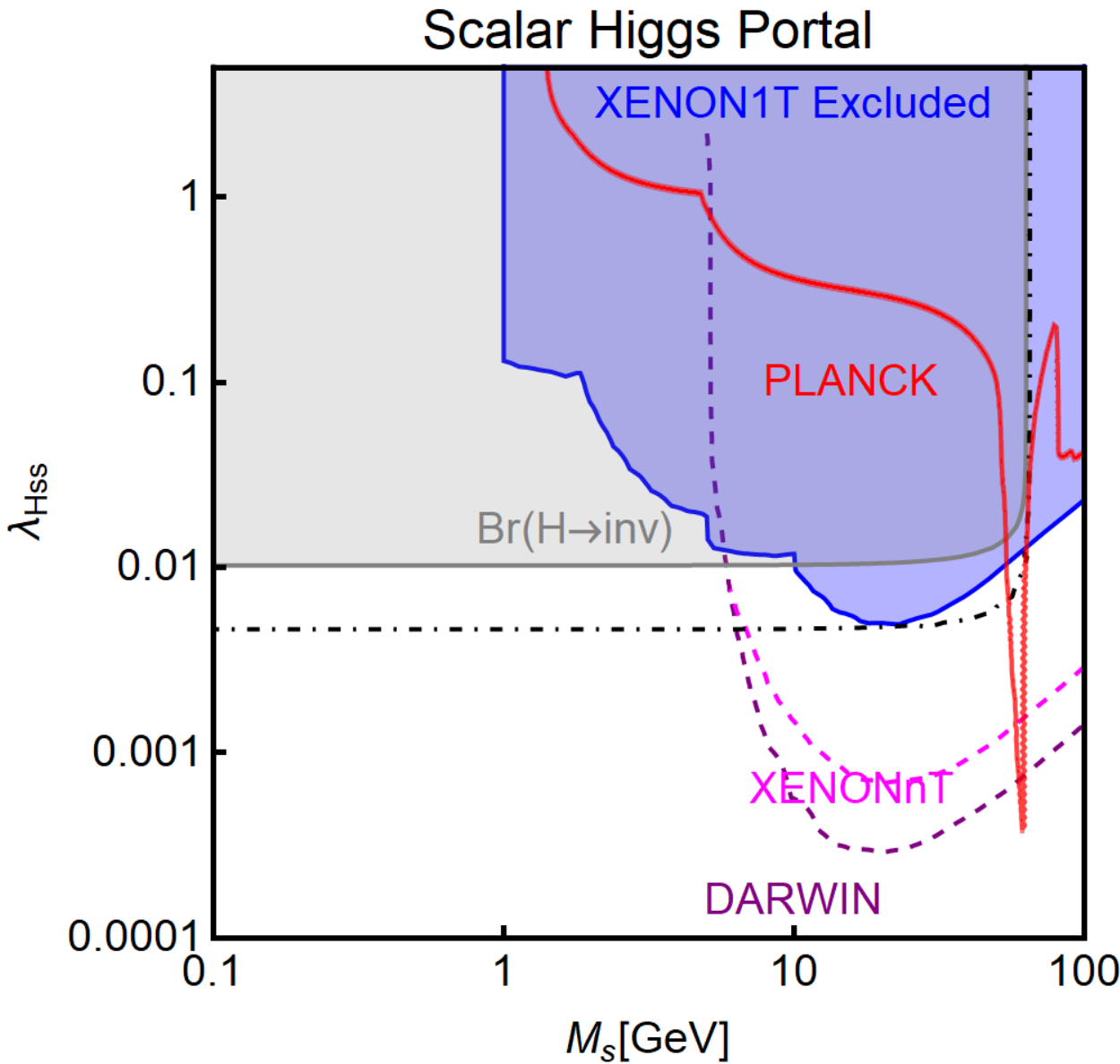}}
\subfloat{\includegraphics[width=0.4\linewidth]{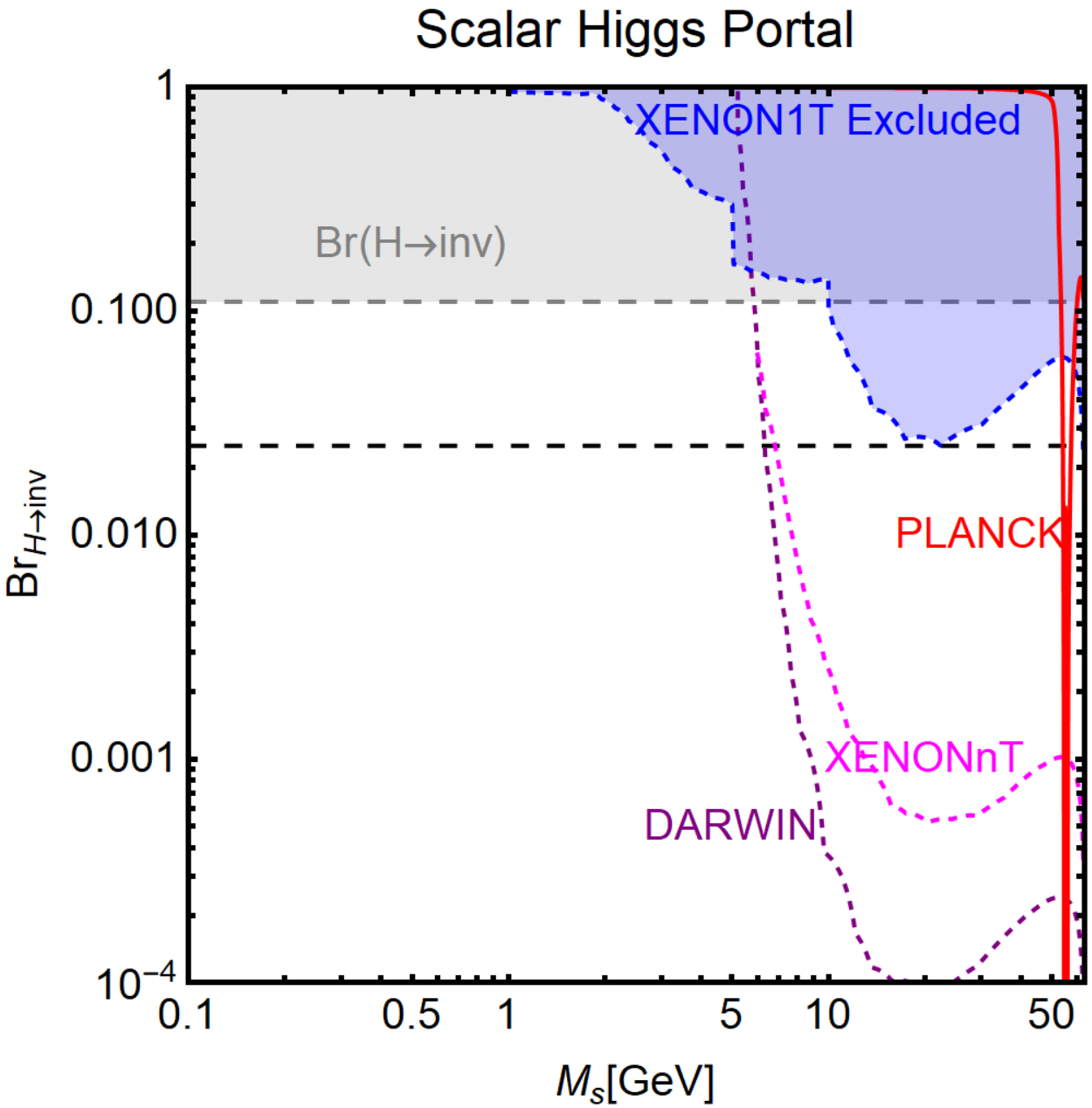}}\\
\subfloat{\includegraphics[width=0.43\linewidth]{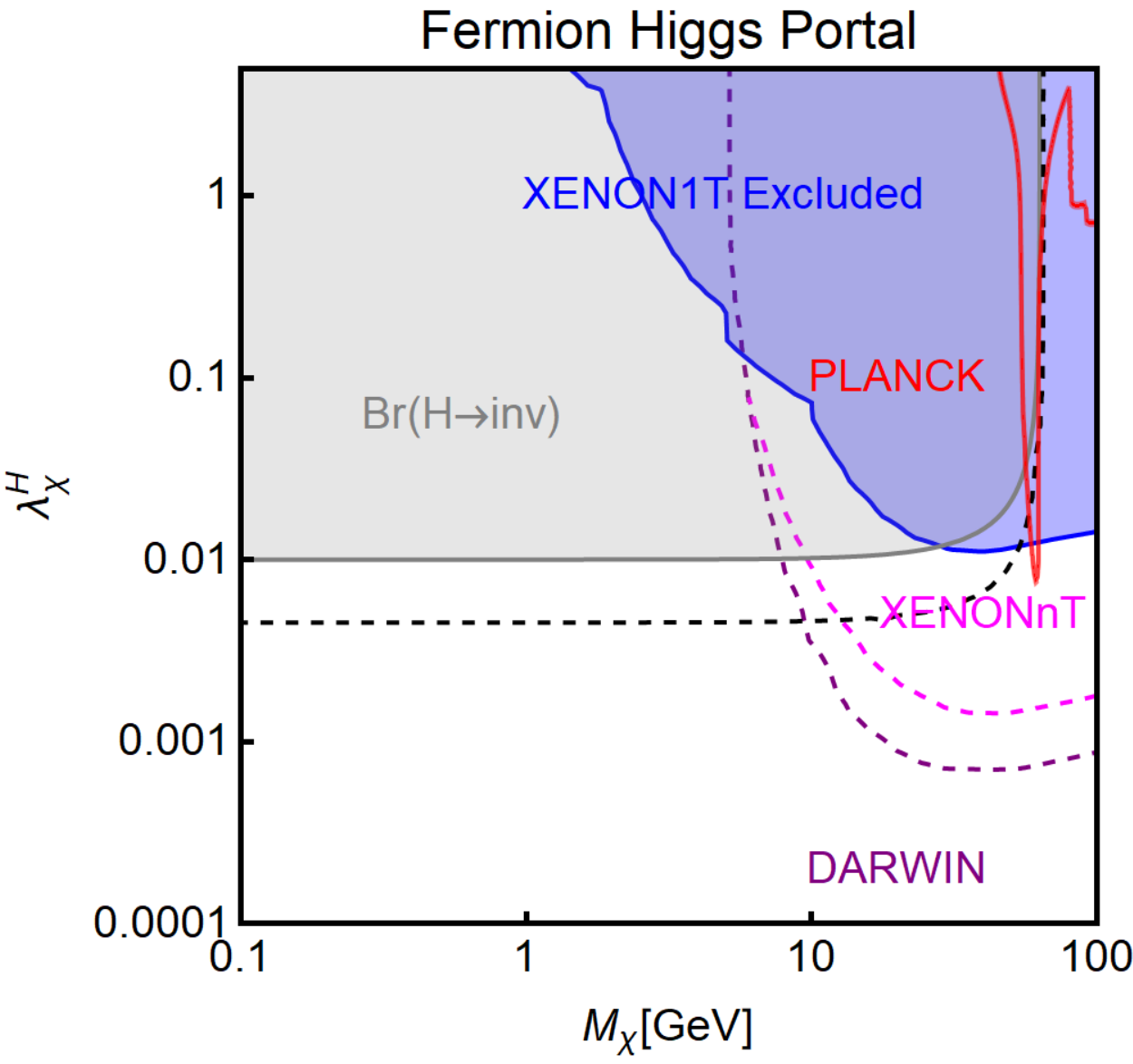}}
\subfloat{\includegraphics[width=0.4\linewidth]{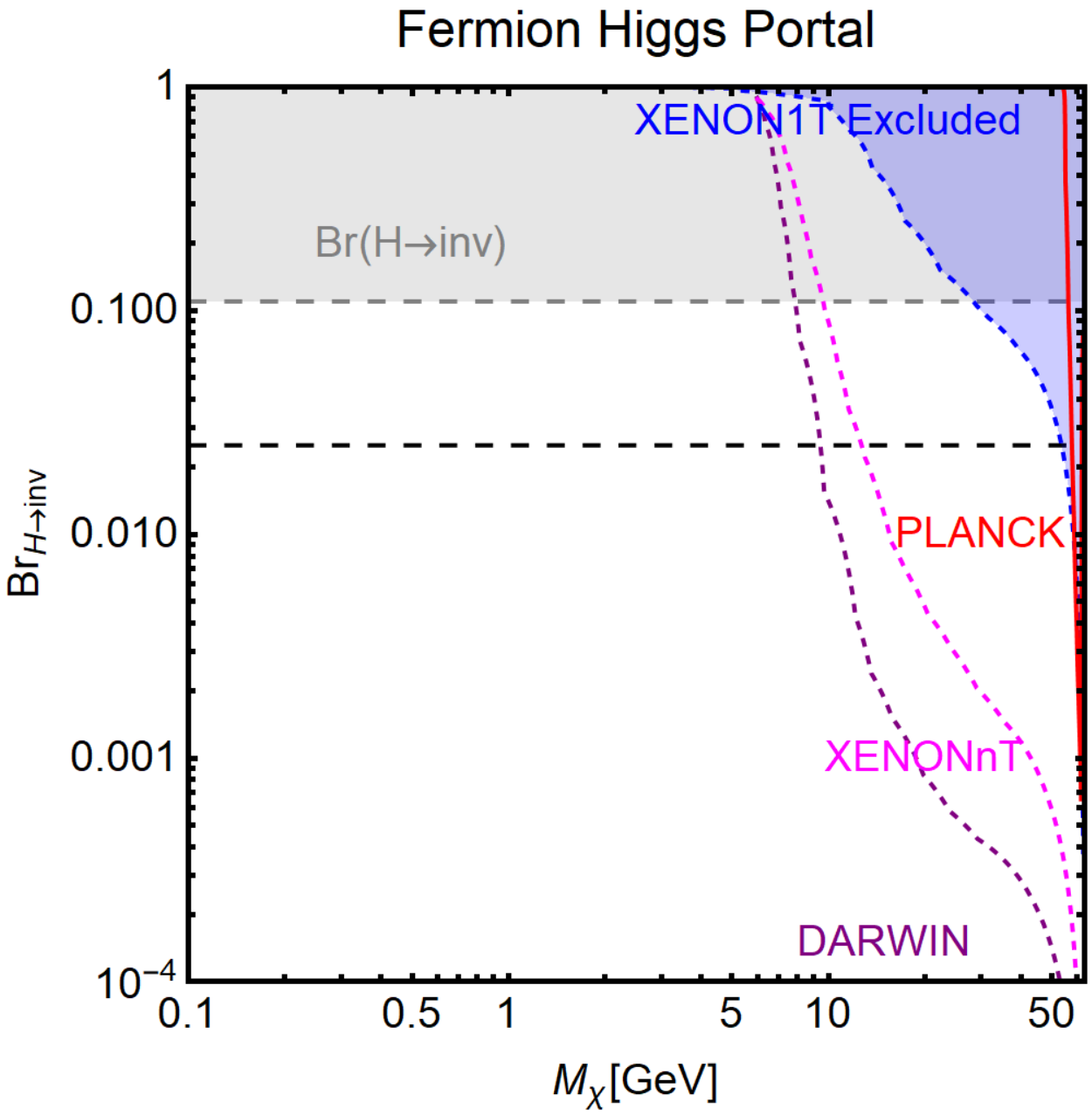}}\\
\subfloat{\includegraphics[width=0.43\linewidth]{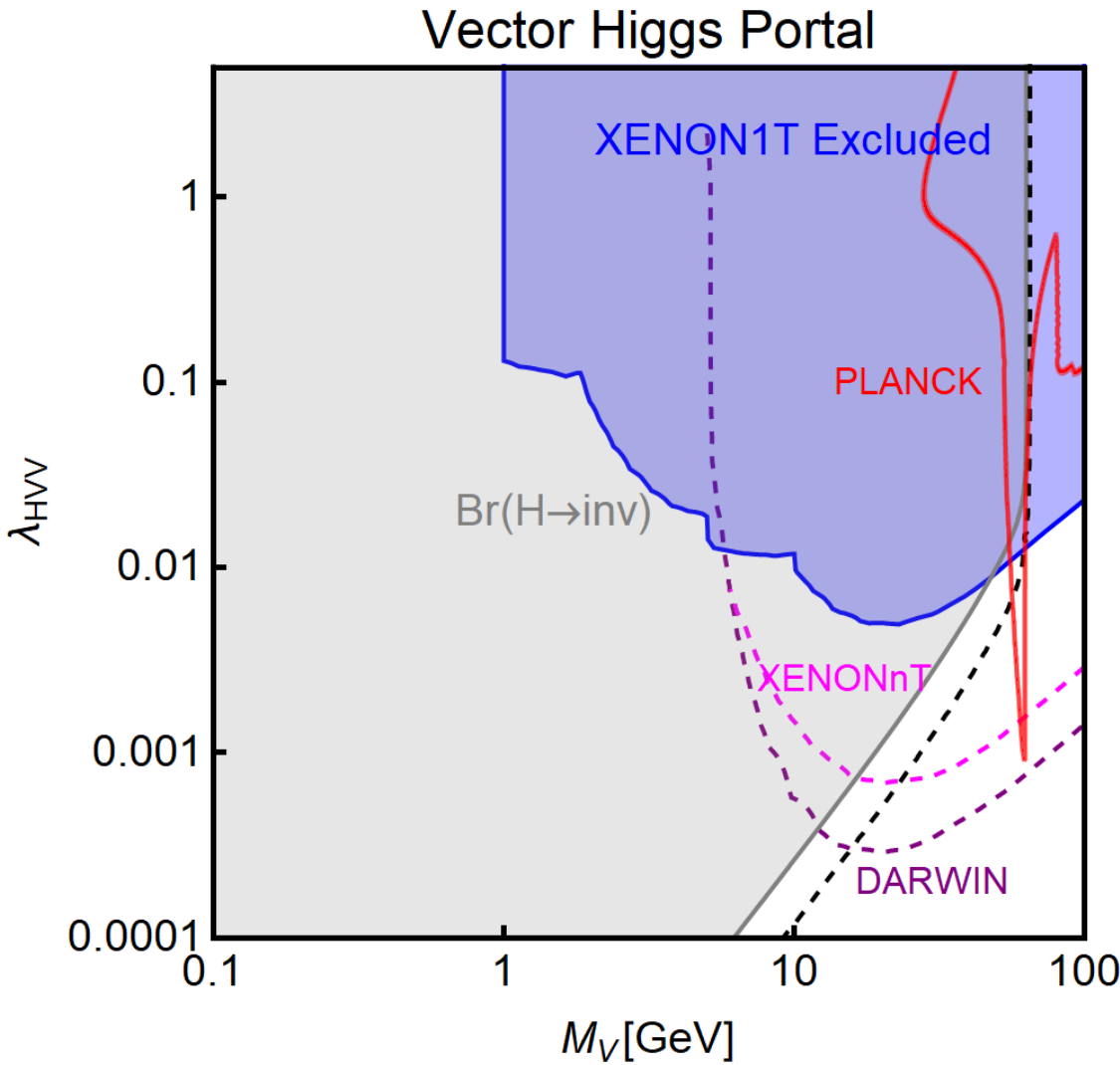}}
\subfloat{\includegraphics[width=0.4\linewidth]{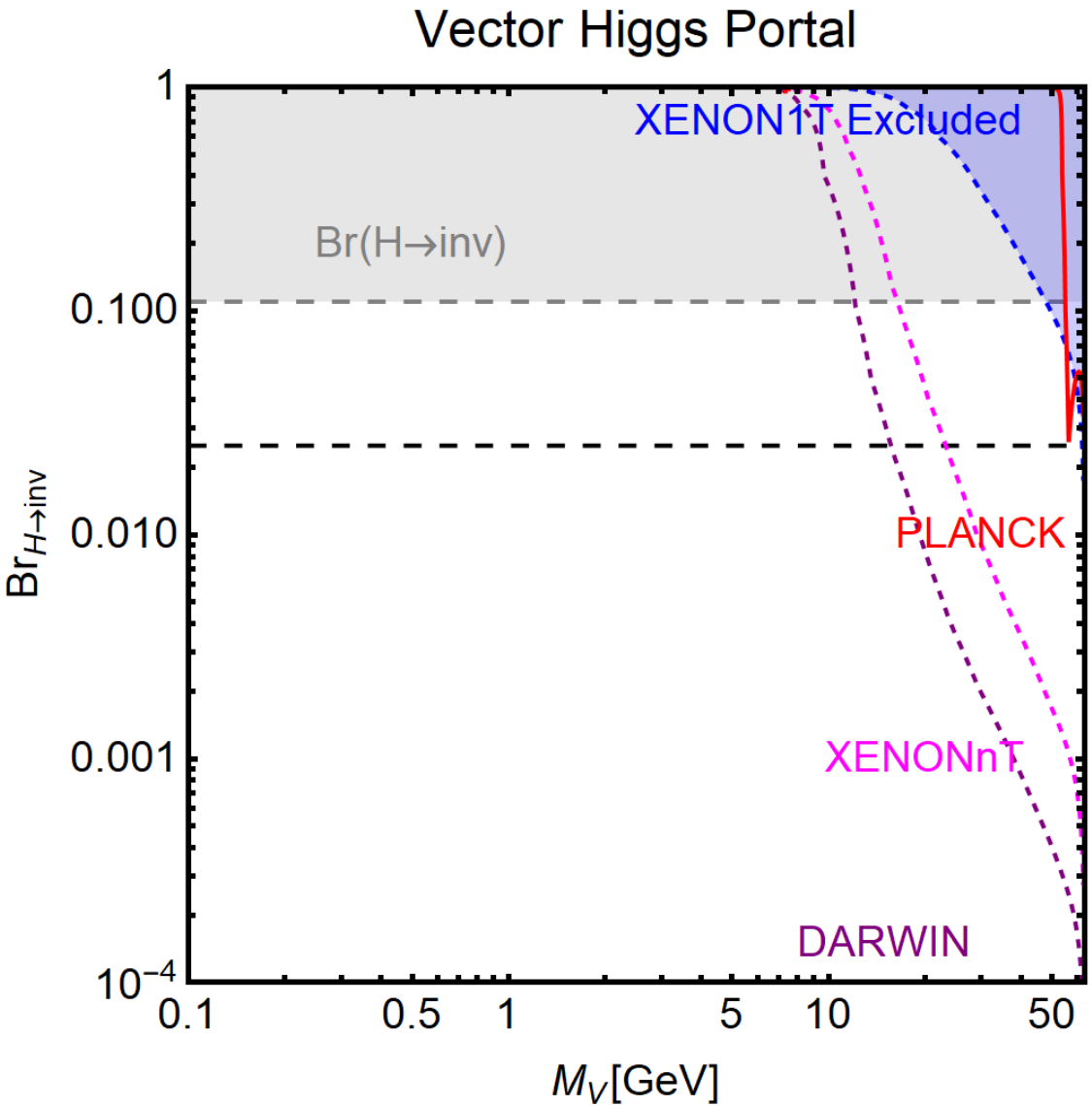}}
\caption{\footnotesize{Constraints in the planes $[M_{X},\lambda_{H XX}]$ (left panels) and $[M_{X}, {\rm BR}(H \to XX)]$ (right panels) for the Higgs-portal DM in the scalar, fermionic  and vector cases. The black contours correspond to the measured DM relic density, the blue and brown regions are excluded by direct detection limits from XENON1T and the invisible Higgs decay branching ratio, respectively. The black contour lines correspond to invisible Higgs branching ratios of $2.5\,\%$ and $0.5\,\%$, while the magenta and purple contours represent the sensitivity reach of the future experiments XENONnT and DARWIN.}}
\label{fig:EFT-all}
\vspace*{-3mm}
\end{figure}

The outcome of our updated numerical analysis is summarized in Fig.~\ref{fig:EFT-all} in which we show the various constraints for the three DM spin cases in two different planes: $[M_X, \lambda_{HXX}]$ and $[M_X, {\rm BR}(H \to XX)]$. In the contours labeled Planck, the DM generates the correct relic density. The blue regions are excluded by DM direct searches: those with masses $M_X \, \gtrsim 5\,\mbox{GeV}$ by XENON1T, supplemented in the  scalar and vector DM cases  by the constraints of the DarkSide--50 experiment \cite{Agnes:2014bvk,Agnes:2018oej} that is effective in the mass range $1\,\mbox{GeV} \leq M_X \leq 5\,\mbox{GeV}$.  Included are the  sensitivity prospects of the forthcoming  LZ/XENONnT (represented by a unique line given the similarities in their expected sensitivities) and DARWIN experiments as well as the regions excluded by the current LHC limit on invisible Higgs decays, BR$(H\! \to \! {\rm inv})<0.11\%$, and the expected sensitivities at future colliders.\smallskip 

From the figure, one sees that for DM masses in the range 5 GeV $<M_X < \frac12 M_H\!=\!62.5$ GeV,  LHC limits from invisible Higgs decays are in general weaker that those obtained  from direct detection. For even smaller DM masses, $M_X \lesssim 5$ GeV, the sensitivity of direct detection experiments is limited by the energy threshold of the detectors and the LHC plays a crucial role in constraining this possibility for sufficiently large DM--Higgs couplings.\smallskip

\begin{figure}[!h]
\centering
\includegraphics[width=0.65\linewidth]{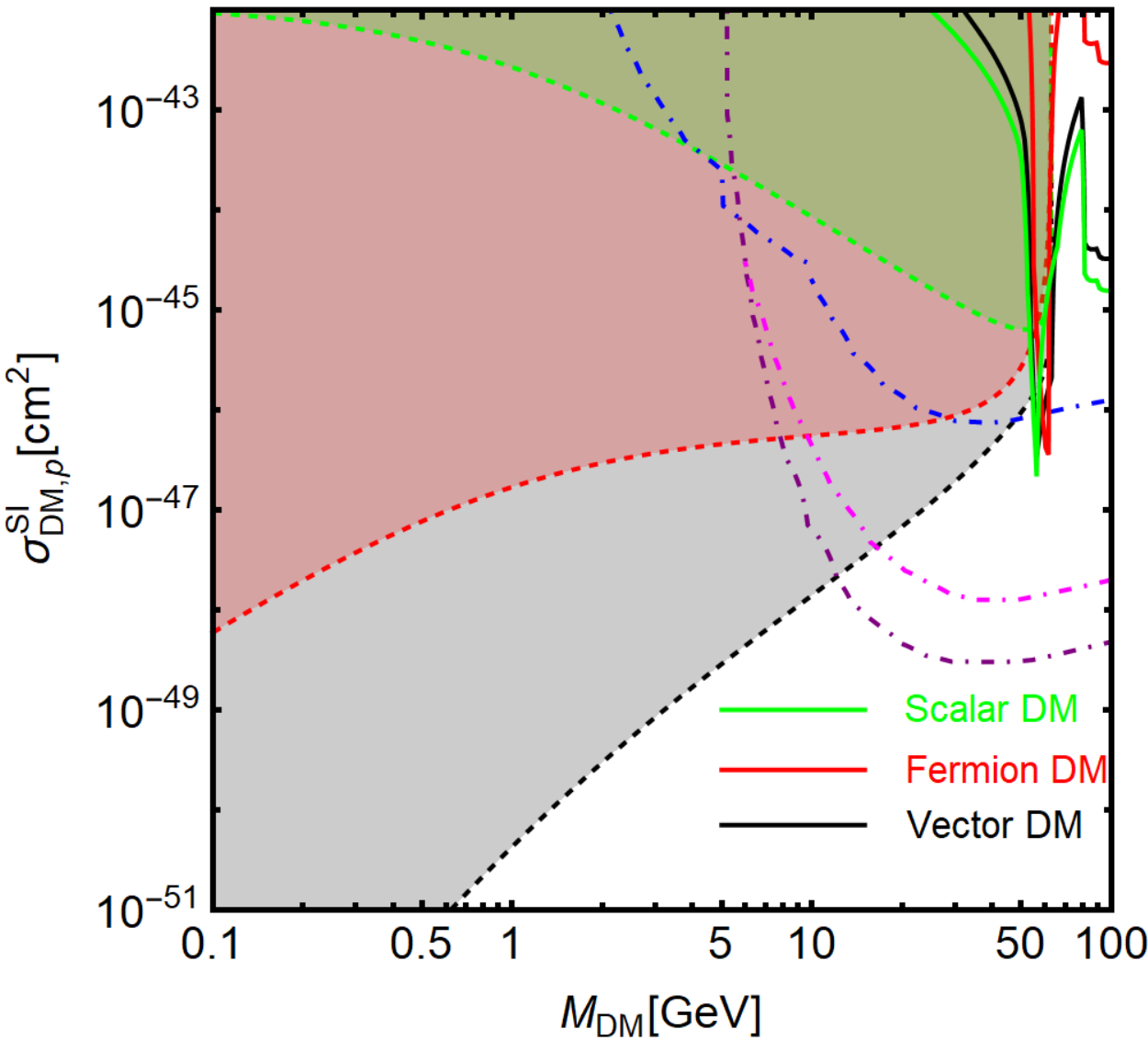}
\vspace*{-.3mm}
\caption{Comparison of the LHC constraints from invisible Higgs decays at the LHC  and limits and sensitivities from direct DM detection experiments. The dot-dashed blue/magenta/purple curves are for the constraints from XENON1T and the prospects from XENONnT and DARWIN respectively. The colored regions above the dashed lines are excluded regions obtained converting the bound $BR(H\rightarrow \mbox{inv})< 11\%$ into an upper bound on the DM scattering cross section. The green/red/black colours refer to scalar/fermionic/vector DM. Finally the solid lines represent the conversion, according the same color code, of the requirement of the correct relic density into, again, a constraint on the DM scattering cross-section.}
\label{LHC-DD-comparison}
\end{figure}

The complementarity between collider and astroparticle physics constraints is further illustrated in Fig.~\ref{LHC-DD-comparison} which summarizes the constraints in the plane $[M_{\rm DM},\sigma_{\rm DM p}^{\rm SI}]$ for the three DM spin hypotheses. For each of them, we have indeed determined, as function of the DM mass, the value of the coupling leading to BR$(H\! \rightarrow \! {\rm inv})\!=\! 0.11$. The values obtained in this way have been used as input for the computation of the DM spin-independent cross section, thus obtaining the green/red/black dashed lines in Fig.~\ref{LHC-DD-comparison}. The colored regions above these lines hence represent values of the DM scattering cross section incompatible with LHC constraints from searches of invisible Higgs decays. This result has been compared with analogous current and projected constraints from dedicated DM experiments, represented by the dot--dashed blue/magenta/purple curves. More precisely, the latter represent the current exclusion from XENON1T (blue dot--dashed line) and expected sensitivities from XENONnT (dot--dashed magenta) and DARWIN (dot--dashed purple). Fig.~\ref{LHC-DD-comparison} finally shows, as solid curves, the spin-independent cross section obtained by fixing the DM--Higgs coupling so that the correct DM relic density is achieved (the curves for scalar and vector DM overlap). Comparing the different lines, it is evident that while direct detection experiments are more sensitive in the high DM mass range, LHC constraints are of utmost importance for the lower mass range. \smallskip

\section{UV-completion through Higgs mixing}
\label{sec:mixing_generics}

\subsection{The singlet Higgs extension of the SM}
 
Although rather simple and elegant, the minimal Higgs portal models suffer from a serious drawback from the theoretical perspective. With the sole exception of the scalar DM case, the coupling between the SM singlet DM and the Higgs bilinear term $|\phi|^2$ is not renormali\-zable. While this is immediately evident in the case of the fermionic DM, the corresponding operator being of dimension-5 and explicitly dependent on the cut-off scale $\Lambda$, the case of a vector DM state is more subtle. Although its coupling to the Higgs boson is of dimension four in energy, it can lead to violation of perturbative unitarity in the Higgs mediated  $VV \rightarrow VV$ self  annihilation process, setting strong conditions on the model parameters \cite{Lebedev:2011iq}. In particular,
an arbitrarily light DM vector particle with arbitrarily high coupling to the Higgs boson is impossible. Such a bound, consequently, impacts directly the region which can be probed by searches of invisible Higgs decays.\smallskip

Hence, both the fermionic and vector Higgs-portal scenarios need the presence of additional degrees of freedom, to be viable at arbitrary energy scales. These extra degrees of freedom might have a significant impact on the astroparticle and collider phenomenology of the DM states \cite{Baek:2012se,Baek:2014jga}, which cannot be described by the EFT approach anymore.\smallskip

In this section, we analyze one of the most economical possibilities of completing the effective Higgs-portal, namely through the introduction of an additional scalar field with mass mixing with the SM Higgs boson. Assuming the additional scalar field to be an ${\rm SU(2)_L \times U(1)_Y}$ singlet, one can generate a renormalizable coupling with the isosinglet DM candidate of any spin. We briefly review below the essential features of such a scenario, following mostly the conventions of Ref.~\cite{Falkowski:2015iwa}. \smallskip

The simplest scalar potential accounting for mass mixing between the SM Higgs field $\phi$ and an additional scalar degree of freedom $S$ is given by
\begin{equation}
\label{eq:general_potential}
    V(\phi,S)=\frac{\lambda_H}{4}\phi^4+\frac{\lambda_{HS}}{4}\phi^2 S^2+\frac{\lambda_S}{4}S^4+\frac{1}{2}\mu_H^2 \phi^2+\frac{1}{2}\mu_S^2 S^2 \, . 
\end{equation}

After spontaneous electroweak symmetry breaking, the fields $\phi$ and $S$ acquire vevs, labeled $v$ and $\omega$ respectively, which read
\begin{equation}
    v^2 \equiv \frac{2 \lambda_{HS}\mu_S^2-4 \lambda_S \mu_H^2}{4\lambda_H \lambda_S-\lambda_{HS}^2},\,\,\,\,\omega^2 \equiv \frac{2 \lambda_{HS}\mu_H^2-4 \lambda_H \mu_S^2}{4\lambda_H \lambda_S-\lambda_{HS}^2} \, .
\end{equation}

In this setup, the following mass matrix for the scalar states is generated
\begin{equation}
    {\mathcal{M}}^2=\left(
    \begin{array}{cc}
    2 \lambda_H v^2 & \lambda_{HS}v \omega \\     \lambda_{HS}v \omega & 2 \lambda_S \omega^2
    \end{array}
    \right) \, , 
\end{equation}
and, under the assumption that the couplings of the scalar potential are real, one obtains  from the requirement that $\omega^2,v^2 >0$, the following constraints \begin{equation}
    \lambda_H > \frac{\lambda_{HS}^2}{4 \lambda_S},\,\,\,\,\lambda_S>0 \, ,
\end{equation}
on the quartic couplings. The matrix ${\mathcal{M}}^2$ can be diagonalized through an orthogonal transformation $O^T {\mathcal{M}}^2 O=\mbox{diag}\left(M_{H_1}^2,M_{H_2}^2\right)$ where
\begin{equation}
    O=\left(
    \begin{array}{cc}
    \cos\theta & \sin\theta  \\
    -\sin\theta & \cos\theta 
    \end{array}
    \right)
~~~{\rm with}~~
\tan 2 \theta = \frac{\lambda_{HS}v \omega}{\lambda_S \omega^2-\lambda_H v^2} \, , 
\end{equation}
and the masses of the two Higgs eigenstates are then given by
\begin{equation}
    M_{H_1,H_2}^2=\lambda_H v^2+\lambda_S \omega^2 \mp \frac{\lambda_S \omega^2-\lambda_H v^2}{\cos 2 \theta} \, , 
\end{equation}
where we identify $H_1$ with the 125 GeV SM-like Higgs boson observed at the LHC. For our phenomenological analysis, we will adopt the set $(M_{H_2},\sin\theta,\lambda_{HS})$ as free parameters. The other two quartic couplings $\lambda_H$, $\lambda_S$ can be written, as functions of the latter, as
\begin{align}
    & \lambda_H=\frac{M_{H_1}^2}{2v^2}+\sin^2 \theta \frac{M_{H_2}^2-M_{H_1}^2}{2v^2} \, , \\
    & \lambda_S=\frac{2\lambda_{HS}^2}{\sin^2 2 \theta}\frac{v^2}{M_{H_2}^2-M_{H_1}^2}\left(\frac{M_{H_2}^2}{M_{H_2}^2-M_{H_1}^2}-\sin^2 \theta\right) \, . 
\end{align}

The magnitude of these quartic couplings is constrained by perturbative unitarity in the $H_i H_i \rightarrow H_j H_j$ annihilation processes which require \cite{Chen:2014ask}
\begin{equation}
    \lambda_i \leq \mathcal{O}\left(4 \pi/ 3\right) \, . 
\end{equation}
This type of constraint will be particularly relevant in the case of vector DM particles.
The couplings of the $H_{1}$ and $H_2$ states with SM particles can be expressed in terms of the set of free parameters that we have adopted, as
\begin{align}
    & \mathcal{L}_{\rm scalar,SM}=\frac{H_1 \cos\theta + H_2 \sin\theta}{v}\left(2 M_W^2 W^{+}_{\mu}W^{-\,\mu}+M_Z^2 Z_\mu Z^\mu -m_f \bar f f\right) \, , 
\label{eq:Lag-mix-SM}
\end{align}
while the trilinear scalar couplings relevant for DM phenomenology are
\begin{equation}
    \mathcal{L}_{\rm scalar,trilinear}=-\frac{\kappa_{111}}{2}v H_1^3-\frac{\kappa_{112}}{2}H_1^2 H_2 v \sin\theta-\frac{\kappa_{221}}{2}H_1 H_2^2 v \cos\theta-\frac{\kappa_{222}}{2}H_2^3 v \, , 
\end{equation}
with the various $\kappa$ factors given by
\begin{align}
\label{eq:trilinear}
    & \kappa_{111}=\frac{M_{H_1}^2}{v^2 \cos\theta}\left(\cos^4 \theta+\sin^2\theta \frac{\lambda_{HS}v^2}{M_{H_1}^2-M_{H_2}^2} \right) , \nonumber\\ 
    & \kappa_{112}=\frac{2 M_{H_1}^2+M_{H_2}^2}{v^2}\left(\cos^2 \theta +\frac{\lambda_{HS}v^2}{M_{H_2}^2-M_{H_1}^2}\right),  \nonumber\\
    & \kappa_{221}=\frac{2 M_{H_2}^2+M_{H_1}^2}{v^2}\left(\sin^2 \theta +\frac{\lambda_{HS}v^2}{M_{H_1}^2-M_{H_2}^2}\right) , \nonumber\\
    & \kappa_{222}=\frac{M_{H_2}^2}{v^2 \sin\theta}\left(\sin^4 \theta+\cos^2\theta \frac{\lambda_{HS}v^2}{M_{H_2}^2-M_{H_1}^2} \right) \, . 
\end{align}
The extended Higgs sector summarized above will act as a double--portal for the DM particles. The interactions between the two sectors are summarized in the next subsection. 

\subsection{DM couplings with the Higgs sector}

In this section, the couplings of the DM particles with different spins to the double-portal  Higgs sector introduced above are discussed, starting with the simplest and renormalizable case of a scalar DM that is also included for the sake of completeness.

\subsubsection{Scalar Dark Matter}

The scalar DM effective the model is not affected by the previously mentioned theoretical problems, at least at the scales relevant for studies at current and future high--energy colliders. We will nevertheless consider the case in which the theory is extended with an additional scalar Higgs degree of freedom. Assuming that both the scalar DM $s$ particle and the additional Higgs mediator $S$ are odd under a suitable discrete symmetry, one can write the following simple interaction Lagrangian
\begin{equation}
\mathcal{L}_{s}=\lambda_{s}^H |\phi|^2 s^2
               +\lambda_{s}^S s^2 |S|^2 , 
\end{equation}
where, for simplicity, the quartic self interaction terms for the scalar fields are omitted. As can be seen, a direct coupling between the DM and the Higgs doublet $\phi$ cannot be a priori forbidden. We nevertheless set it to zero, i.e. $\lambda_{X}^H=0$, in order to reduce the number of free parameters. After the breaking of the electroweak symmetry, the coupling of the DM particle with the physical states $H_1$ and $H_2$ will be simply
\begin{align}
    & \mathcal{L}_s=\lambda_s^S \omega (-\sin\theta H_1 +\cos\theta H_2)s^2\nonumber\\
    & + \lambda_s^S \cos^2 \theta H_1^2 s^2 
 -2 \sin\theta \cos\theta \lambda_s^S H_1 H_2 s^2+\lambda_s^S \cos^2 \theta H_2^2
 s^2 \, , 
\end{align}
where the expression of the vev $\omega$ has been given in the previous subsection. Notice that the previous Lagrangians describe a real scalar DM state. One could nevertheless write analogous Lagrangians for a complex scalar state just by replacing $s^2 \rightarrow |s|^2$ and considering a global U(1) symmetry, rather than a discrete $Z_2$, to stabilize the DM particle. We also remark that one can accommodate a variant of this scenario in which the DM is a pseudo-goldstone boson of the aforementioned global U(1) symmetry \cite{Mambrini:2015nza,Huitu:2018gbc,Abe:2020iph}. Despite of its interesting phenomenology, especially for what concerns DM direct detection \cite{Karamitros:2019ewv,Arina:2019tib,Alanne:2020jwx,Glaus:2020ihj}, we will not consider this scenario here.

\subsubsection{Fermionic Dark Matter}

The isosinglet fermionic DM case can, in principle, be coupled with an isosinglet scalar with arbitrary coupling strength. We will nevertheless assume that the DM mass is originating from the vacuum expectation value $\omega$ of the new Higgs field. We will then consider a Dirac fermion DM state described by the following Lagrangian,
\begin{equation}
    \mathcal{L}_{\chi}=-y_\chi \bar \chi \chi S \, , 
\end{equation}
where the coupling is related to the DM mass by
\begin{equation}
    y_\chi={M_\chi}/{\omega} \, . 
\end{equation}
By using the two relations
\begin{align}
& \omega^2=\frac{M_{H_1}^2 \sin^2 \theta +M_{H_2}^2 \cos^2 \theta}{2 \lambda_S} \, , \nonumber\\
& \lambda_S=\frac{2\lambda_{HS}^2}{\sin^2 2 \theta}\frac{v^2}{M_{H_2}^2-M_{H_1}^2}\left(\frac{M_{H_2}^2}{M_{H_2}^2-M_{H_1}^2}-\sin^2 \theta\right) \, , 
\end{align}
one can restrict the set of new free parameters of the model to  $(M_\chi,M_{H_2},\sin\theta,\lambda_{HS})$.

Notice that, in the case of fermionic Dark Matter, it would potentially interesting to consider as well the possibility of a CP violating scalar sector. As well known this would lead, from a phenomenological perspective, to very interesting scenarios, see e.g. \cite{Ipek:2014gua,Goncalves:2016iyg,Bauer:2017ota,Tunney:2017yfp,Abe:2018bpo,Arcadi:2017wqi,Arcadi:2018pfo,Arcadi:2019lka}. For simplicity, we will not consider here this kind of scenarios.

\subsubsection{Vector Dark Matter}

In order to properly address the bounds from perturbative unitarity in $VV \rightarrow VV$
scattering through Higgs boson exchange, we introduce an explicit mechanism for the generation of the DM mass based on the spontaneous breaking of a dark gauge symmetry.In this subsection, we consider the simplest case, namely the dark U(1) scenario in which the DM content is minimal; the discussion of more complicated scenarios will be postponed to the next section. The dark U(1) model has been already studied in e.g. Refs.~\cite{Arcadi:2020jqf,Lebedev:2011iq,Baek:2012se,Gross:2015cwa}, its main features are summarized below.  The newly introduced fields are described by a Lagrangian of the type
\begin{equation}
    \mathcal{L}_{\rm hidden}=-\frac{1}{4}V_{\mu \nu}V^{\mu \nu}+{\left(D^\mu S \right)}^{\dagger}\left(D_\mu S\right)-V(S,\phi) \, , 
\end{equation}
where $V_{\mu \nu}=\partial_\mu V_\nu-\partial_\nu V_\mu$ is the field strength of the DM and $D_\mu=\partial_\mu +i \tilde{g} V_\mu$ with $\tilde{g}$ being the new gauge coupling. The potential $V(S,\phi)$ has the same form as the one given in eq.~(\ref{eq:general_potential}). Upon spontaneous breaking of the U(1) gauge symmetry, the DM particle acquires a mass term $M_V= \tilde{g} \omega /2$. Its interactions are described by the following Lagrangian
\begin{equation}
    \mathcal{L}_{\rm DM}=\frac{\tilde{g}^2}{4}\omega\rho V_\mu V^\mu+\frac{\tilde{g}^2}{8}\rho^2 V_\mu V^\mu=\frac{\tilde{g}}{2}M_V \rho V_\mu V^\mu+\frac{\tilde{g}^2}{8}\rho^2 V_\mu V^\mu \, , 
\end{equation}
where $\rho$ is the physical degree of freedom of $S=\frac{1}{\sqrt{2}}\left(\omega + \rho\right)$. The complete interaction Lagrangian that is relevant for DM phenomenology can be then written as
\begin{align}
    & \mathcal{L}=\frac{\tilde{g}M_V}{2}\left(-H_1 \sin\theta + H_2 \cos \theta\right)V_\mu V^\mu \nonumber\\
    & +\frac{\tilde{g}^2}{8}\left(H_1^2 \sin^2 \theta -2 H_1 H_2 \sin\theta \cos\theta+H_2^2 \cos^2 \theta\right)V_\mu V^\mu \nonumber\\
    & +\frac{H_1 \cos\theta+H_2 \sin\theta}{v}\left(2 M_W^2 W_\mu^{+}W^{\mu -}+M_Z^2 Z_\mu Z^\mu -m_f \bar f f\right) \nonumber\\
    & -\frac{\kappa_{111}}{2}v H_1^3-\frac{\kappa_{112}}{2}H_1^2 H_2 v \sin\theta-\frac{\kappa_{221}}{2}H_1 H_2^2 v \cos\theta-\frac{\kappa_{222}}{2}H_2^3 v \, , 
\end{align}
where the expressions of the effective couplings $\kappa_{iij}$ are given in eq.~(\ref{eq:trilinear}).   The portal coupling $\lambda_{HS}$ can be traded with the dark gauge coupling $\tilde{g}$ using the following relation
\begin{equation}
\label{eq:DU1quartic}
    \lambda_{HS}=\tilde{g}\sin 2 \theta \frac{M_{H_2}^2-M_{H_1}^2}{4 v M_V} \, , 
\end{equation}
and, hence, the set of free parameters will be composed of $\left(M_V, \tilde{g}, \sin\theta ,M_{H_2} \right)$. 
\subsection{Dark Matter phenomenology}

When comparing dedicated astrophysical DM searches and searches of invisible Higgs decays at colliders, re-expressed as limits on the DM elastic scattering cross section on nucleons, some important aspects should not be overlooked.\smallskip 

$i)$ When more realistic models than  the effective one are considered, DM observables might be affected by the presence of additional degrees of freedom so that the results obtained in terms of the effective Higgs-portal might not be straightforwardly applied to these more realistic and sophisticated scenarios.\smallskip 

$ii)$ The sensitivity of direct detection experiments diminishes as the DM mass drops below values of the order of 10 GeV, because of the energy threshold of the detectors, in particular those based on Xenon. On the contrary, searches for invisible Higgs decays can probe arbitrary low DM particle masses.\smallskip 
 
$iii)$ The limits from direct detection experiments crucially depend on astrophysical inputs like the local DM density and, in particular, are determined under the assumption that the particle scattering with nuclei represents the total DM component of the universe with a relic density compatible with the experimental determination.  In contrast, searches at colliders probe simply the coupling of the DM candidate that is  possibly produced in Higgs boson decays, without any assumption on the DM local density and, consequently, on its abundance. Considering, on the contrary, a specific mechanism for DM production in the early universe, might strongly limit or even completely rule out the DM mass range which can be explored through searches of invisible Higgs decays.\smallskip  

In this section, we will review the phenomenology of the DM particles in the UV-complete scenarios introduced previously, taking into account the first two aspects above. A discussion that includes the third point will be relegated to the next section. \smallskip

In the case of scalar, fermionic, and vector particles, the DM scattering cross sections on the proton are respectively given by
\begin{align}
    & \sigma_{s p}^{\rm SI}=\frac{\mu_{s p}^2}{16\pi}\frac{(\lambda_{s}^S)^2 \sin^2 \theta \cos^2 \theta m_p^2 \omega^2}{v^2 M_s^2}\mathcal{F}\left(M_s,M_{H_1},M_{H_2},v_s\right)f_p^2 \, , \nonumber\\
    & \sigma_{\chi p}^{\rm SI}=\frac{\mu_{\chi p}^2}{\pi}\frac{y_{\chi}^2 \sin^2 \theta \cos^2 \theta m_p^2}{v^2}\mathcal{F}\left(M_\chi,M_{H_1},M_{H_2},v_\chi\right)f_p^2 \, , \nonumber\\
    & \sigma_{V p}^{\rm SI}=\frac{\mu_{V p}^2}{4 \pi}\frac{{\tilde{g}}^2 \sin^2 \theta \cos^2 \theta m_p^2}{v^2}\mathcal{F}\left(M_V,M_{H_1},M_{H_2},v_V\right)f_p^2 \, ,
\end{align}
where $m_p$ is the proton mass, $f_p \approx 0.3$ the nucleon form factor while $\mu_{Xp} = {M_X m_p}/{(M_X+m_p)}$ is the DM--proton reduced mass. The impact of the additional degree of freedom, i.e. $H_2$ in the extended Higgs case discussed here, is encoded in the function $\mathcal{F}$ whose general expression is given by \cite{Baek:2014jga}
\begin{equation}
\mathcal{F}\!=\!\frac{1}{4 M_X^2 v_X^2}\left[ \sum_i \left(\frac{1}{M_{H_i}^2}\!-\!\frac{1}{4 M_X^2 v_X^2\!+\!M_{H_i}^2} \right)\!-\!\frac{2}{\left(M_{H_2}^2\!-\!M_{H_1}^2\right)}\sum_i (-1)^{i\!-\!1}\log\left(1\!+\!\frac{4 M_X^2 v_X^2}{M_{H_i}^2}\right) \right] \, ,
\end{equation}
with $v_X$ being the DM velocity in the laboratory frame. In the limit $M_{H_2} \gg 2 M_X v_X$, the function $\mathcal{F}$ simplifies to the known expression
\begin{equation}
    \mathcal{F}\simeq {\left(\frac{1}{M_{H_1}^2}-\frac{1}{M_{H_2}^2} \right)}^2 \, . 
\end{equation}
As mentioned in the previous section the DM scattering cross section can be directly related to the invisible Higgs branching fraction. For the three DM spins, we can hence write
\begin{align}
  &  \left. \sigma_{s p}^{\rm SI}\right \vert_{\rm mix}=2 \cos^2 \theta \frac{\mu_{\chi p}^2 M_{H_1}}{M^2_{s}\beta_{s H_1}}{\left(\frac{m_p^2}{v^2}\right)}\, \mathcal{F}\, {\rm BR}(H_1 \rightarrow \mbox{inv})\Gamma_{H_1}^{\rm tot}f_p^2 \, ,\nonumber\\  
  &  \left. \sigma_{\chi p}^{\rm SI}\right \vert_{\rm mix}=8 \cos^2 \theta \frac{\mu_{\chi p}^2}{M_{H_1}\beta_{\chi H_1}^3}{\left(\frac{m_p^2}{v^2}\right)}\, \mathcal{F}\, {\rm BR}(H_1 \rightarrow \mbox{inv})\Gamma_{H_1}^{\rm tot}f_p^2 \, ,  \nonumber\\
   & \sigma_{Vp}^{\rm SI}|_{\rm mix}=32 \cos^2 \theta  \mu_{Vp}^2 \frac{M_V^2}{M_{H_1}^3 \beta_{VH_1}\eta_{VH_1}} \mathcal{F} {\rm BR}(H_1 \rightarrow \mbox{inv})\Gamma_{H_1}^{\rm tot}  \frac{m_p^2}{v^2}|f_p|^2 \, , 
\end{align}
where $\beta_{XH_1}\!=\!\sqrt{1\!-\!4 {M_X^2}/{M_{H_1}^2}}$ and $\eta_{XH_1}\!=\! \left(1\!-\!{4 M_V^2}/{M_{H_1}^2}\!+\!12 {M_V^4}/{M_{H_1}^4}\right)$.
By compa\-ring the latter expressions with the corresponding ones obtained for the EFT Higgs portals:
\begin{align}
    & \left. \sigma_{s p}^{\rm SI}\right \vert_{\rm EFT}=2 \frac{\mu_{s p}^2 M_{H}}{M_{s}^2\beta_{\chi H}^3}{\left(\frac{m_p^2}{v^2}\right)}{\rm BR}(H_1 \rightarrow \mbox{inv})\Gamma_{H_1}^{\rm tot}f_p^2\, , \nonumber\\
    & \left. \sigma_{\chi p}^{\rm SI}\right \vert_{\rm EFT}=8 \frac{\mu_{\chi p}^2}{M_{H}\beta_{\chi H}^3}{\left(\frac{m_p^2}{v^2}\right)}{\rm BR}(H_1 \rightarrow \mbox{inv})\Gamma_{H_1}^{\rm tot}f_p^2 \,\nonumber\\
     & \sigma_{Vp}^{\rm SI}|_{\rm EFT}=32  \mu_{Vp}^2 \frac{M_V^2}{M_H^3 \beta_{VH} \eta_{VH}} {\rm BR}(H_1 \rightarrow \mbox{inv})\Gamma_{H_1}^{\rm tot} \frac{1}{M_H^4}\frac{m_p^2}{v^2}|f_p|^2 \,
\end{align}
one can see that the prediction of the EFT approach and the one of the renormalizable Higgs-portal model coincide in the limit $\cos^2 \theta \mathcal{F}\rightarrow 1$. This means that, for a sufficiently heavy $H_2$ state which will then decouple from phenomenology, the EFT limit is recovered. This possibility is, nevertheless, potentially troublesome in the case of vector DM since, as already noted, an excessive hierarchy between the DM mass and the mass of the $H_2$ state is forbidden by perturbative unitarity.\smallskip

Following a strategy analogous to the one discussed in Ref.~\cite{Arcadi:2020jqf}, and  remaining agnostic for the time being on the production mechanism of the DM in the early universe, a parameter scan of the three DM models discussed above is carried out considering the following ranges for the three input parameters of the Higgs sector
\begin{align}
     M_{H_2} \in \left[M_{H_1},3\,\mbox{TeV}\right] , \
     \sin\theta \in \left[10^{-4},0.3\right] ,\
     \lambda_{HS} \in \left[10^{-2},10\right] , \
     \end{align}
and, for the following additional parameters,
\begin{align}
     & M_s \in \left[0.1,1000\right]\,\mbox{GeV}\ , \ \  \lambda_X^S \in \left[10^{-4},1\right] \, , \nonumber \\
     & M_\chi \in \left[0.1,1000\right]\,\mbox{GeV} \, , \nonumber \\
     & M_V \in \left[0.1,1000\right]\,\mbox{GeV} \ , \ \ \tilde{g} \in \left[10^{-2},10\right] \, , 
  \end{align}
in the scalar, fermionic and vector DM cases, respectively. Only the model points satisfying theoretical constraints such as the perturbativity of the coupling of the DM to the Higgs boson as well as perturbative unitarity in the $H_i H_i \rightarrow H_j H_j$ processes as discussed previously, have been kept in our multidimensional parameter scan. \smallskip

Additional heavy neutral Higgs bosons can lead to visible signals at the LHC and have been constrained in various ATLAS and CMS dedicated searches. In the scenarios discussed here, and in view of the Lagrangian eq.~(\ref{eq:Lag-mix-SM}), the partial decay widths of the new $H_2$ state into SM fermions and gauge bosons are simply those of the SM-like Higgs boson but with a mass $M_{H_2}$ damped by the mixing factor $\sin^2\theta$ so that in the absence of additional channels, the branching ratios are the same as in the case of a heavy  SM-like Higgs state \cite{Djouadi:2005gi,Djouadi:2005gj}. If the $H_2$ mass is large enough, $M_{H_2} > 200$ GeV, the decays $H_2 \to WW,ZZ$ will largely dominate. There are, however, additional decay modes to be expected: in addition to the invisible Higgs decay channel $H_2 \to XX$, there is also the cascade Higgs decay $H_2 \to H_1 H_1$ which can be extremely important. In our analysis, the following constraints from the searches of a heavy CP--even Higgs boson at the LHC  are accounted for \footnote{Besides the searches mentioned in the main text, the search of the $H_2$ state produced through the vector boson fusion process and decaying into invisible particles, see e.g. Ref.~\cite{Aaboud:2018sfi}, could be also relevant. We have nevertheless not included it in the present analysis.}

-- $H_2 \rightarrow WW$ \cite{Sirunyan:2019pqw}, 

-- $H_2 \rightarrow ZZ$, $H_2 \rightarrow H_1 H_1 \rightarrow \gamma \gamma WW$ \cite{Aaboud:2018ewm}, 

-- $H_2 \rightarrow H_1 H_1 \rightarrow \gamma \gamma bb$ \cite{Aaboud:2018ftw}, 

-- $H_2 \rightarrow H_1 H_1 \rightarrow \bar b b \bar b b$ \cite{Aaboud:2018knk}, 

and the model points incompatible with these constraints are discarded. \smallskip

\begin{figure}[!h]
\vspace*{-1mm}
    \centering
    \subfloat{\includegraphics[width=0.48\linewidth]{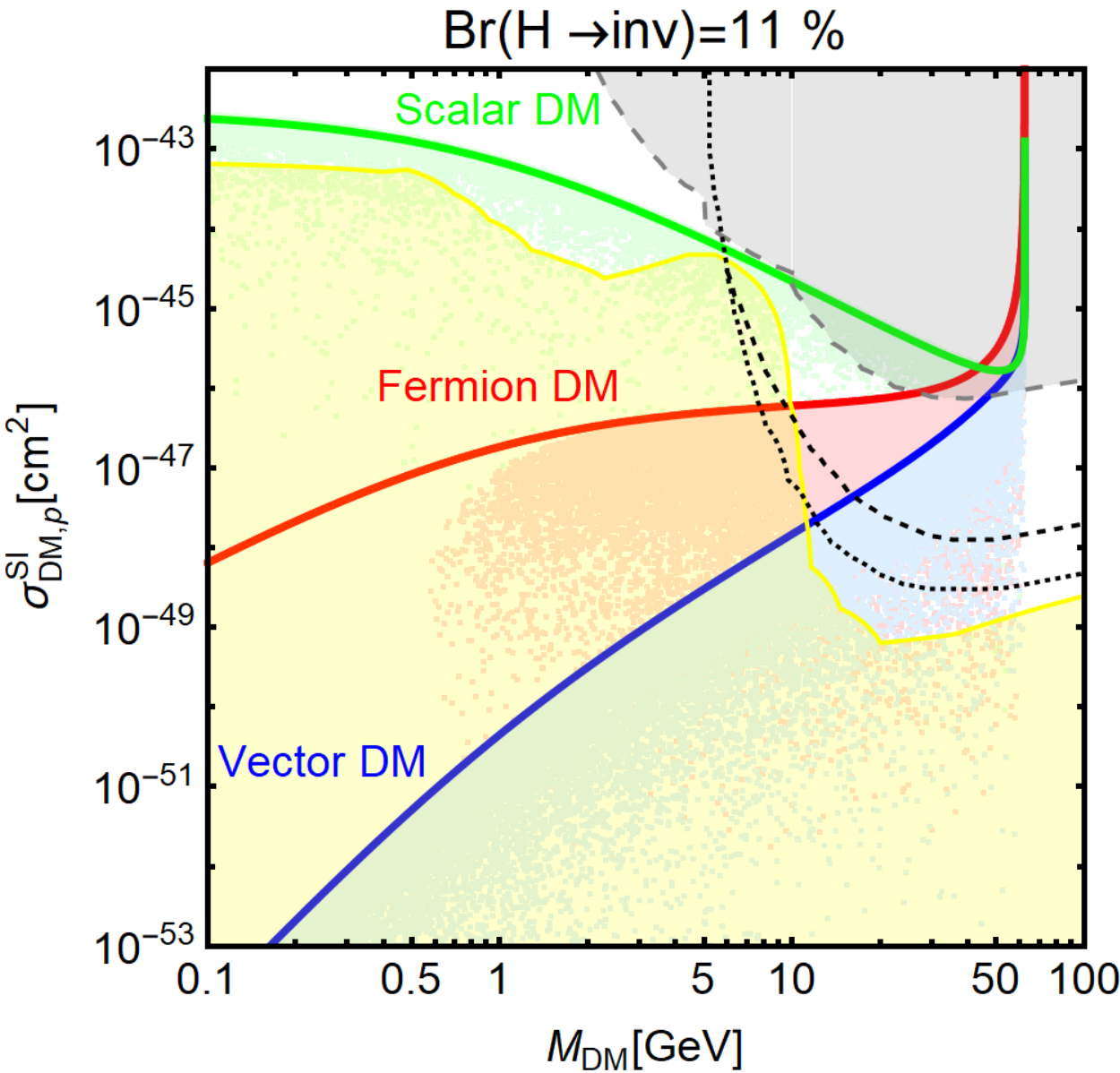}}~~
    \subfloat{\includegraphics[width=0.48\linewidth]{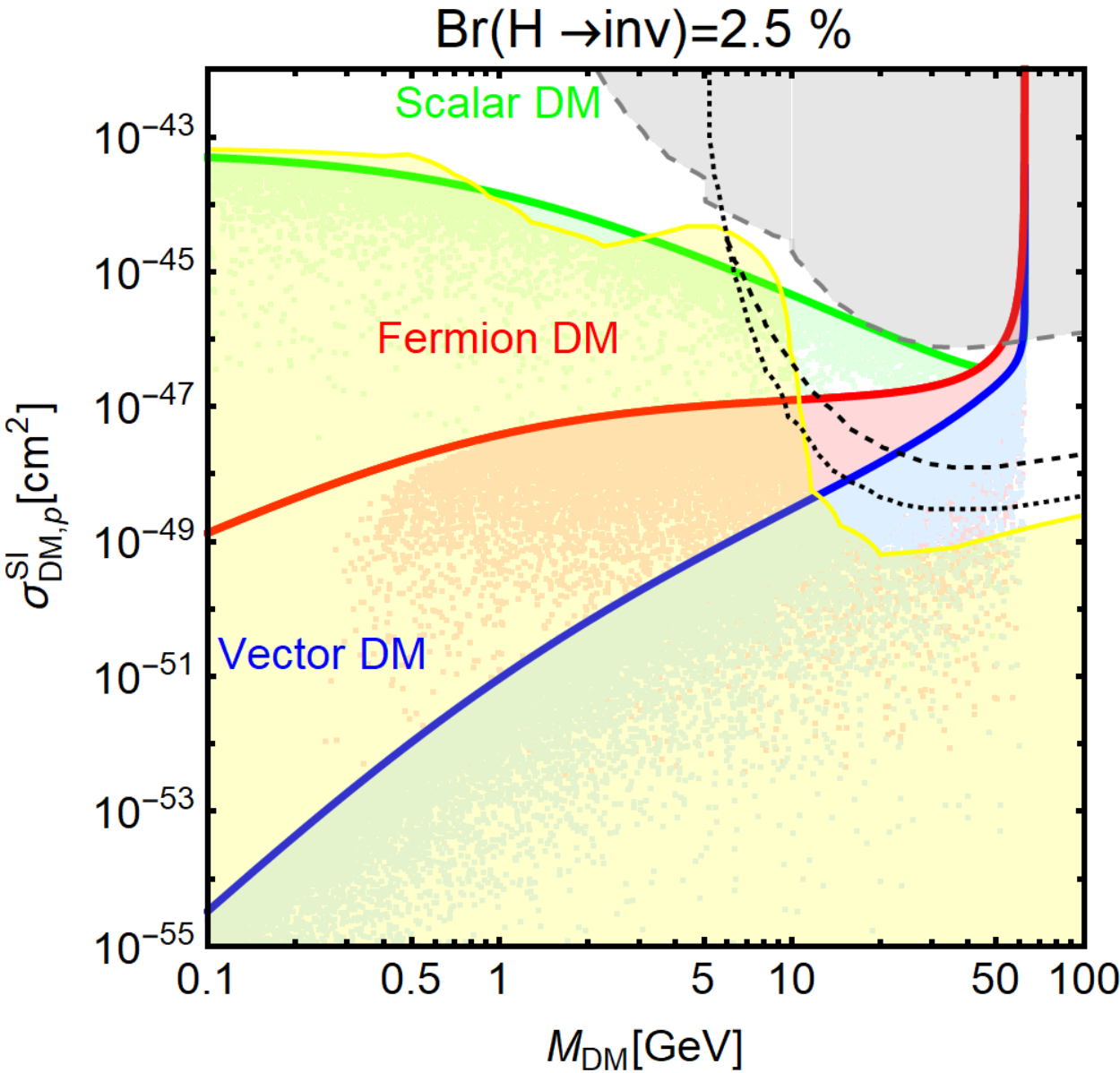}}
\vspace*{-1mm}
    \caption{Predictions for the DM scattering cross section in the scalar (green line), fermionic (red line), and vector (blue line) effective Higgs-portal scenarios, as functions of the DM mass, requiring a value of the invisible branching fraction of the SM-like Higgs boson to be $11\,\%$ (left plot) and $2.5\,\%$ (right plot). The green, red and blue points represent the value of $\sigma_{\rm DM\, p}^{\rm  SI}$ obtained by varying the parameters of the renormalizable complete models with mixing of the SM-like Higgs field with an additional Higgs singlet, of the scalar, fermionic and vector  Higgs-portals. The gray areas show the regions excluded by XENON1T, while the additional lines illustrate the sensitivities associated to XENONnT and DARWIN; the region corresponding to the so-called neutrino floor is shown in yellow.}
    \label{fig:pBrmulti}
\vspace*{-1mm}
\end{figure}

The results of our scan are summarized in the two panels of Fig.~\ref{fig:pBrmulti} which show the expected scattering cross sections of the DM particle over protons, corresponding to the model points for which the invisible branching ratio of the SM-like Higgs particle is constrained to be BR$(H \rightarrow \,\mbox{inv})=11\,\%$ (left panel) and BR$(H \rightarrow \,\mbox{inv})=2.5\,\%$ (right panel).  We remind again the reader that these two values represent, respectively, the present experimental sensitivity at the LHC and the one expected for the high-luminosity upgrade of LHC. The figure also shows, as green/red/blue contours, the expected cross sections for the effective scalar/fermionic/vector Higgs-portals. As already pointed out in Ref.~\cite{Arcadi:2020jqf} (see also Ref.~\cite{Baek:2014jga}), the second scalar $H_2$ causes a destructive interference in the DM scattering cross section, making it smaller with respect to the prediction of the effective Higgs-portal for the same DM mass. This interference becomes negligible as $M_{H_2}$ increases and the effective limit is recovered for $M_{H_2} \gtrsim 1\,\mbox{TeV}$. \smallskip

Fig.~\ref{fig:pBrmulti} shows also current limits and future prospects from dedicated direct DM searches. More precisely, the gray region above the dashed gray line represents the exclusion from searches by XENON1T presented in Ref.~\cite{Aprile:2018dbl,Aprile:2019xxb}. Furthermore, the expected sensitivities from the XENONnT~\cite{Aprile:2020vtw} and DARWIN~\cite{Aalbers:2016jon} experiments are shown as magenta and purple contours, respectively. The yellow regions correspond, finally, to the so called neutrino-floor, i.e. the irreducible background due to the coherent scattering of solar and atmospheric neutrinos over nucleons. While direct detection experiments will eventually be the most sensitive probe for $M_{\rm DM} \gtrsim 10\,\mbox{GeV}$, for lighter DM state, collider searches for invisible Higgs decays allow to probe regions of the parameters space well inside the neutrino floor.\smallskip

As will be discussed in the next section, achieving the correct relic density for the DM state will require, in the simplest models, the presence of a light $H_2$ boson. We have therefore extended the range of the scan to the $[0.1\,\mbox{GeV},M_{H_1}]$ region. Similarly to the case $M_{H_2}>M_{H_1}$, constraints on the extra Higgs boson should be taken into account. For a light $H_2$ boson, the following limits are accounted for:\smallskip 

$i)$ searches of low mass Higgs bosons at LEP \cite{Abreu:1990bq} and particularly DELPHI \cite{Barate:2003sz};\smallskip

$ii)$ constraints from the decays of $B$ and $K$ mesons \cite{Aaij:2012vr,Lees:2013kla,Wei:2009zv,Artamonov:2008qb,Krnjaic:2015mbs};\smallskip

$iii)$ constraints from beam dump experiments \cite{Bergsma:1985qz}.\smallskip

\begin{figure}[!t]
\vspace*{-1mm}
    \centering
    \subfloat{\includegraphics[width=0.5\linewidth]{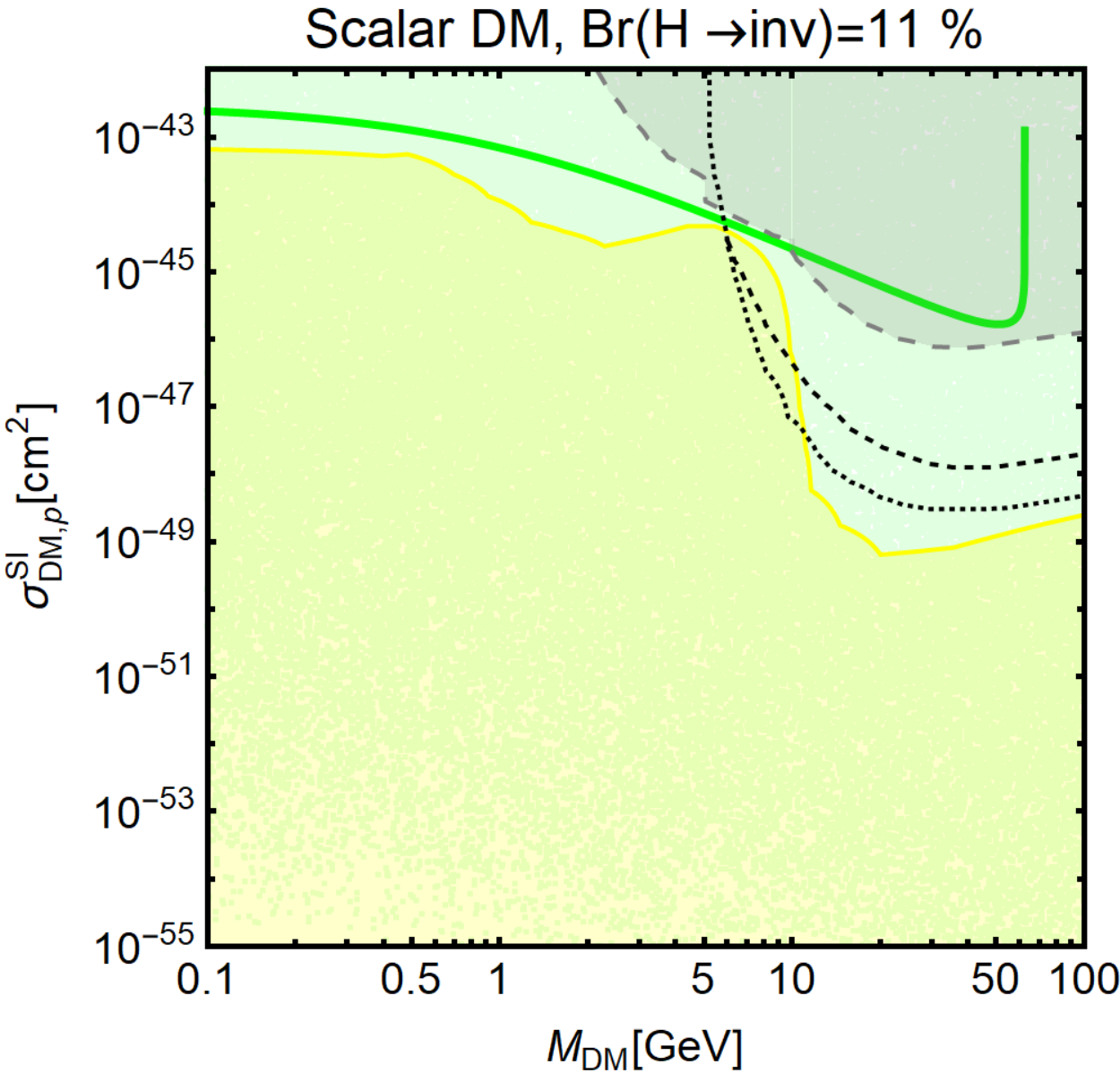}}
    \subfloat{\includegraphics[width=0.5\linewidth]{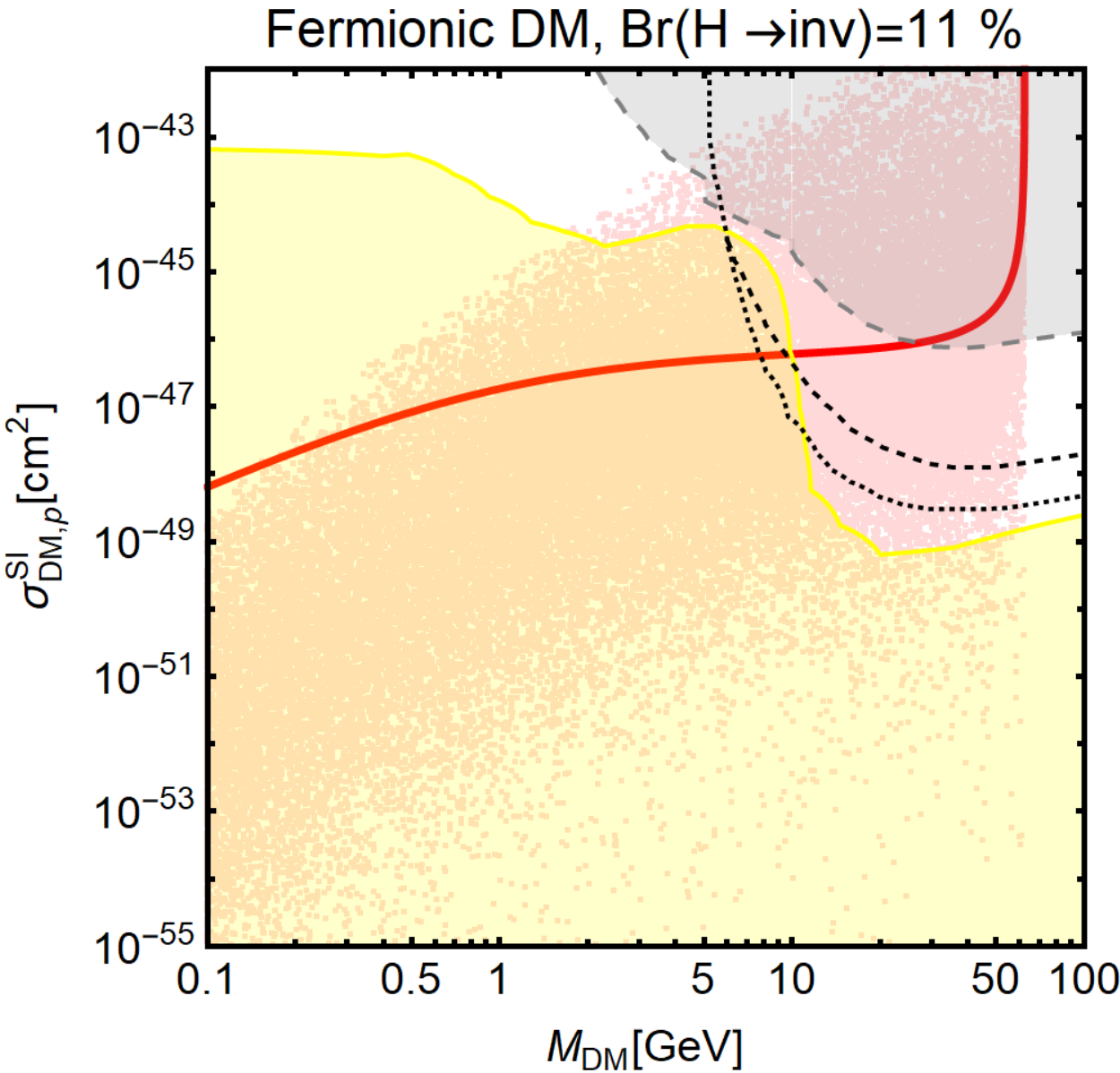}}\\
    \subfloat{\includegraphics[width=0.5\linewidth]{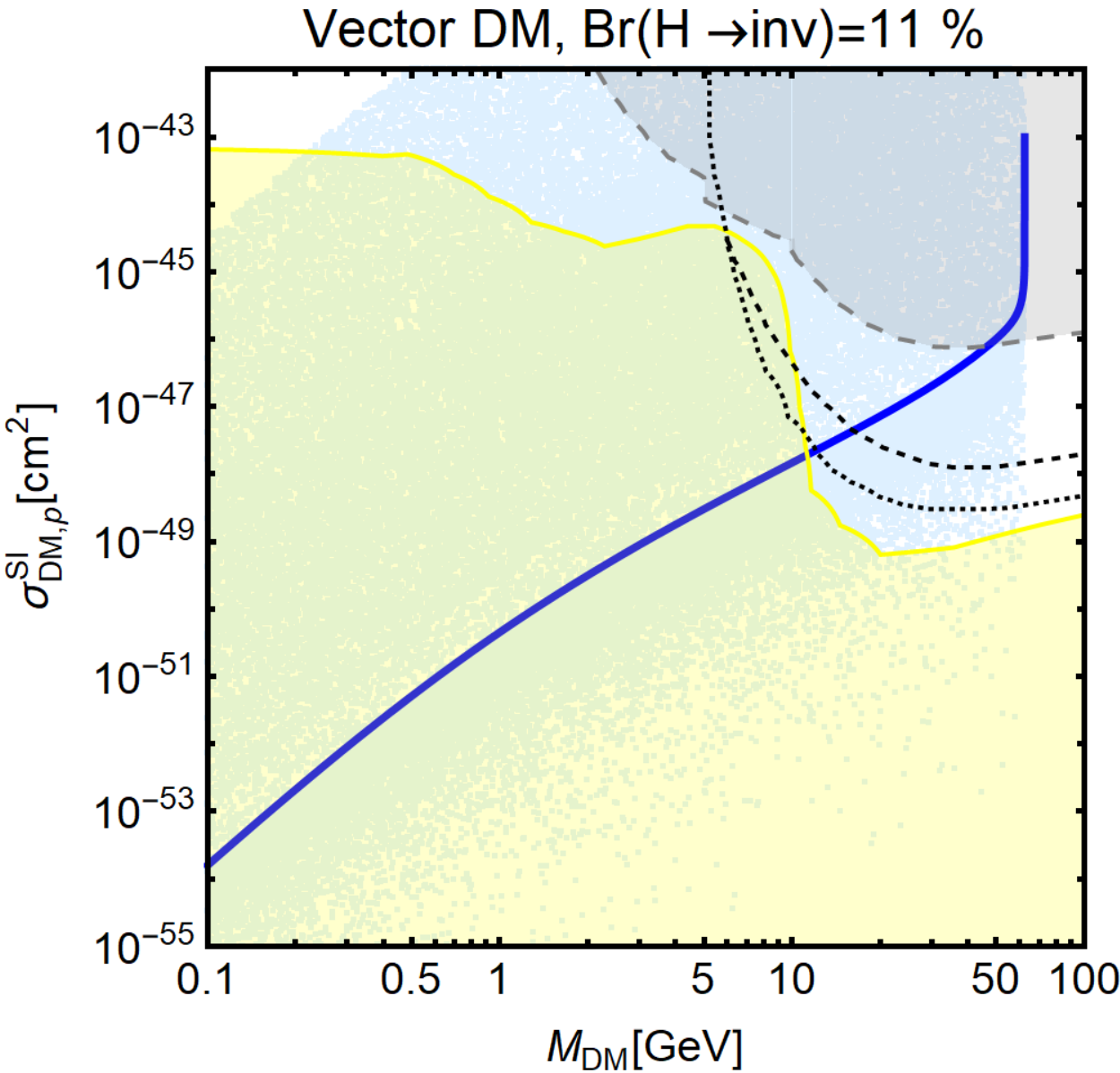}}
\vspace*{-1mm}
    \caption{Same as in Fig.~\ref{fig:pBrmulti} considering in addition the case of a light $H_2$ boson. The three panels are for the scalar, fermionic and vector DM cases. Only the case of an invisible Higgs branching ratio of BR$(H_1 \rightarrow \mbox{inv})=0.11$ are displayed.}
    \label{fig:pBrmulti_light}
\vspace*{-1mm}
\end{figure}

The results of the scan over the extended $M_{H_2}$ range are shown for the scalar, fermionic and vector DM cases individually, in the three panels of Fig.~\ref{fig:pBrmulti_light}. Here, only  the limit on the invisible Higgs decay rate from the LHC of BR$(H_1 \rightarrow \,\mbox{inv})=11\,\%$ are considered. As the mass of the $H_2$ state decreases, the corresponding contributions will dominate the DM scattering cross section, which will then deviate from the predictions of the effective Higgs portal models. Furthermore, the $H_1 \rightarrow H_2 H_2$ channel, if kinematically accessible, can affect the invisible width of the heavy Higgs state\footnote{Note that if this $H_1 \to H_2 H_2$ possibility occurs, the LHC limit on the invisible SM-like Higgs branching ratio (in particular the one derived indirectly from the signal strengths in the various $H_1$ final state searches performed by the ATLAS and CMS collaborations) might be weakened and values larger than the adopted limit of BR$(H_1 \rightarrow \,\mbox{inv})=11\,\%$ might be possible.}. \smallskip

When comparing direct detection limits with those from invisible Higgs decays, it is important to note that, in principle, the latter probe arbitrarily low values of the DM particle masses while the sensitivity of direct detection experiments is limited by the energy threshold cutoffs. We nevertheless limit the range of considered DM masses to 0.1 GeV based on more general cosmological considerations. Indeed, throughout most of this paper, the DM candidates are assumed to be thermal relics, i.e. that the DM particle was in thermal equilibrium in the early universe prior to a decoupling temperature $T_{\rm dec}$. In this case, the very stringent CMB bound on the extra relativistic degrees of freedom \cite{Aghanim:2018eyx} excludes, at the $95\,\%$ confidence level, a decoupling temperature for beyond the SM particles, below the temperature of the QCD phase transition, of the order of 100 MeV. According to the details of the freeze-out, we can convert this bound into a lower bound on the DM mass. Considering non-relativistic freeze-out, namely $T_{\rm dec} \lesssim M_X$, we obtain a generic lower bound $T_{\rm dec} \simeq M_X \gtrsim 0.1\,\mbox{GeV}$, which has then been adopted in our scans\footnote{In this framework, an additional and potentially more stringent bound can be applied to the DM particle mass as CMB anisotropies can be affected by energy injection from residual late time annihilation processes of the DM \cite{Galli:2009zc,Slatyer:2015jla,Slatyer:2015kla}. Planck has set a very strong bound on the DM annihilation cross section as a function of the DM mass \cite{Aghanim:2018eyx}, which excludes the value favored by the WIMP paradigm, namely $3 \times 10^{-26}\,{\mbox{cm}}^3\, {\mbox{s}}^{-1}$, for masses below 10 GeV. Contrary to the one from $\Delta N_{\rm eff}$, this bound is more model dependent since the DM features sizable late time annihilations only if its cross section is $s$-wave dominated.}.

\subsection{DM phenomenology including the relic density constraint}

In this section, the phenomenology of the DM particles including the requirement that they generate the correct relic density within the conventional freeze-out paradigm is discussed. The DM states therefore initially had sufficient interactions with the primordial thermal bath to be in thermal equilibrium. At a later stage, identified with the so-called freeze-out temperature $T_{f.o.}={M_{\rm DM}}/{20}-{M_{\rm DM}}/{30}$, it went out of chemical equilibrium. The DM final relic density $\Omega_{\rm DM}h^2$, which can be probed through CMB measurements, is entirely determined by the thermally averaged DM annihilation cross section $\langle \sigma v \rangle$ through the expression \cite{Gondolo:1990dk}
\begin{equation}
    \Omega_{\rm DM}h^2 \approx 8.76 \times 10^{-11}\,{\mbox{GeV}}^{-2}{\left[\int_{T_0}^{T_{f.o.}}\langle \sigma v \rangle \frac{dT}{M_{\rm DM}}\right]}^{-1} \, , 
\end{equation}
where $T_0$ is the present time temperature. In our study, in  order to address the constraints from the correct DM relic density in a precise manner, we have implemented all the models discussed in the numerical package micrOMEGAs-5 \cite{Belanger:2014vza}.

\subsubsection{Scalar Dark Matter}

We start by addressing the phenomenological constraints in the case of a scalar DM.  We have numerically computed the DM relic density for the model points that pass the requirements of the scan discussed in the previous subsection and kept only the parameter assignments that lead to a density  $\Omega_{\rm DM}h^2$ within $3 \sigma$ of the experimentally determined value and that lead to a DM scattering cross section over nucleons below the present limits set by the  XENON1T experiment. Furthermore, since scalar DM particles have an $s$-wave dominated annihilation cross section, we have included indirect detection constraints, mostly determined by the FERMI experiment in the range of interest for DM masses, from searches of DM annihilation in dwarf-spheroidal galaxies \cite{Ahnen:2016qkx,Hoof:2018hyn} taking into account, whenever kinematically accessible, the $ss \rightarrow H_2 H_2 \to \bar f f \bar f' f'$ processes~\cite{Clark:2017fum}, as well as CMB constraints on residual DM annihilations \cite{Slatyer:2015jla,Slatyer:2015kla,Aghanim:2018eyx}. The latter constraints can serve as a useful complement in the low mass region, namely $M_s<10\,\mbox{GeV}$.\smallskip

\begin{figure}[!h]
\vspace*{-1mm}
    \centering
    \includegraphics[width=0.55\linewidth]{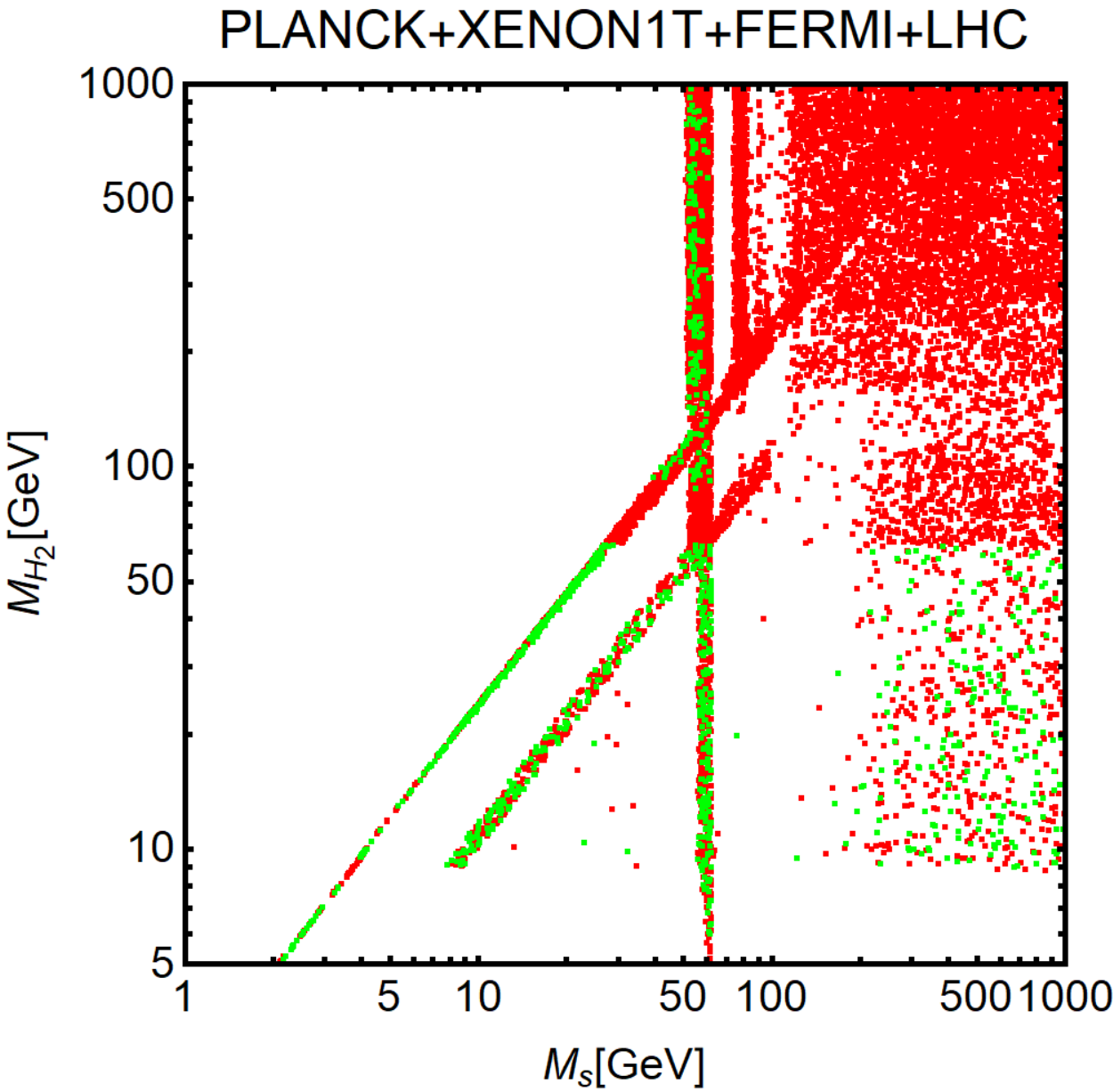}
\vspace*{-1mm}
    \caption{Model points, in the $(M_s,M_{H_2})$ plane, for the mixing model with scalar DM passing all the theoretical and experimental constraints. The model points featuring $2.5\,\% \leq {\rm BR} (H_1 \rightarrow \,\mbox{inv}) \leq 11\,\%$ have been highlighted in green.}
    \label{fig:scalar_DM_scan}
\vspace*{-1mm}
\end{figure}

The model points passing all these constraints are displayed in  Fig.~\ref{fig:scalar_DM_scan} in the $[M_s,M_{H_2}]$ plane. Among them, the green points feature an invisible Higgs branching fraction of $2.5\,\% \leq {\rm BR}(H_1 \rightarrow \,\mbox{inv})\leq 11\,\%$ while the red points are for the situation in which this  invisible branching fraction is suppressed or even null,  as is the case when the corresponding decay channels are kinematically forbidden. As can be seen, there are model points that can be probed by LHC searches of invisible Higgs decays only if one of these conditions is met: $M_{s}\sim M_{H_2}$, $M_s \sim \frac12 M_{H_2}$ and $M_s \sim \frac12  M_{H_1}$. We also remark that on the $y$-axis of the figure, one has $M_{H_2} \geq 5\,\mbox{GeV}$ since below this mass value, the bounds on the mixing angle in the Higgs sector $\theta$, as illustrated in the previous subsection, are too strong to allow for a viable relic density according to the WIMP paradigm. \smallskip

\begin{figure}[!h]
\vspace*{-1mm}
    \centering
    \subfloat{\includegraphics[width=0.48\linewidth]{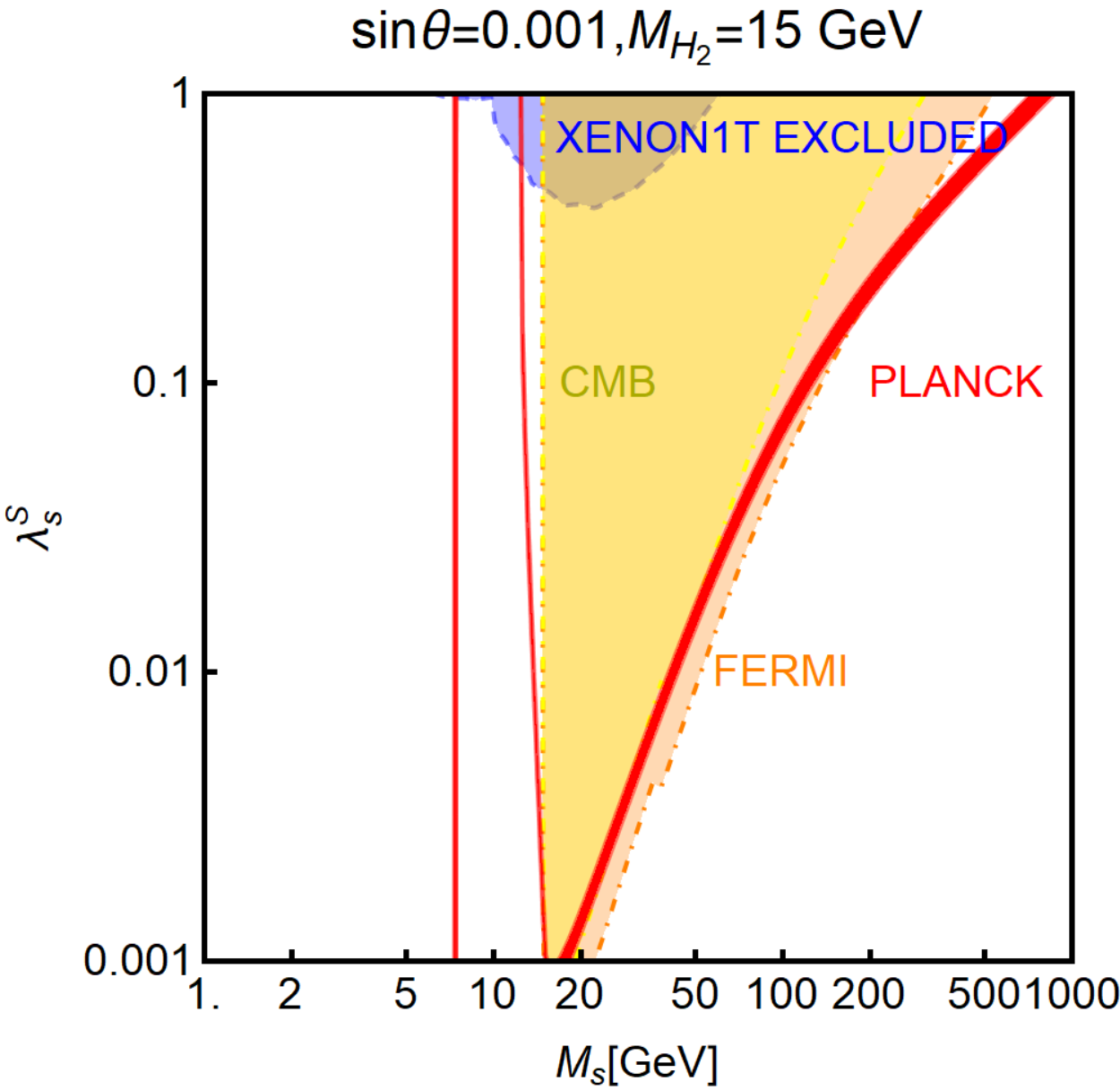}}
    \subfloat{\includegraphics[width=0.48\linewidth]{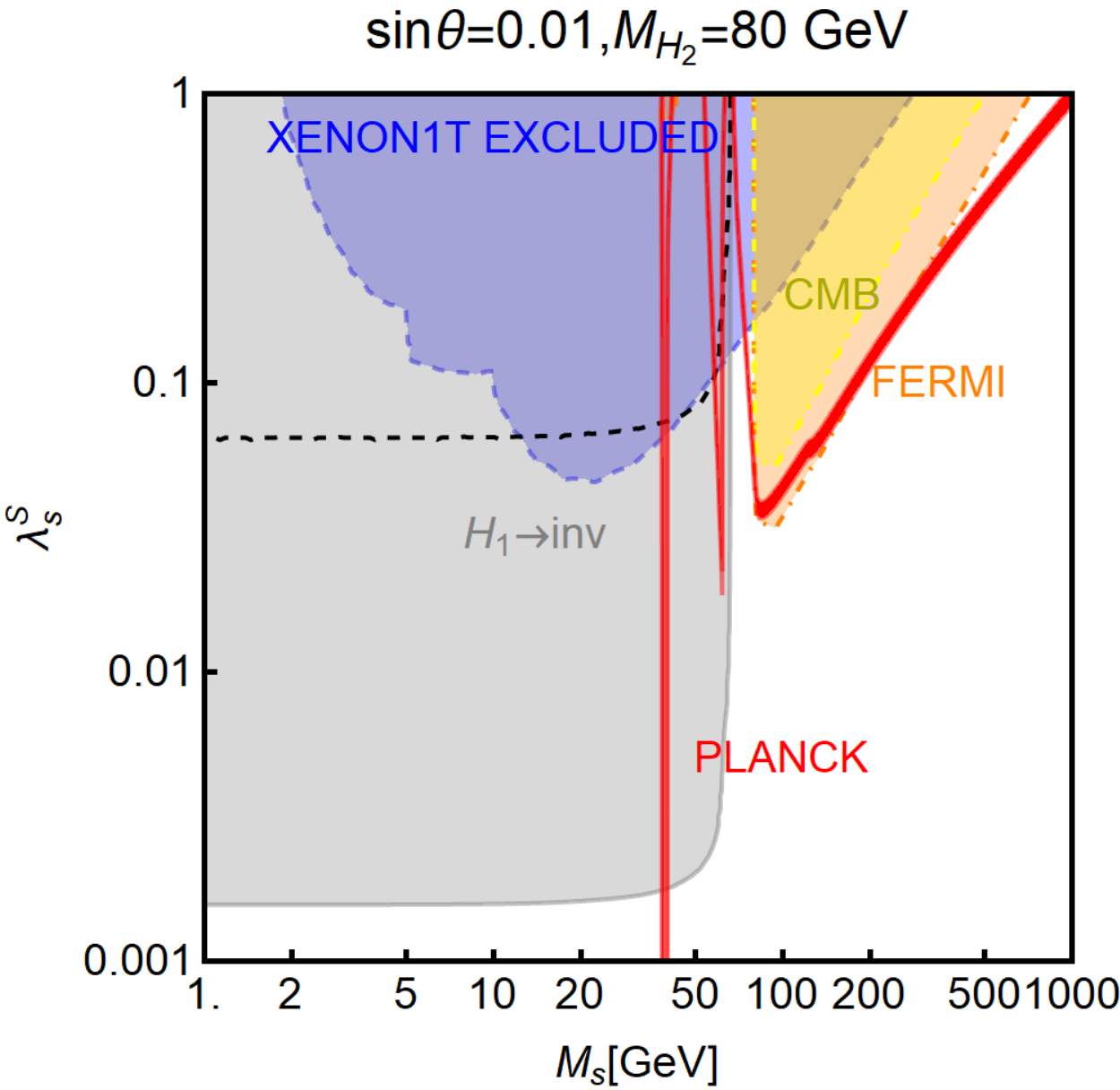}}\\
    \subfloat{\includegraphics[width=0.48\linewidth]{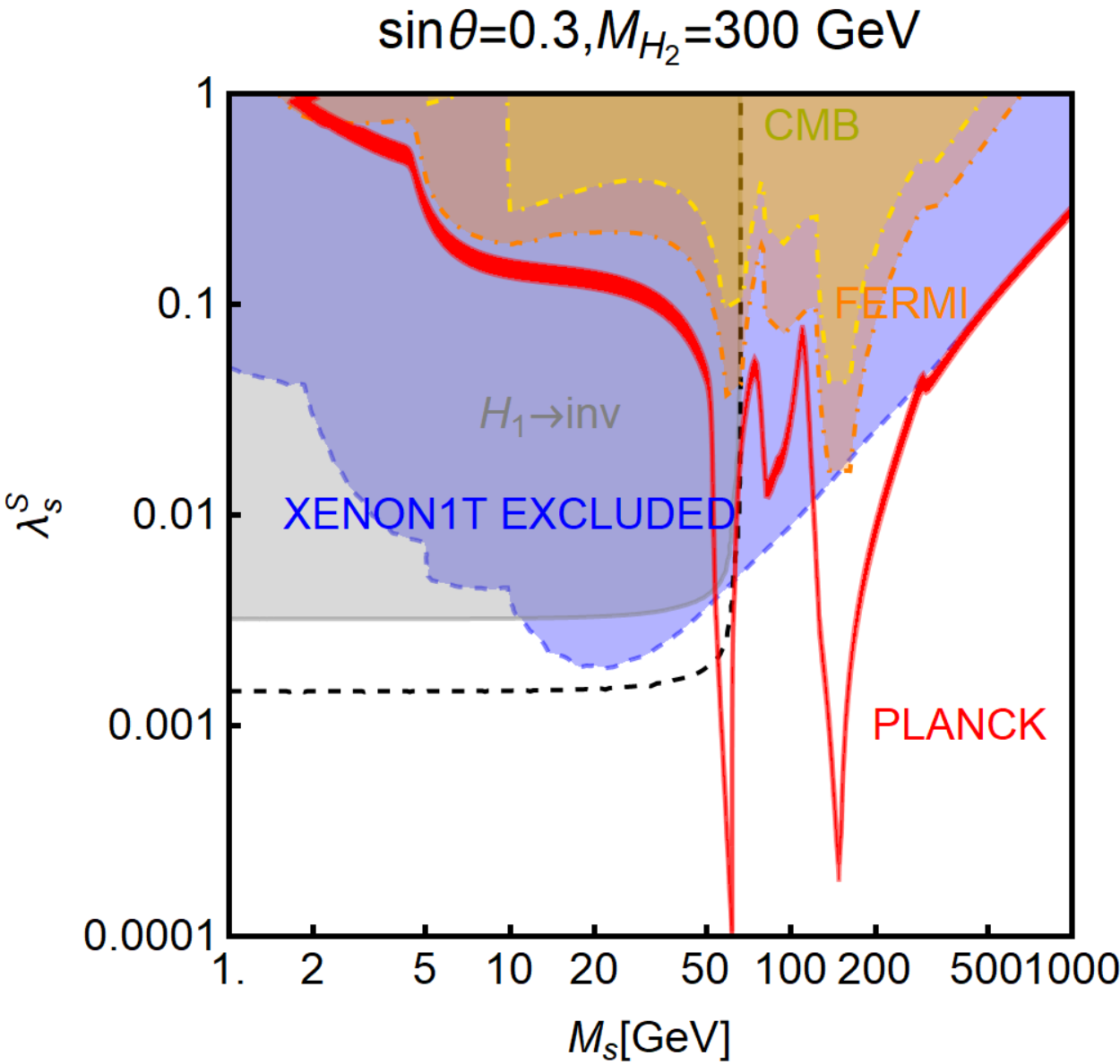}}
    \subfloat{\includegraphics[width=0.48\linewidth]{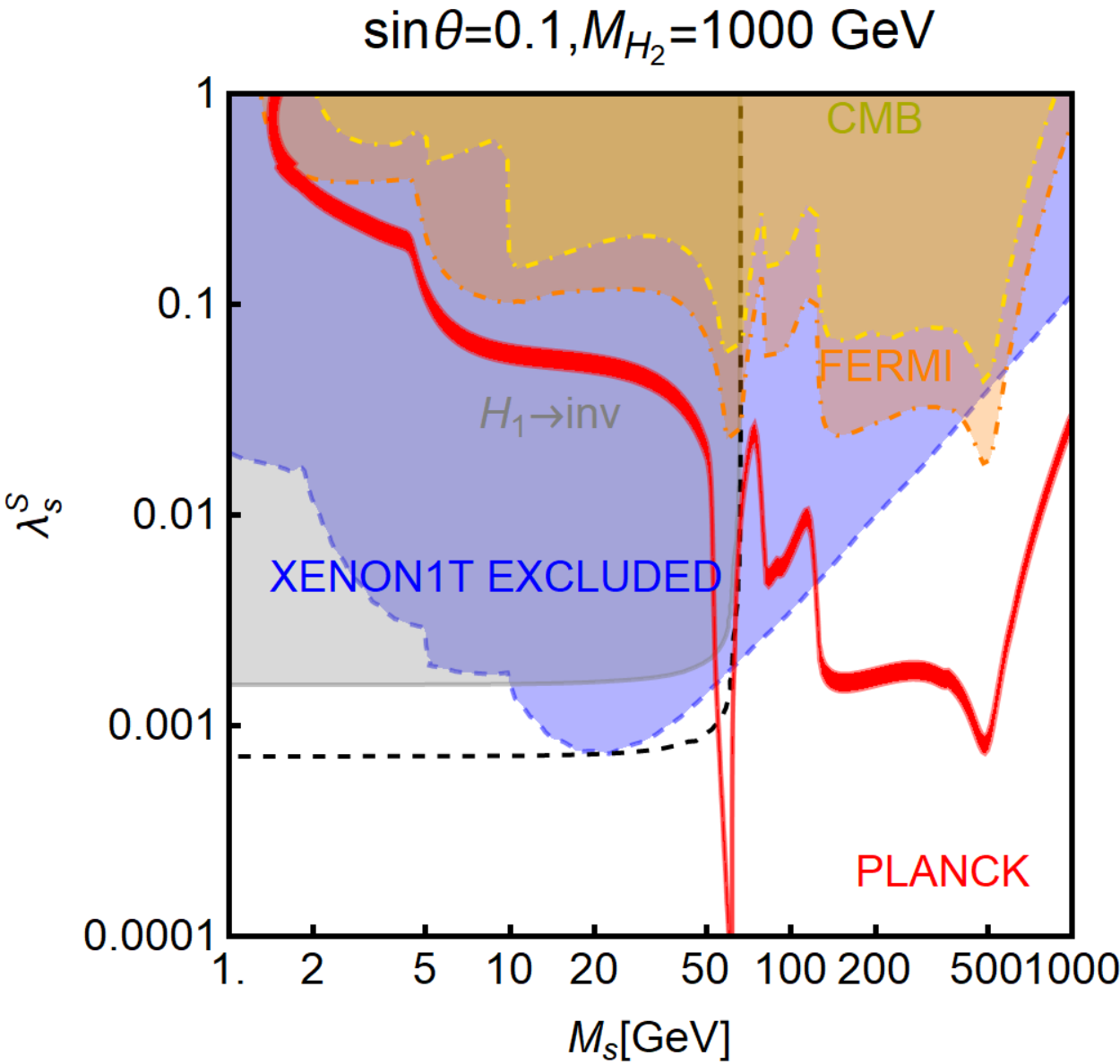}}
\vspace*{-1mm}
    \caption{Summary of constraints for a scalar DM particle $s$ interacting with the SM through the mixing portal, in the $[M_s,\lambda_{s}^{S}]$  plane, for specific values of ($\sin\theta$, $M_{H_2}$)  of $\left(0.001,15\,\mbox{GeV}\right)$, $\left(0.01,80\,\mbox{GeV}\right)$, $\left(0.3,300\,\mbox{GeV}\right)$ and $\left(0.1,1000\,\mbox{GeV}\right)$. The red contours correspond to the correct relic density, the blue region is excluded by constraints from DM direct detection (XENON1T) and the gray region corresponds to the bounds from the Higgs invisible branching fraction.}
    \label{fig:scalar_DM_ben}
\vspace*{-1mm}
\end{figure}

In order to better understand this feature, we display in Fig.~\ref{fig:scalar_DM_ben} the outcome of a more focused analysis of some benchmark models. The various panels in the figure show the combination of the different constraints in the $[M_s,\lambda_s^S]$  plane when considering increasing values of the Higgs mass $M_{H_2}$. The different experimental exclusions are represented as colored regions. The blue region corresponds to the one excluded by the DM direct detection results of XENON1T, the gray region is excluded by the requirement that BR$(H_1 \rightarrow \mbox{inv})\leq 0.11$, while the orange and yellow regions correspond to the constraints from DM indirect detection as given by the FERMI experiment and again, the PLANCK experiment, referring this time to the bound on effects on residual DM annihilations on CMB. The figure finally exhibits as black dashed contours, the contour for the value BR$(H_1\rightarrow \mbox{inv})=0.025$, i.e. the expected constraint on the invisible Higgs decays from HL--LHC.\smallskip

The upper left panel of Fig.~\ref{fig:scalar_DM_ben} shows the case of a very light $H_2$ state and a rather small value of the $H$--$S$ mixing angle. In this configuration, the correct relic density can be achieved only in correspondence of the very narrow $\frac12 M_{H_2}$ Higgs pole and when the $ss \rightarrow H_2 H_2$ annihilation process is kinematically allowed. As has been shown in Ref.~\cite{Arcadi:2016qoz}, this configuration ensures a low impact from direct detection constraints. Indirect detection constraints, however, probe efficiently the low mass regime. This panel does not show contours corresponding to the invisible decay of the SM-like Higgs boson. This is due to the fact that the invisible branching fraction is entirely dominated by the $H_1 \rightarrow H_2 H_2$ process whose rate is controlled by model parameters which are not varied in the plot. The parameter assignment of this particular benchmark corresponds to BR$(H_1 \rightarrow \mbox{inv})\simeq {\rm BR} (H_1 \rightarrow H_2 H_2)=0.07$. \smallskip

The second benchmark still corresponds to a light $H_2$ state but above the kinematical threshold of the $H_1 \rightarrow H_2 H_2$ cascade decay process. Again, besides the $s$-channel Higgs poles, a  viable relic density requires $M_s > M_{H_2}$ and indirect detection provides the most stringent constraints, ruling out DM masses up to 100 GeV.\smallskip

Finally, the last two benchmarks feature high values of the $H_2$ mass and high values of the mixing angle, close to the experimental limits from the measurement of Higgs signal strengths at the LHC. Here, direct detection turns out to be the strongest constraint, ruling out the possibility of DM particles with masses below a few hundreds GeV,  with the only exception being when it is close to the $\frac12 M_{H_1}$ pole. \smallskip

\begin{figure}[!h]
\vspace*{-1mm}
    \centering
    \includegraphics[width=0.55\linewidth]{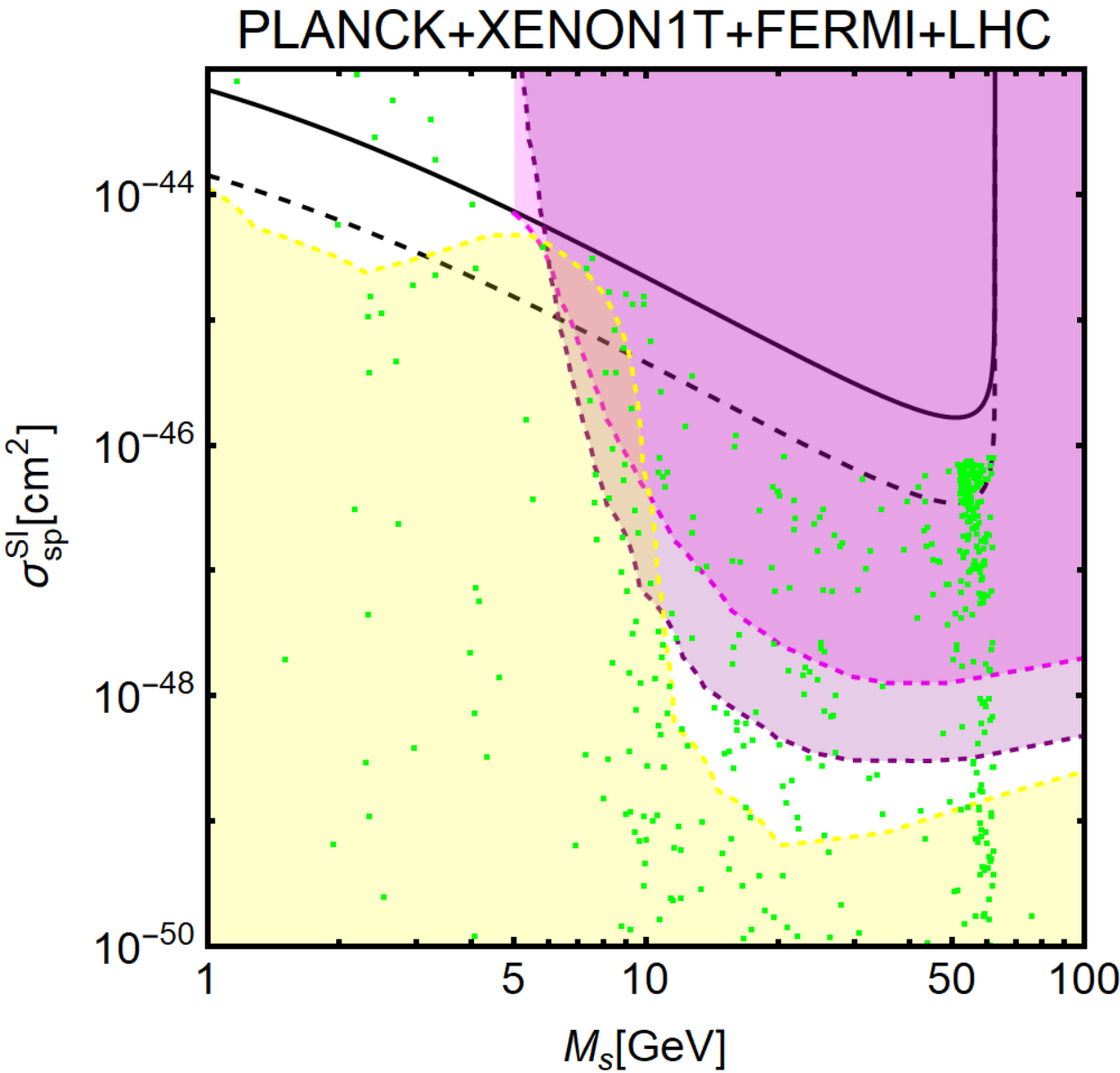}
\vspace*{-1mm}
    \caption{{Model points, from the general parameter scan illustrated in the main text, complying with all DM constraints, namely correct relic density, DM-nucleon cross-section below XENON1T limit and DM annihilation cross-section compatible with ID and CMB limits, ad leading to an invisible Higgs branching ratio of  $2.5\,\% \leq {\rm BR}(H_1 \rightarrow XX) \leq 11\,\%$}. The magenta and purple regions correspond to the sensitivity of the XENONnT and DARWIN experiments while the yellow region represents the neutrino floor. Finally  the solid and dashed black curves correspond to the SI DM scattering cross-section, as computed in the EFT scalar Higgs-portal, requiring that the value of the DM/SM Higgs coupling leads to values of the invisible Higgs branching fraction of $11\%$ (solid black line) and $2.5\%$ (dashed black line).}
    \label{fig:scalar_DM_corr_plot}
\vspace*{-1mm}
\end{figure}

We finally recast the direct detection versus the Higgs invisible decays analysis, once all the phenomenological constraints are accounted for, including in particular the requirement of a correct DM relic density. Fig.~\ref{fig:scalar_DM_corr_plot} shows then in the $[M_s,\sigma_{Xp}^{\rm SI}]$ plane, the model points that fulfilled all the phenomenological constraints and, in addition, lead to an invisible Higgs branching ratio in the range $2.5\,\% \leq {\rm BR}(H_1 \rightarrow ss) \leq 11\,\%$, i.e. which can be probed at the HL--LHC. The figure shows also two iso--contours, in solid and dashed black, corresponding to the expected cross sections, as function of the DM mass, for the effective scalar Higgs portal with coupling of the DM to the Higgs set such that the invisible branching fraction is equal to $11\,\%$ and $2.5\,\%$, respectively. If the distribution, in the $[M_s,\sigma_{Xp}^{\rm SI}]$ plane, of the model points fell between the two contours and had a similar shape, one could state that the model under consideration has the effective Higgs--portal as a viable limit and can, hence, be probed by the correlation plots shown by the experimental collaborations. In the opposite case instead, a full recast of the experimental limits would be needed. As shown by Fig.~\ref{fig:scalar_DM_corr_plot}, the distribution of the points of the scalar mixing model, which comply with DM phenomenology, is substantially different from the contours corresponding to the effective Higgs--portal. This is due to the fact that, in the majority of cases, to have the correct relic density for a light DM, one needs comparatively a light $H_2$, with a mass possibly smaller than $\frac12 M_{H_1}$. As a consequence, the invisible Higgs width is dominated by the $H_1 \rightarrow H_2 H_2$ decay process. For comparison, the figure also shows the prospects at the next generation XENONnT and DARWIN experiments and the yellow region corresponding to the neutrino floor. As can be easily seen, a sensitive portion of model points fall inside the neutrino floor below the sensitivity of the DARWIN experiment. Search of invisible Higgs decays are then a very efficient complement for dedicated DM searches.

\subsubsection{Fermionic Dark Matter}

\begin{figure}[!h]
\vspace*{-1mm}
    \centering
    \subfloat{\includegraphics[width=0.55\linewidth]{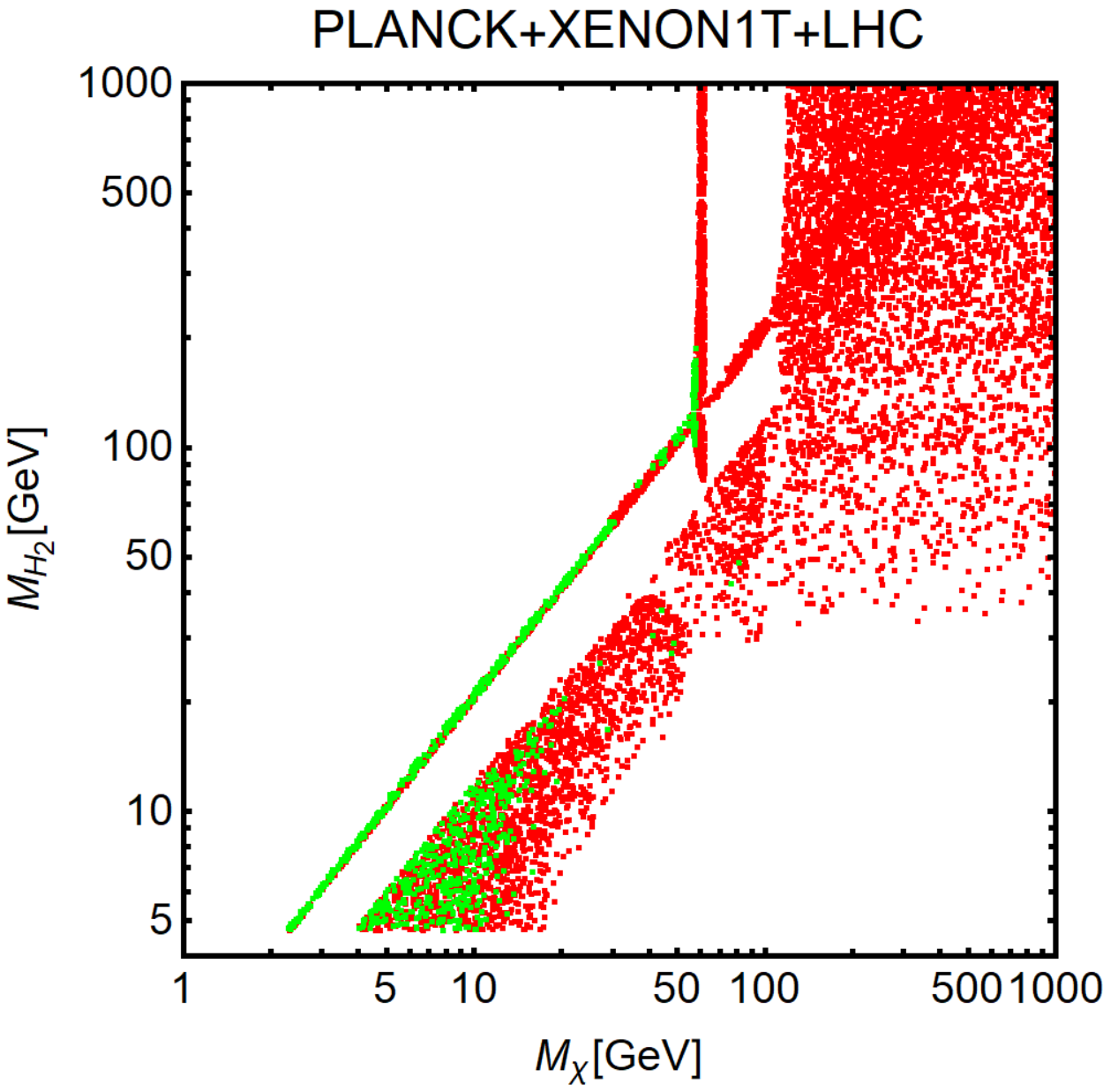}}
\vspace*{-1mm}
    \caption{Parameter survey for the fermionic DM model in the $[M_\chi,M_{H_2}]$ plane when the requirement of a correct relic density is applied. The green points feature an invisible Higgs branching fraction 
in the range $2.5\% \leq {\rm BR}(H_1 \rightarrow \chi \chi) \leq 11\,\%$. }
    \label{fig:scanF}
\vspace*{-1mm}
\end{figure}

We have repeated the same analysis discussed before in the case of a (Dirac) fermionic DM.  The result of applying the requirement of a viable phenomenology for the spin--$\frac12$ DM particle, including  its compatibility with the measured relic density (within 3$\sigma$), to the survey of model parameters performed above  is summarized in Fig.~\ref{fig:scanF} in the $[M_\chi,M_{H_2}]$ plane. The green points in the figure, which feature invisible branching fractions of the SM-like Higgs state between $2.5\,\%$ and $11\,\%$,  cover a wider region of the parameter space at low DM masses, compared to the case of a scalar DM state. This is due to the $p$-wave suppression in the  annihilation cross section for fermionic DM. Consequently, the CMB and indirect detection constraints have a negligible impact. This is the case as the value of the annihilation cross section at CMB decoupling and at present times is suppressed with respect to its value at thermal freeze-out. Nevertheless, the same general conclusion previously stated is  still valid: a viable DM requires a light $H_2$ state which, consequently, significantly impacts the collider phenomenology. \smallskip

\begin{figure}[!h]
\vspace*{-1mm}
    \centering
    \subfloat{\includegraphics[width=0.48\linewidth]{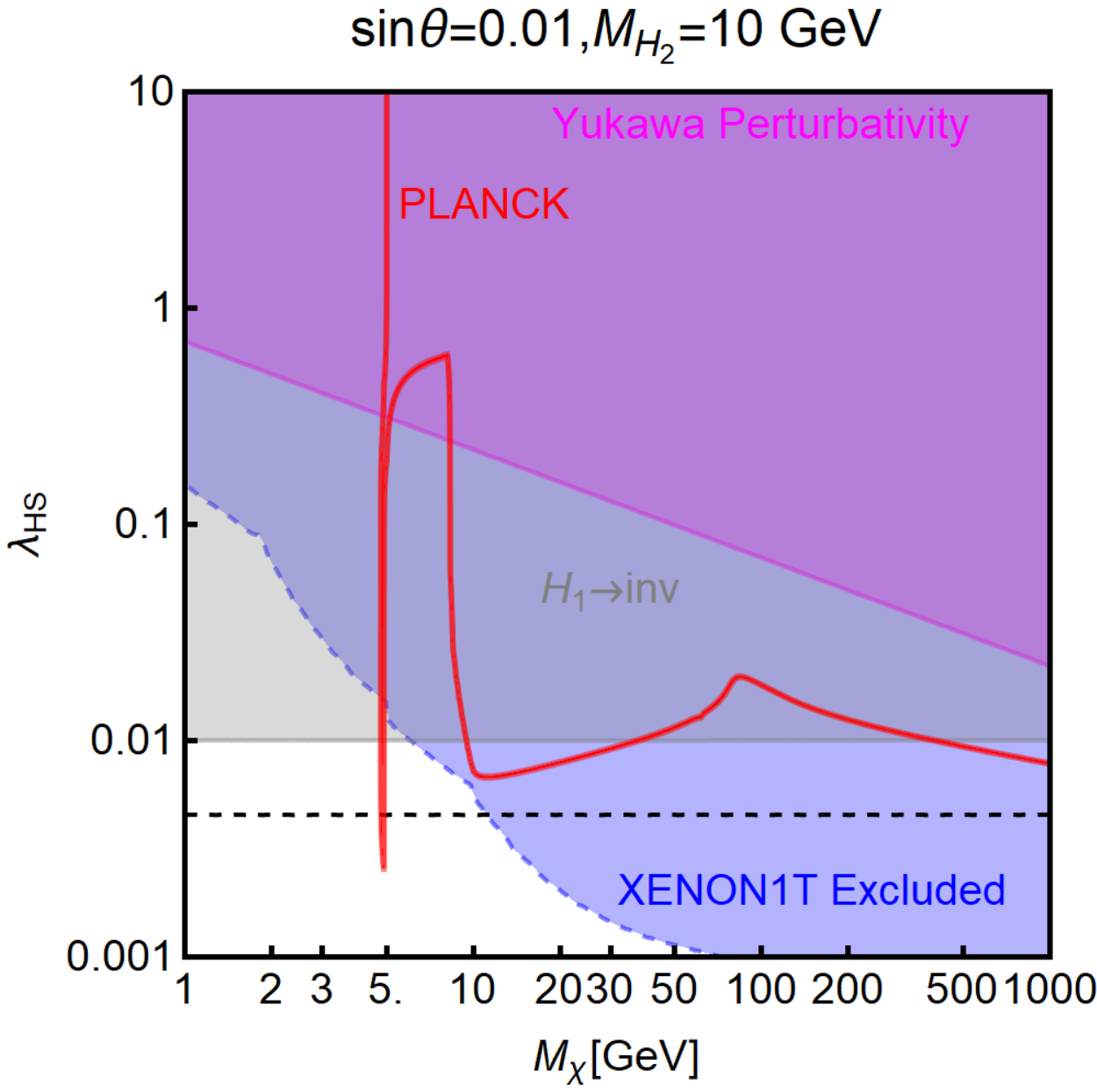}}
    \subfloat{\includegraphics[width=0.48\linewidth]{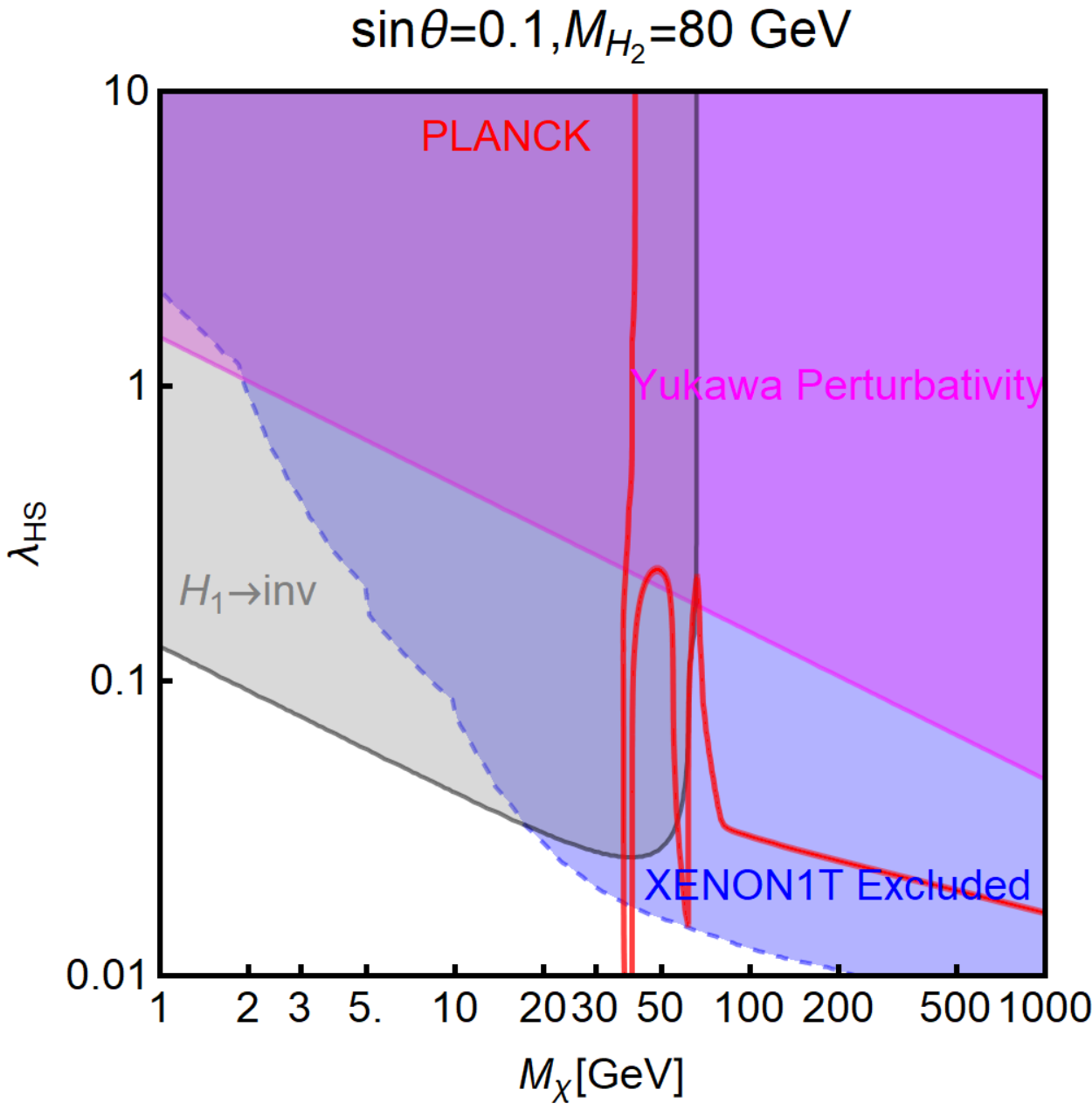}}\\
    \subfloat{\includegraphics[width=0.48\linewidth]{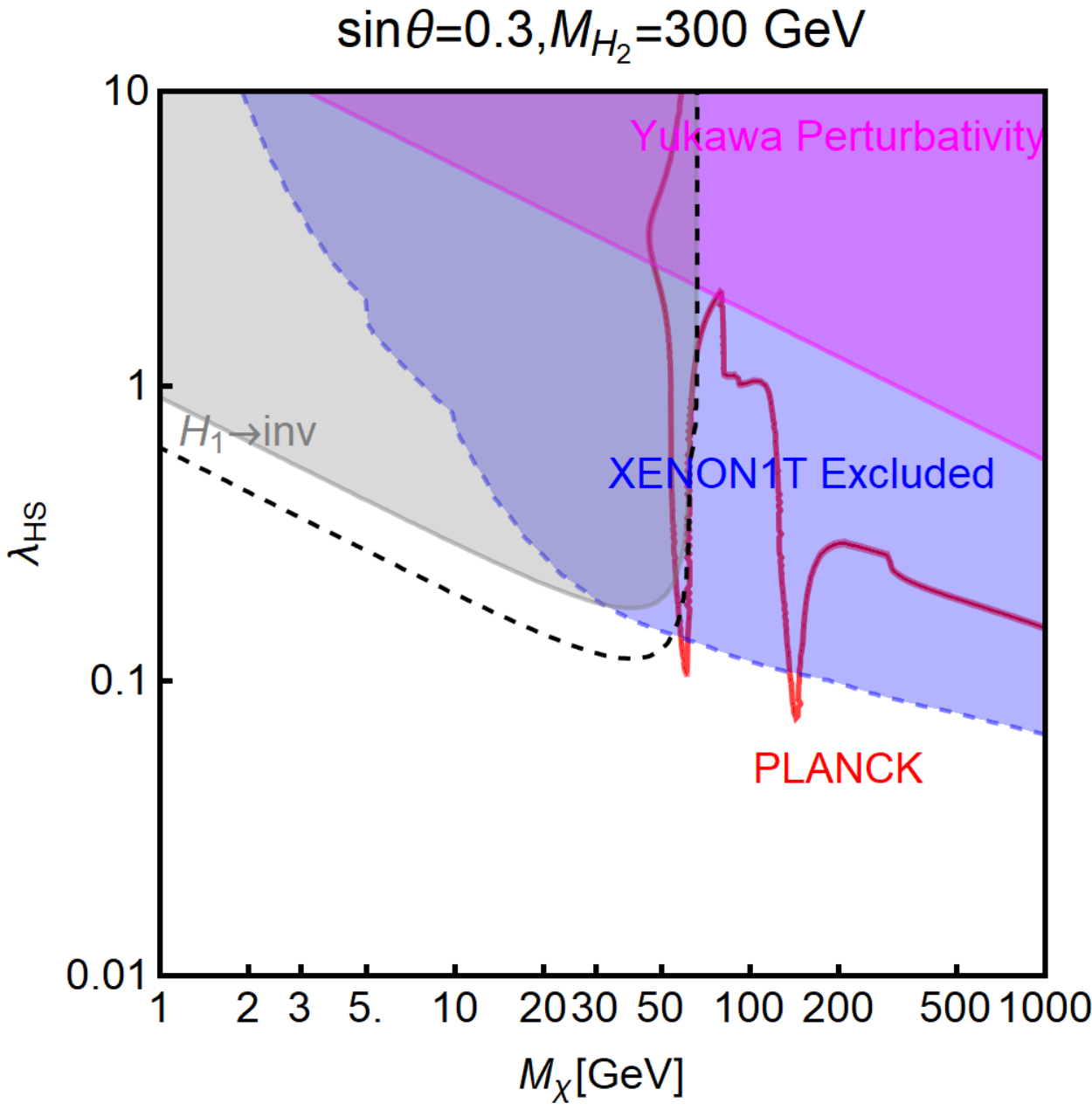}}
    \subfloat{\includegraphics[width=0.48\linewidth]{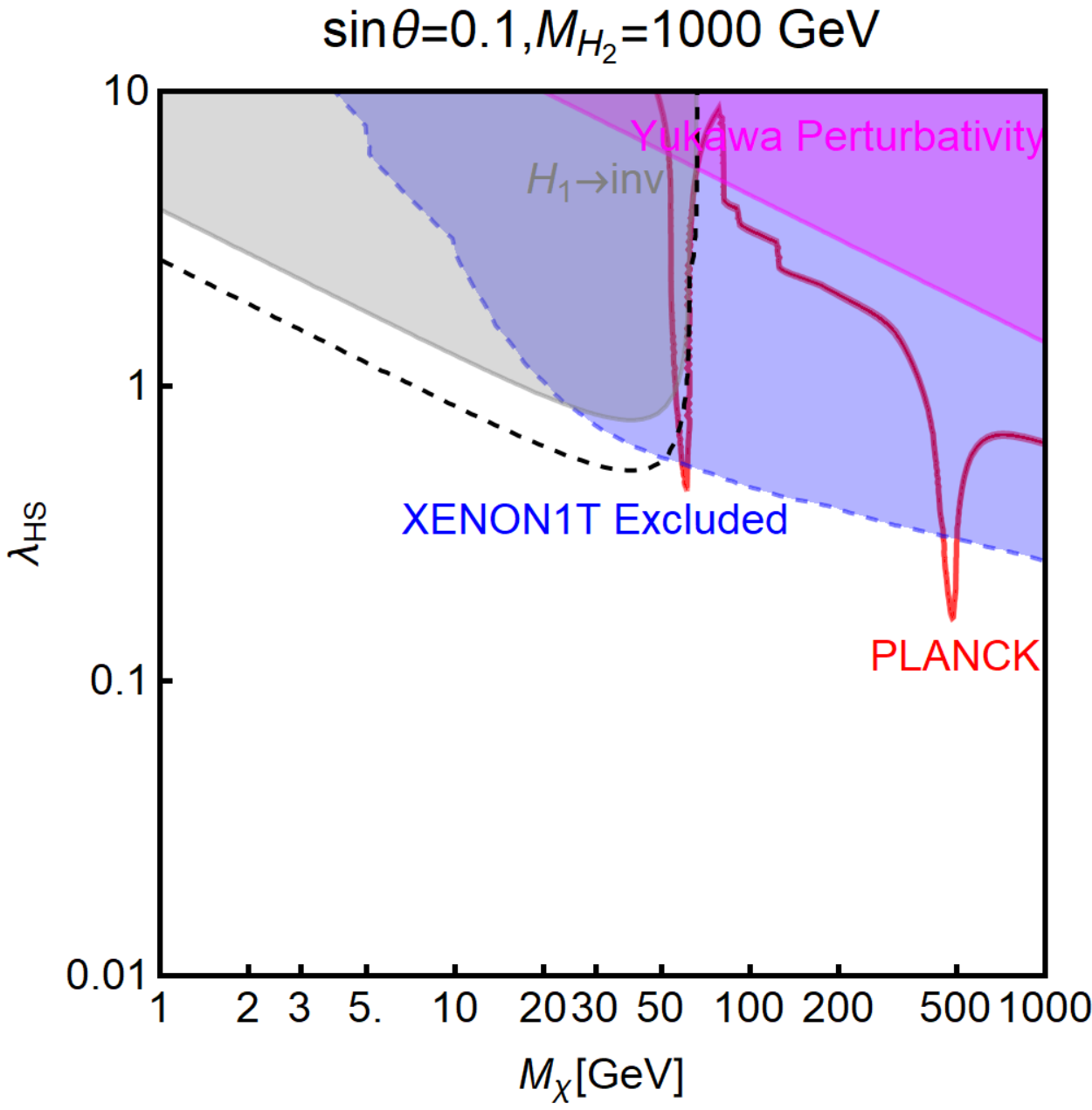}}
\vspace*{-1mm}
    \caption{Summary of constraints for the renormalizable fermionic Higgs-portal scenario in the $[M_\chi,\lambda_{HS}]$  plane, for three assignments of the $(\sin\theta,M_{H_2})$ pair, namely $\left(0.01,60\,\mbox{GeV} \right)$, $\left(0.3,300\,\mbox{GeV}\right)$ and $\left(0.1,1000\,\mbox{GeV} \right)$. The red contours correspond to the correct relic density, the blue region is excluded by constraints from DM direct detection (XENON1T), the gray region corresponds to the bounds from the Higgs invisible branching fraction and in the magenta region, the Yukawa coupling of the DM state becomes non-perturbative.}
    \label{fig:fermionic_DM}
\vspace*{-1mm}
\end{figure}

To ease the interpretation of the outcome of Fig.~\ref{fig:scanF}, we have summarized in Fig.~\ref{fig:fermionic_DM} the various constraints in the plane $[M_\chi, \lambda_{HS}]$ for some benchmark assignments of the $M_{H_2}$ and $\sin\theta$ parameters. In each panel, we display in red the region of parameter space corresponding to the correct DM relic density. The other phenomenological constraints are the exclusion limits from XENON1T shown in blue and the ones from searches for invisible Higgs decays at the LHC shown in gray. The green regions are excluded by unitarity and perturbativity constraints.\smallskip

The benchmarks that are illustrated in Fig.~\ref{fig:fermionic_DM} represent the following scenarios. The first one is the case of a very light $H_2$ state, close to the experimental bounds on extra Higgs bosons. Given the presence of a light Higgs mediator, the DM scattering cross section over nucleons is very high, so that most of the $[M_\chi,\lambda_{HS}]$ parameter space is ruled out by current constraints. There is, nevertheless a narrow window at low DM masses, in which all constraints are evaded and direct detection as well as the invisible Higgs branching ratio are  compatible with the sensitivity of future experiments. We notice, however, that the contour for the invisible Higgs branching fraction is independent of the DM mass and is still present for $M_\chi > \frac12 M_{H_1}$. This is due to the fact that invisible Higgs decays are dominated by the $H_1 \to H_2 H_2$ cascade process. \smallskip

The second benchmark still features the $M_{H_2} < M_{H_1}$ mass region but this time, above the $H_1 \rightarrow H_2 H_2$ threshold. As can be seen, we do not have  the correct DM relic density for light DM states, where constraints from invisible decays of the Higgs boson should have an impact. This is due to the fact that
DM annihilations are not effective there unless $M_\chi \approx \frac12 M_{H_2} \approx 450\,\mbox{GeV}$ or $M_\chi > M_{H_2}$. This particular benchmark is completely ruled out with the exception of a very tiny region around the $\frac12 M_{H_2}$ pole. \smallskip

The third benchmark corresponds to a high value of the Higgs mixing angle, $\sin\theta=0.3$, which is the present experimental constraints from the SM-like Higgs signals strengths, for the considered mass range of the $H_2$ state. The latter has a mass above that of the SM-like Higgs boson but is still light enough to be within the reach of near future collider searches. As can be clearly seen, in the third panel of Fig.~\ref{fig:fermionic_DM} we have no viable relic density in the region where the invisible $H_1 \rightarrow \chi \chi$ process is kinematically accessible. A viable DM particle is also ruled out at higher mass values by constraints from direct detection, again with the exception of the $\frac12 M_{H_1}$ and $\frac12 M_{H_2}$ poles. \smallskip

The fourth and last panel of Fig.~\ref{fig:fermionic_DM} illustrate the benchmark in which the  assignment of the $(\sin\theta,M_{H_2})$ pair of parameters is taken to be $(0.1,1\,\mbox{TeV})$. In this case, the effect of the extended Higgs sector on the DM scattering cross section, as compared to the effective Higgs-portal, becomes marginal and the $H_2$ boson decouples. The impact on DM constraints is analogous to the previous benchmark.\smallskip

\begin{figure}[!h]
\vspace*{-1mm}
    \centering
    \subfloat{\includegraphics[width=0.55\linewidth]{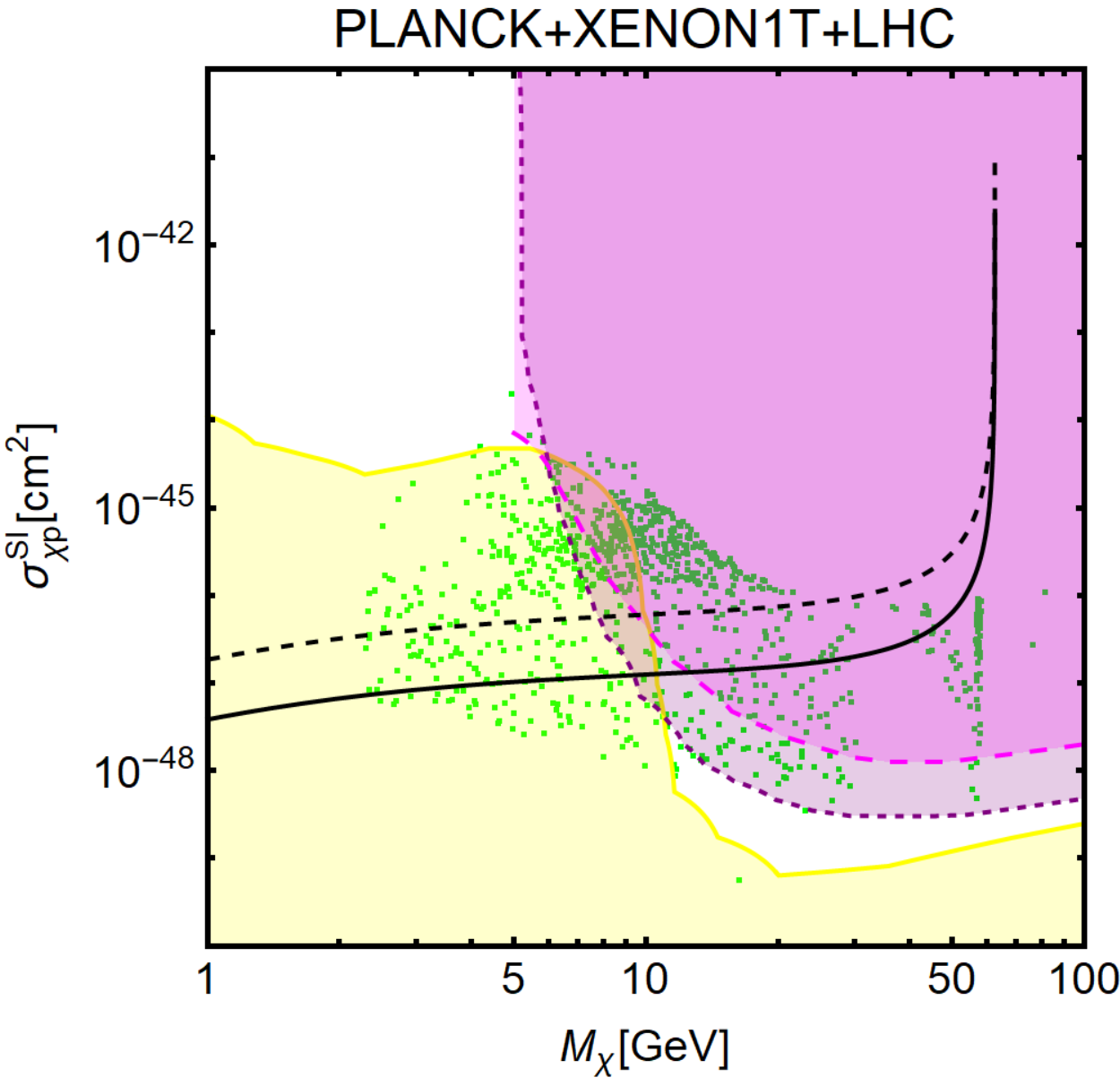}}
\vspace*{-1mm}
    \caption{
    The same as Fig. \ref{fig:scalar_DM_corr_plot} but for fermionic DM.}
    \label{fig:scanF2}
\end{figure}

Finally, we  illustrate in Fig.~\ref{fig:scanF2} the predicted DM-nucleon elastic scattering cross section as a function of the DM mass, but considering only the model points with an invisible Higgs branching fraction in the range $2.5\,\leq {\rm BR}(H_1 \rightarrow \mbox{inv}) \leq 11\,\%$. \smallskip

As usual, the distribution of points in the plane can be compared with the isocontours corresponding to the effective fermionic Higgs-portal. Since present constraints from XENON1T have been automatically included in the scan, the figure shows simply the projected sensitivities from XENONnT and DARWIN, as well as the region corresponding to the neutrino floor. As can be seen, the distribution of the model points significantly deviates from the case of the effective Higgs-portal. The reason for this outcome is twofold. First of all, as already pointed out, with the exception of the $H_1$ pole, a viable DM phenomenology requires a light $H_2$ state which dominates the DM scattering cross section, exceeding for the same value of the couplings, the prediction of the effective Higgs-portal scenario. The second reason is that the light Higgs states actually dominate, through the $H_1 \rightarrow H_2 H_2$ process, the invisible decay width of the SM-like Higgs boson.\smallskip 

In summary, one can conclude from the previous discussion that once the requirement of the correct relic density for the DM particle is enforced through the standard freeze-out paradigm,  the effective Higgs-portal cannot represent a viable limit of the more UV-complete model illustrated in this work.

\subsubsection{Vector Dark U(1) model}

We finally come to the case of the vector DM hypothesis and, as this case has been discussed already in the recent analysis of Ref.~\cite{Arcadi:2020jqf}, we simply give in this subsection some complementary material and information.

\begin{figure}[!h]
    \centering
    \includegraphics[width=0.55\linewidth]{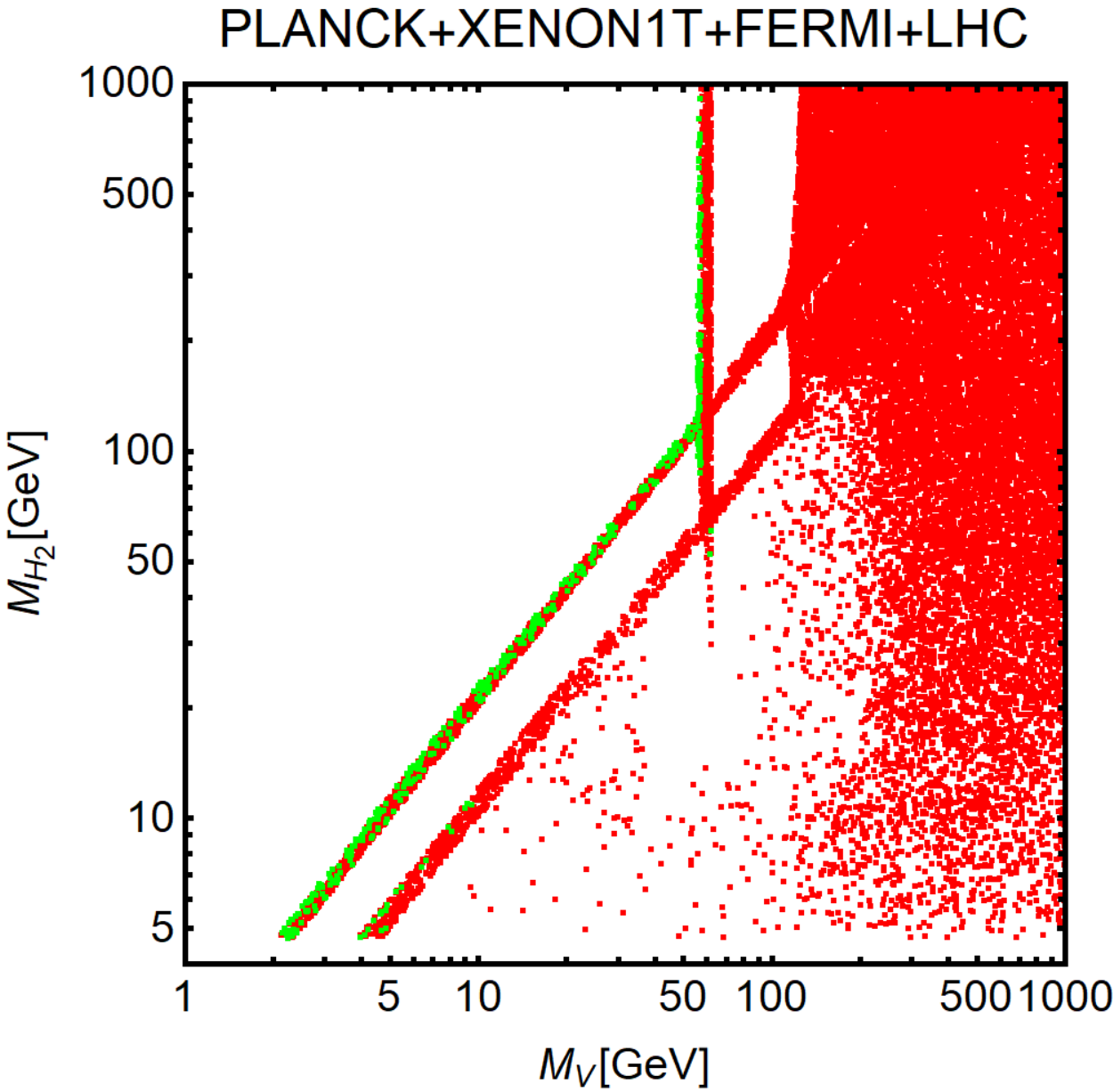}
\vspace*{-1mm}
    \caption{Parameter survey for the dark U(1) model in the $[M_V,M_{H_2}]$ plane. The red points correspond to those leading to a correct relic density. The green points feature, in addition, an invisible SM-like Higgs branching fraction between $2.5\,\%$ and $11\,\%$.}
    \label{fig:scan_U1}
\end{figure}

Applying the DM constraints to the survey over the [$M_V,M_{H_2},\tilde{g},\sin\theta$] parameter set introduced earlier yields the result shown in Fig.~\ref{fig:scan_U1} in the $[M_V,M_{H_2}]$ plane. Again, the red points correspond to model parameters that pass all present phenomenological constraints and lead to the correct DM cosmological relic density. We also display in green the area of parameters where the invisible branching ratio of the SM-like Higgs boson is within the reach of near future experiments. Again, a viable vector DM candidate in the kinematical reach of invisible decays requires a comparatively light $H_2$. Also, vectorial DM features an $s$-wave dominated cross section and, therefore, most of the unexcluded points lie in correspondence with the $M_V \sim \frac12 M_{H_2} $ resonance pole and near the threshold of the $VV \rightarrow H_2 H_2$ annihilation process.  \smallskip

\begin{figure}[!h]
\vspace*{-1mm}
    \centering
    \subfloat{\includegraphics[width=0.45\linewidth]{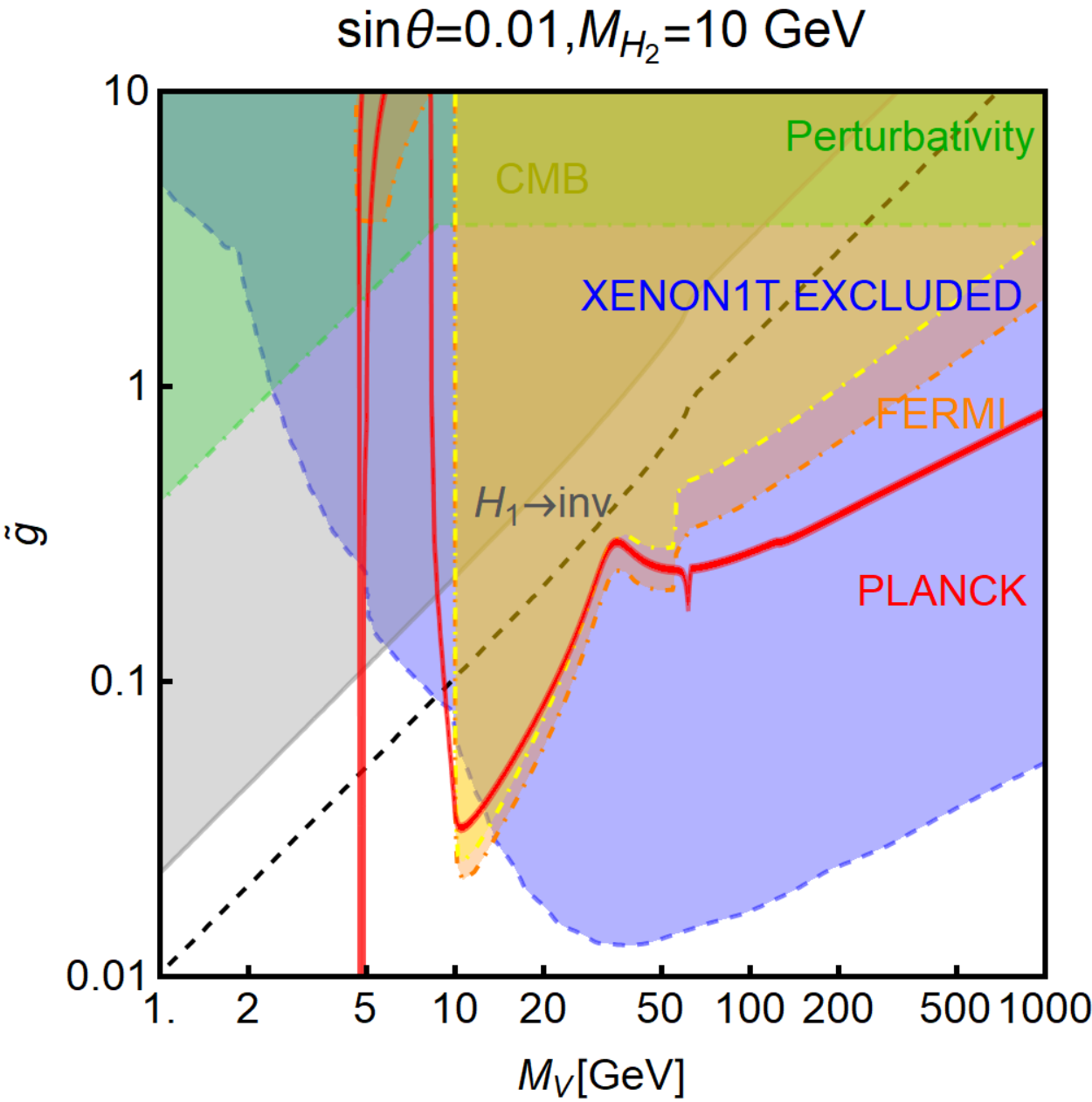}}
    \subfloat{\includegraphics[width=0.45\linewidth]{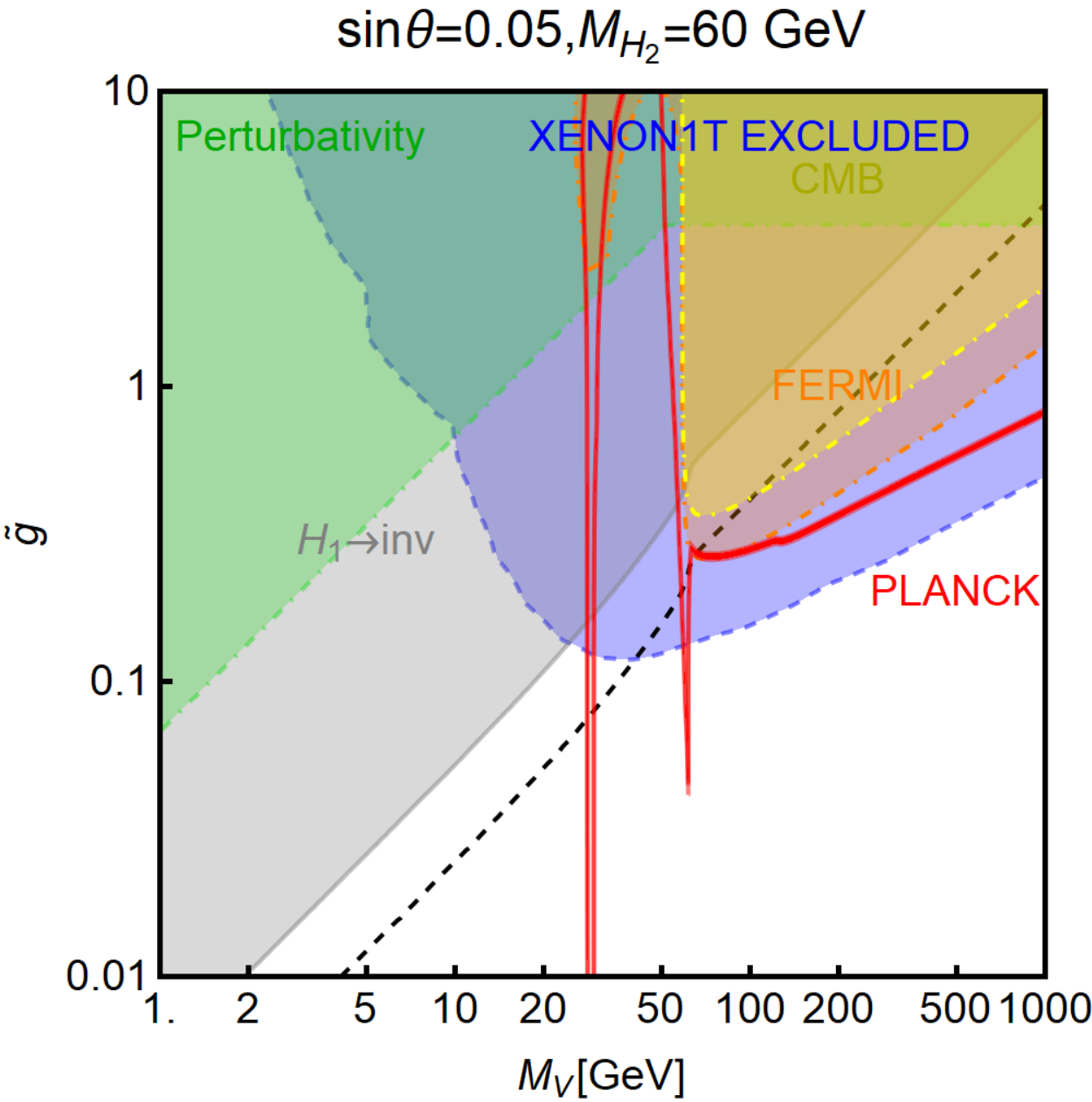}}\\
    \subfloat{\includegraphics[width=0.45\linewidth]{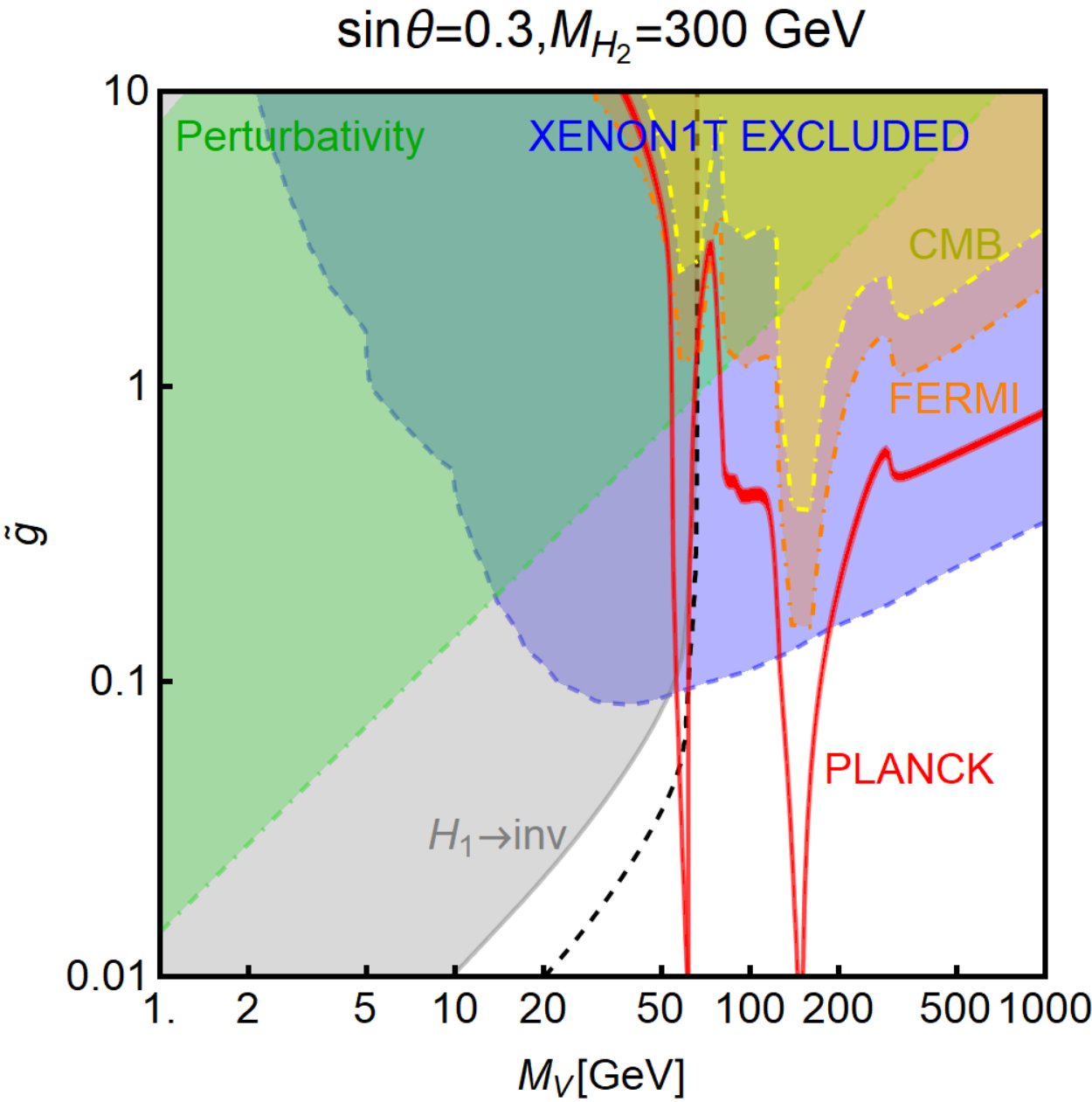}}
    \subfloat{\includegraphics[width=0.45\linewidth]{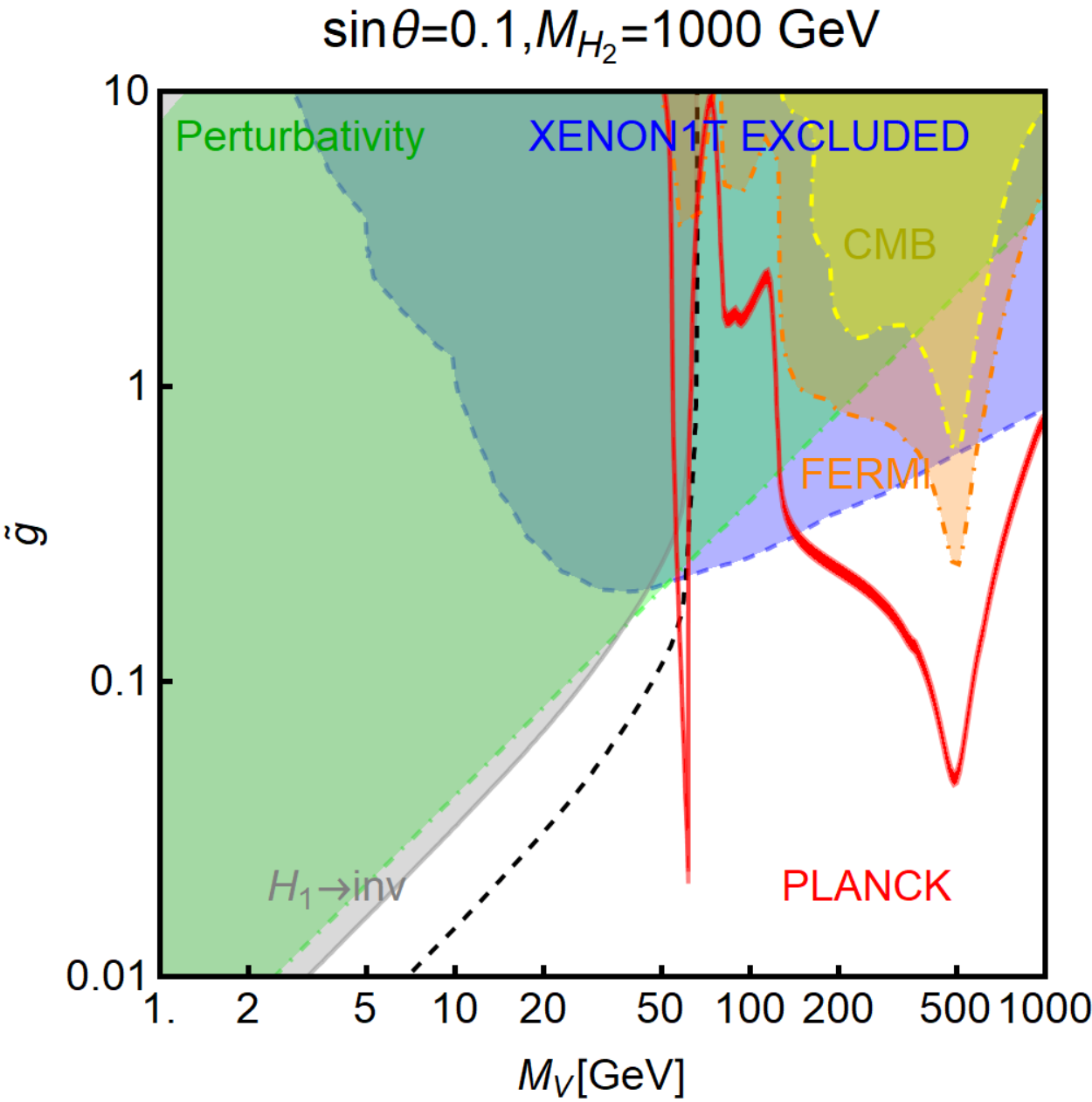}}
\vspace*{-1mm}
    \caption{Summary of constraints for benchmark points of the dark U(1) model. The red contours correspond to the correct relic density, the blue region is excluded by constraints from DM direct detection by XENON1T, the orange/yellow regions correspond to exclusions from FERMI/CMB and the gray region corresponds to the bounds from the Higgs invisible branching fraction. Finally, the green region is excluded by the constraints on the scalar self-couplings from perturbative unitarity.} 
    \label{fig:U1heavy}
\vspace*{-1mm}
\end{figure}

After the general survey of the parameter space of the model performed above, we have conducted again a more detailed  analysis focusing on four benchmarks scenarios identified by specific assignments of the $M_{H_2}$ and $\sin\theta$  parameters. While the outcome is very similar to the one obtained in the previous models, it is nevertheless interesting to note the strong impact of the perturbative unitarity bound as the mass of the $H_2$ state increases. In particular, for $M_{H_2}=1 \,\mbox{TeV}$, the constraint from the unitarity bound becomes comparable to that of the invisible decay rate of the SM-like Higgs boson. \smallskip

\begin{figure}[!h]
\vspace*{-1mm}
    \centering
    \subfloat{\includegraphics[width=0.55\linewidth]{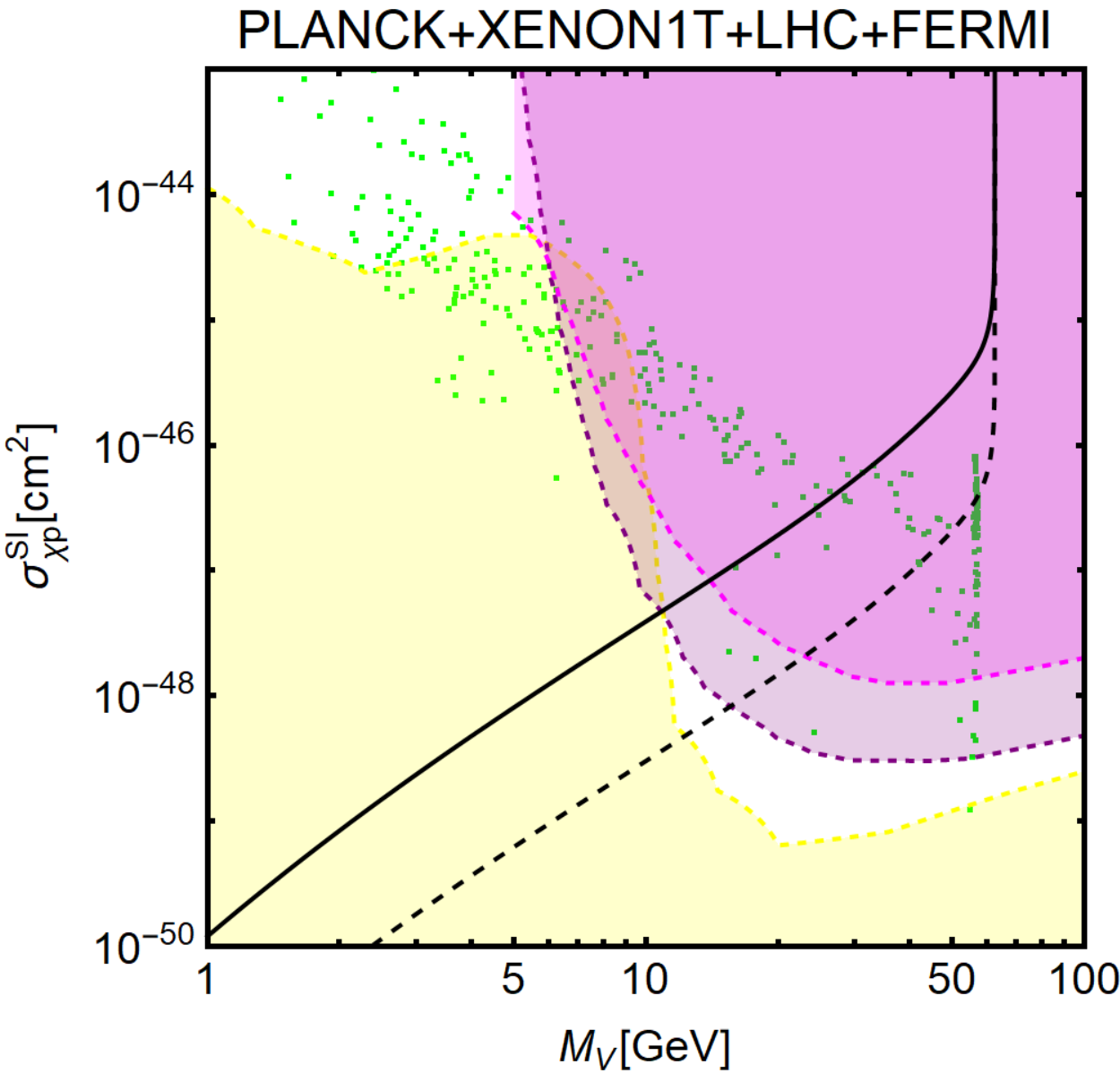}}
\vspace*{-1mm}
    \caption{The same as Fig.~\ref{fig:scanF2} for the dark U(1) model.}
    \label{fig:scan_U1_bis}
\vspace*{-1mm}
\end{figure}

Finally, a more detailed characterization of the region of parameter space in which invisible decays of the $H_1$ boson are kinematically allowed is made. The allowed region in the $[M_V,\sigma_{\chi p}^{\rm SI}]$ plane is shown in Fig.~\ref{fig:scan_U1_bis} where the same color code as in Fig.~\ref{fig:scanF2} has been adopted. We see again that the required presence of a light $H_2$ boson strongly influences the DM scattering cross section and the invisible decays of the $H_1$ state. While not dominant, as it was the case for a fermionic DM, the $H_1 \rightarrow H_2 H_2$ process still substantially contributes, with a rate around $50\,\%$, to the invisible decays of the SM-like Higgs boson.\smallskip 

Hence, it appears again that the conventional interpretation of the LHC results does not properly describe scenarios in which the experimentally measured relic density of the WIMP DM is achieved. However, and again, if a different production mechanism for the DM particle were to be assumed, a completely different conclusion would be reached.

\section{Extended Dark Matter sectors}

In this section, we will further extend our analysis and consider models in which the
spin 0, $\frac12$ and 1 DM states, and possibly the mediator of their interactions, are not the only particles added to the SM spectrum and address scenarios in which the DM sectors are enlarged.

\subsection{The scalar inert doublet model}

The so-called inert Higgs doublet model  \cite{Deshpande:1977rw,LopezHonorez:2006gr,Barbieri:2006bg,Ma:2006km,Arhrib:2013ela}
is probably the simplest and best example of a concrete and UV-complete scenario in which a
scalar DM is present that is not a singlet under the electroweak gauge group, while the Higgs sector is SM-like. In this framework, the scalar sector of the SM is extended with an additional SU(2) doublet $\phi^{'}$, leading to the following scalar potential 
\begin{eqnarray}
V \! = \!  \mu^2 |\phi|^2\! +\! \mu'^2 |\phi'|^2\! + \! \lambda_1 |\phi|^4 \! \! \! +\! \lambda_2 |\phi'|^4 \!  +\!  \lambda_3 |\phi|^2 |\phi'|^2 \! + \! \lambda_4 |\phi^{\dagger}\phi'|^2 \!  + \!  \frac{\lambda_5}{2}\left[ (\phi^{\dagger}\phi')^2 \! +\!  \mbox{h.c.} \right]\, . 
\label{eq:VIDM}
\end{eqnarray}
Contrary to the SM doublet $\phi$, the new field does not participate to the breaking of the electroweak  symmetry,  as it does not develop a vev. Furthermore, an {\it ad hoc} $\mathbb{Z}_2$ symmetry is introduced, forbidding a coupling of the $\phi^{'}$ field with pairs of SM fields.  After spontaneous symmetry breaking, it can be decomposed into the following physical states
\begin{equation}
\Phi'^ T  = \begin{pmatrix} H^+ \ , \  \frac{1}{\sqrt{2}} (H^0+ iA^0 ) \end{pmatrix}~,
\end{equation}
where, in terms of the SM vev $v$, the SM-Higgs field has a mass  $M_H^2=\mu^2+3 \lambda_1^2 v^2$, while the two charged  $H^{\pm}$ and the two neutral $H^0$ and $A^0$ states have masses given by 
\begin{eqnarray}
\label{eq:IDM_masses}
 M_{H^{\pm}}^2  = \mu'^2+\frac12 \lambda_3 v^2\, , \ \  
 M_{H^0}^2 =\mu'^2+\frac{1}{2}\lambda_L v^2\, , \ \ 
 M_{A^0}^2 =\mu'^2+\frac{1}{2}\lambda_R v^2 \, ,
\end{eqnarray}
where the two combinations of the quartic couplings which, respectively, correspond to the couplings of the SM-$H$ state to $H^0H^0$ and $A^0A^0$ pairs are given by 
\begin{eqnarray}
\lambda_{L/R}= \frac12 ( \lambda_3 + \lambda_4 \pm \lambda_5).
\label{eq:cplg:labdaL}
\end{eqnarray}
Similarly to the usual 2HDM, it is possible to identify, as free input parameters of the model, the four physical masses $M_H,M_{A^0},M_{H^0},M_{H ^{\pm}}$. As a consequence of the discrete $\mathbb{Z}_2$ symmetry, the lightest among the electrically neutral scalar bosons $H^0$ and $A^0$ will be the DM candidate and,  in our study, we consider only the possibility that it is the CP-even scalar $H^0$. Similarly to the case of the Higgs-portal for scalar DM discussed previously, $H^0$ interacts with the ordinary states through the SM-like Higgs field $H$ with a coupling given by $\lambda_L$.\smallskip

Being part of an SU(2) multiplet, the DM state can interact, with gauge strength, with the $W$ and $Z$ bosons as well. The prediction for the invisible branching fraction of the SM-like Higgs boson might nevertheless differ from the case of the effective Higgs-portal scenario since the pseudoscalar state $A^0$, if light enough, can also contribute to the SM Higgs invisible width through the process $H \rightarrow A^0A^0$. \smallskip

Following the usual strategy, we have performed a scan over the following parameters of the inert model in the ranges:
\begin{align}
     M_{H^0}, M_{A^0} \in \left[0.1,100 \right]\,\mbox{GeV}\, , \ 
     M_{H^{\pm}} \in \left[80,300\right]\,\mbox{GeV}\, , \
     |\lambda_i| < 4 \pi \, , 
\end{align}
and retained only the model points passing the following perturbativity and unitarity constraints on the scalar potential:
\begin{equation}
     \lambda_{1,2} >0,\,\,\, \lambda_3,\,\,\, \lambda_3+\lambda_4-|\lambda_5| >-2 \sqrt{\lambda_1 \lambda_2} \ . \end{equation}

We then applied the various experimental constraints, first  from electroweak precision tests which constrain the mass splitting between the new scalar states and some collider bounds on these new bosons \cite{Pierce:2007ut}, namely $M_{H^0}+M_{A^0}> M_Z$ from the invisible width of the $Z$ boson as well as the LEP2 constraints $M_{H^{\pm}}> 70\!-\!90\,\mbox{GeV}$ on the charged and $M_{A^0}> 100\,\mbox{GeV},M_{H^0} > 80\,\mbox{GeV}$ on the neutral states. Furthermore, we require a DM elastic scattering cross section on nucleons below the current XENON1T limit and, finally, a Higgs boson with an invisible branching fraction,  BR$(H \rightarrow \mbox{inv})={\rm BR}(H \rightarrow H^0 H^0)+{\rm BR}(H \rightarrow A^0A^0) <11\%$. We also indicate the regions in which this invisible branching ratio could be measured at HL--LHC and, thus, lies in the range between $2.5\,\%$ and $11\,\%$. We will also, subsequently, impose  the requirement that the DM state reproduces the measured cosmological relic density. \smallskip 

\begin{figure}
\vspace*{-1mm}
    \centering
    \includegraphics[width=0.55\linewidth]{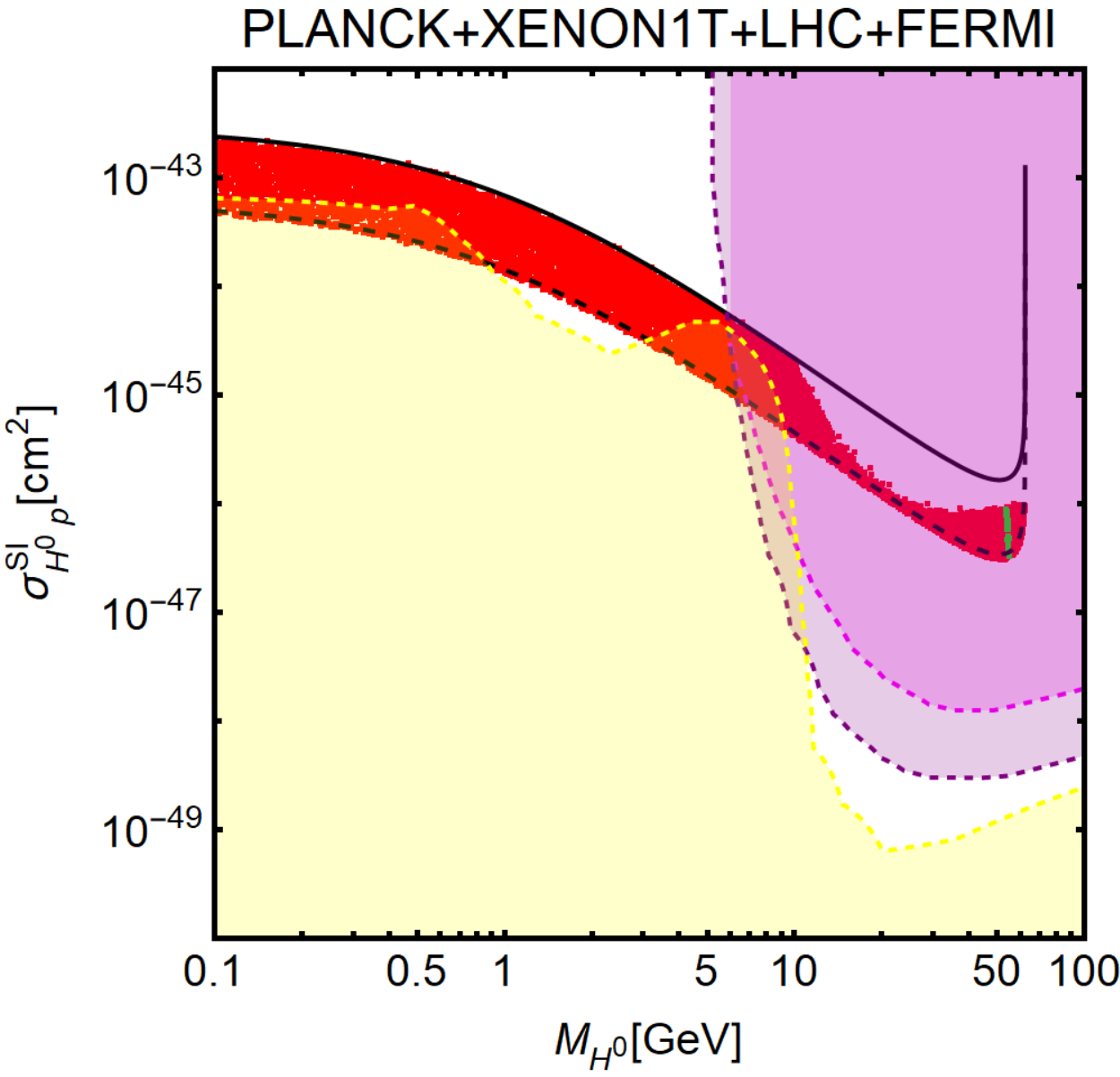}
\vspace*{-1mm}
    \caption{Model points (red) of the inert doublet model satisfying all the experimental constraints with the exception of the DM relic density, and with a Higgs invisible branching fraction comprised between $2.5\%$ and $11\%$. The solid and dashed lines illustrate the expected contours obtained with the above two values of the branching ratio in the effective Higgs-portal model for a scalar DM. The small area with the green points complies in addition with the correct relic density. The magenta/purple and yellow regions represent the sensitivity regions of the XENONnT and DARWIN experiments and the neutrino floor.}
    \label{fig:inert_doublet_DM}
\vspace*{-1mm}
\end{figure}

These model points are displayed in red in Fig.~\ref{fig:inert_doublet_DM} in the $[M_{H^0},\sigma_{H^0\, p}^{\rm SI}]$ plane. The additional model points which feature the correct DM relic density according to the WIMP paradigm are indicated in green. All the model points are lying within the isocontours corresponding to the predictions of the effective Higgs-portal which, therefore, represents an excellent limit. The picture once more substantially changes if the correct DM relic density is required, leaving only a very narrow strip marked in green in the figure, of unexcluded model parameter points which correspond to the $M_{H^0}\sim \frac12 M_H$ pole. Outside this region, the value of the DM coupling required to comply with the other experimental constraints is suppressed, leading to an overabundance of DM. Once again, the latter constraint could be avoided if a modified cosmological history of the universe is assumed.\\

\subsection{Singlet-doublet fermion model}

An alternative to extending the Higgs sector with a singlet field, to avoid the issue of a non renormalizable SM singlet fermion DM, is to keep the Higgs sector of the theory minimal and consider a fermionic dark sector made by different ${\rm SU(2)_L}$ such that the difference of their isospins is $\Delta I=\frac{1}{2}$ \cite{Lopez-Honorez:2017ora}, so that a Yukawa-like interaction for the DM state can be written. In this work we will just consider the minimal scenario, dubbed singlet--doublet model~\cite{Cohen:2011ec,Cheung:2013dua,Calibbi:2015nha}. Indeed, in more complicated models, like the doublet-triplet \cite{Freitas:2015hsa,Dedes:2014hga,Beneke:2016jpw} or the triplet-quadruplet \cite{Tait:2016qbg}, gauge interactions have a prominent impact on DM phenomenology. In particular, the correct relic density, assuming as usual the thermal paradigm, can be achieved only for DM masses in the multi-TeV range, hence far from kinematical reach of invisible Higgs decays.

In the singlet--doublet model, the SM spectrum is extended by introducing a Weyl fermion $S$, singlet with respect to the SM gauge group, and a pair $D_{L,R}$ of Weyl fermions transforming, instead, as doublets under SU(2) and with hypercharges $\pm \frac{1}{2}$. For these fields, one can assume the following decomposition
\begin{equation}
D_L=\left( \begin{array}{c} N_L \\ E_L \end{array} \right), \,\,\,\,\,
D_R=\left( \begin{array}{c} -E_R \\ N_R \end{array} \right), \,\,\, S \; . 
\end{equation} 
Given their quantum numbers, the new states will be coupled to the SM Higgs doublet according to the following Lagrangian
\begin{equation}
\label{eq:SD_lagrangian}
\mathcal{L}=-\frac{1}{2}M_S S^2-M_D D_L D_R -y_1 D_L \phi S-y_2 D_R \widetilde{\phi} S+\mbox{h.c.},
\end{equation}
where $\tilde{\phi}=i \sigma_2 \phi^{*}$ and, for completeness, we have also written explicitly the mass terms of the new fields. These new fields can thus also be coupled with the SM leptons through the Higgs doublet. In order to avoid this possibility, which would be dangerous for the stability of the DM particle, the presence of a discrete $\mathbb{Z}_2$ symmetry is assumed. After electroweak symmetry breaking, the masses of the neutral components of the new states are obtained by diagonalizing, with a unitary transformation $U$, the following mass matrix
\begin{equation}
\label{eq:SD_mass_matrix}
M=\left(
\begin{array}{ccc}
M_S & {y_1 v}/{\sqrt{2}}~~ & {y_2 v}/{\sqrt{2}}~~ \\
{y_1 v}/{\sqrt{2}}~~ & 0 & M_D \\
{y_2 v}/{\sqrt{2}}~~ & M_D & 0
\end{array}
\right)\, ,
\end{equation}
leading to three Majorana fermions
\begin{equation}
\chi_i=S U_{i1}+D_L U_{i2}+D_R U_{i3}\, .
\end{equation}
the lightest of which is the DM candidate. The new fermionic spectrum is completed by an electrically charged Dirac state $\psi^{\pm}$ with a mass $M_\psi \simeq M_D$. The relevant interaction Lagrangian finally reads \cite{Arcadi:2017kky}
\begin{align}
\label{eq:physical_SD}
\mathcal{L}&=g_{H \chi_i \chi_j}H \bar \chi_i \chi_j +\mbox{h.c.}\nonumber\\
&+\bar \chi_i \gamma^\mu \left(g_{Z \chi_i \chi_j}^V -g_{Z \chi_i \chi_j}^A \gamma_5\right) \chi_j Z_\mu+\bar \psi^{-}\gamma^\mu \left(g_{W^{\mp} E^{\pm} N_i}^V-g_{W^{\mp} E^{\pm} \chi_i}^A \gamma_5 \right) W^{-}_\mu \chi_i \nonumber\\
& -e \bar \psi^- \gamma^\mu \psi^- A_\mu -\frac{g}{2 \cos^2 \theta_W}(1-2 \sin^2\theta_W) \bar \psi^- \gamma^\mu \psi^- Z_\mu  \; , 
\end{align}
with $g$ being the SU(2) gauge coupling constant while $\theta_W$ is the Weinberg angle\footnote{This scenario can be seen as a limit of the minimal supersymmetric SM where the extra Higgs bosons, in addition to the SM-like one $h$ which acts as a portal, are heavy; see e.g.~Refs~\cite{Djouadi:2005dz,Baer:2004xx}.}.\smallskip

As it can be easily seen, the DM is coupled also with the gauge bosons as a consequence of its mixing with the SU(2) doublet. When writing the general expression of the coupling of DM pairs with the neutral bosons of the SM, one then obtains \cite{Calibbi:2015nha,Arcadi:2017kky}
\begin{align}
     g_{h \chi_1 \chi_1}&=-\frac{(2 y_1 y_2 M_D +(y_1^2+y_2^2) M_{\chi_1})v}{M_D^2+\frac{v}{2}(y_1^2+y_2^2)+2 M_S M_{\chi_1}-3 M_{\chi_1}^2}\ , \nonumber \\
     g_{Z \chi_1 \chi_1}&=-\frac{M_Z v (y_1^2-y_2^2)\left(M_{\chi_1}^2-M_D^2\right)}{2 (M_{\chi_1}^2-M_D^2)^2+v^2 (4 y_1 y_2 M_{\chi_1} M_D+(y_1^2+y_2^2)(M_{\chi_1}^2+M_D^2))}
\, , 
\end{align}
and taking the limit $M_D > y_{1,2}v \gg M_S$, one  obtains
\begin{equation}
   g_{h \chi_1 \chi_1} = -\frac{2 y_1 y_2 v}{M_D}\, ,   \ \ \ 
   g_{Z\chi_1 \chi_1} \simeq  \frac{M_Z v \left(y_1^2-y_2^2\right)}{2 M_D^2} \, .
\end{equation}
In other words, once the states $\chi_2,\chi_3$ and $\psi$ become much heavier than the DM particle, their impact on the phenomenology diminishes, the singlet-doublet model will resemble the effective Higgs-portal for a fermionic DM, and the new physics scale $\Lambda$ can be identified with the mass term $M_D$ of the new fermion doublet. It is  worth noting that the coupling of the DM with the $Z$ boson exactly vanishes for $|y_1|=|y_2|$. \smallskip 

It should also be mentioned that another model, analogous to the singlet-doublet one, can be realized also with vector-like fermions \cite{Yaguna:2008hd,Arcadi:2019lka}. However, in this case, DM-related observables are mostly determined by the DM interactions with the Z boson. Consequently, the latter scenario does not represent a viable limit of the EFT Higgs-portal scenario discussed here. This possibility will therefore be ignored in our analysis. \smallskip 

To study the singlet-doublet fermionic model, we have performed a scan of the $(M_S,M_D,y,\theta)$ parameter set over the following ranges
\begin{align}
    M_S \in \left[0.1,3000\right]\,\mbox{GeV} \, , \ 
    M_D \in \left[100,3000\right]\,\mbox{GeV} \, , \
    y \in \left[10^{-3},10\right]\, , \ 
    & \tan\theta \in [-20,20] \, , 
\end{align}
and, again, retained only the model points with an invisible Higgs branching fraction of BR$(H \rightarrow {\rm inv}) < 11\,\%$. We also excluded points in which the $Z$ boson decays into invisible DM particles, by enforcing the constraint $\Gamma(Z \rightarrow {\rm inv}) < 2.3 \,\mbox{MeV}$ from precision LEP measurements at the $Z$ boson resonance \cite{Zyla:2020zbs}.\smallskip 

\begin{figure}
    \centering
    \subfloat{\includegraphics[width=0.5\linewidth]{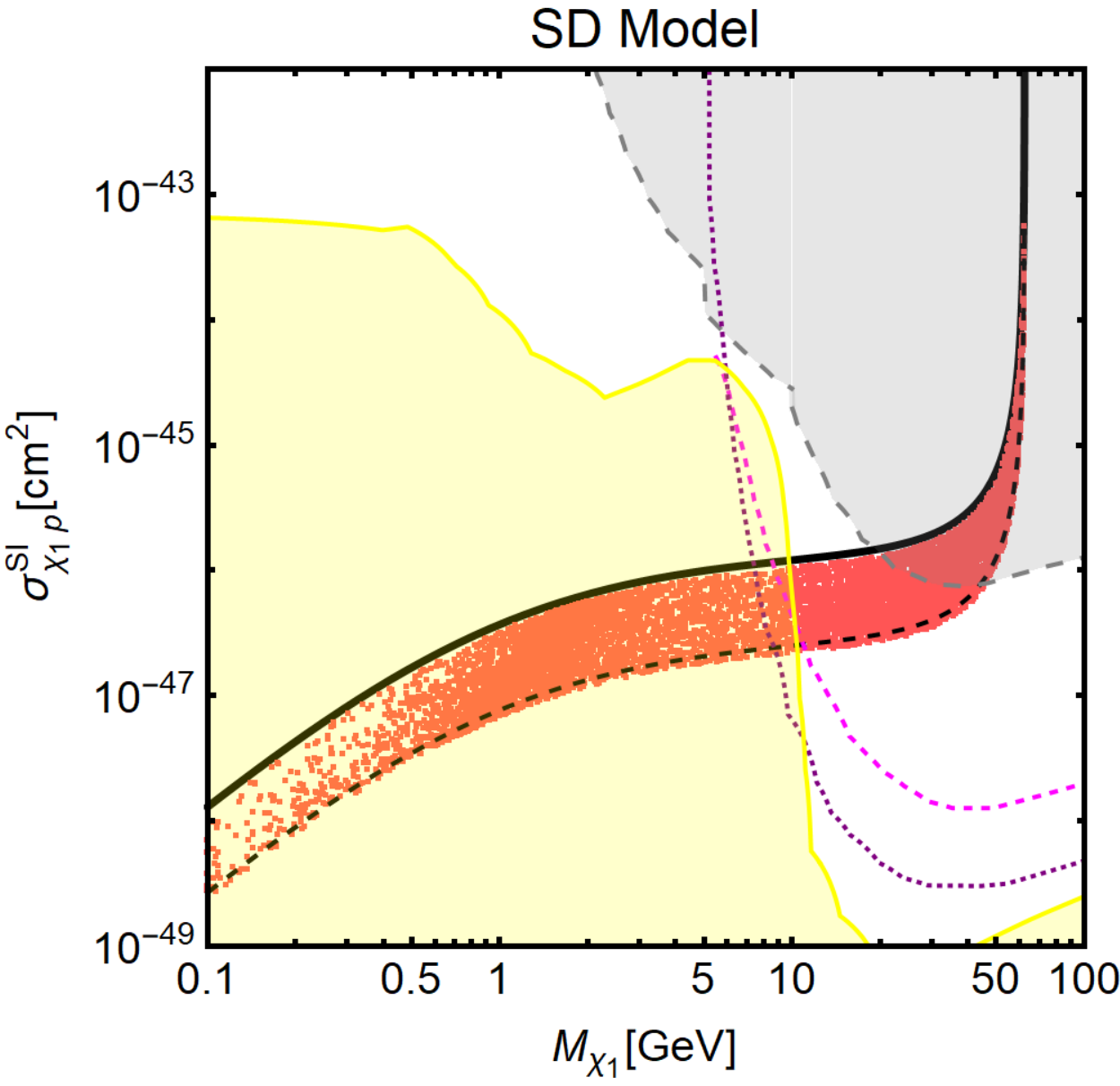}}
    \subfloat{\includegraphics[width=0.5\linewidth]{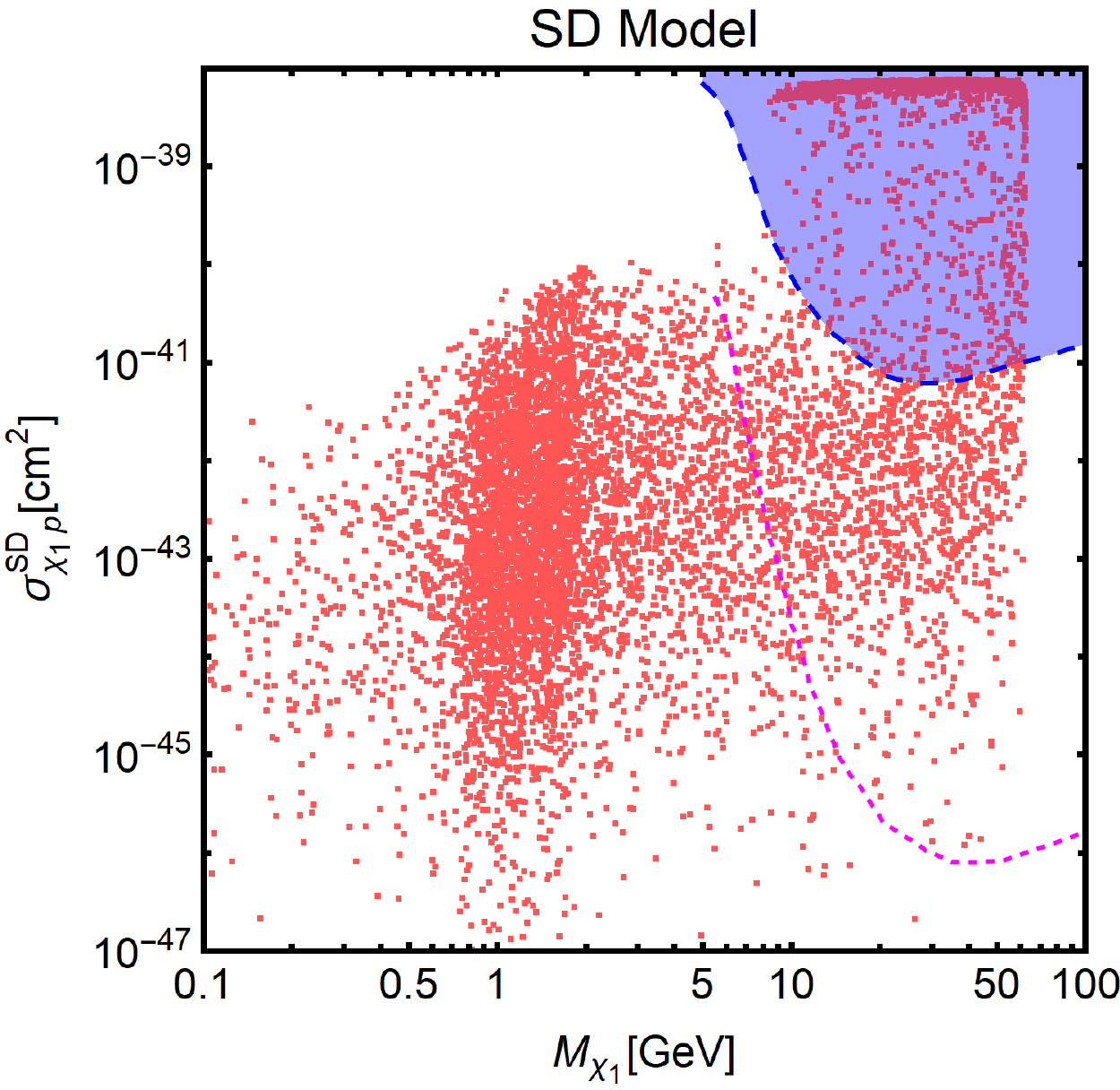}}
    \caption{Model points in the planes $[M_{\chi_1},\sigma_{\chi_1 p}^{\rm SI}]$ 
(left panel) and $[M_{\chi_1},\sigma_{\chi_1 p}^{SD}]$ (right panel) obtained from a survey of the parameter space of the fermionic singlet-doublet model, with the requirement of an invisible Higgs branching ratio of $2.5 \,\% < \mbox{BR}(H \rightarrow {\rm inv}) < 11\,\%$ (red points). The gray/blue regions are excluded by the direct detection bound from XENON1T on DM spin dependent and independent cross sections, the magenta and purple lines are the projected sensitivities from the XENONnT and DARWIN experiments, respectively, while the yellow region in the left panel corresponds to the so-called neutrino floor.}
    \label{fig:pBrSD}
\end{figure}

The result of the scan on the model parameters is summarized in Fig.~\ref{fig:pBrSD}.  
The left panel shows in red the model points in the $[M_{\chi_1},\sigma_{\chi_1 p}^{\rm SI}]$ plane with an Higgs invisible branching fraction smaller than the LHC limit, namely  BR$(H \rightarrow \mbox{inv}) = 0.11$, but above the expected sensitivity of the HL-LHC upgrade, i.e $2.5\%$. The plot also shows, as black solid and dashed contours, the corresponding expected spin-dependent cross section for the effective Higgs-portal fermionic model.\smallskip

The results are totally analogous to what already obtained in the inert doublet scalar model, with the model points entirely lying within the isocontours corresponding to the SM-like Higgs boson have been also compared with the ones coming from dedicated DM searches. The gray region in the left panel of Fig.~\ref{fig:pBrSD} represents, indeed, the area excluded by DM searches by the XENON1T experiment \cite{Aprile:2018dbl,Aprile:2019xxb}. Furthermore, the expected sensitivity from next future Xenon-based experiments,  XENONnT and DARWIN, as well as the so-called neutrino floor (yellow region) are also depicted.\smallskip 

As can be easily seen, for $M_{\chi_1} \gtrsim 10\,\mbox{GeV}$, DM direct detection is the most efficient probe. In contrast, for lower DM masses, searches of invisible Higgs decays can probe DM masses lying deeply inside the neutrino floor area. As the DM particle is also coupled to the $Z$-boson, a spin dependent cross section is also present. We have thus shown, in the right panel of Fig.~\ref{fig:pBrSD}, the predicted spin dependent  cross section as a function of the DM mass and compared it with the exclusion bound, this time shown as a blue region, also  given by XENON1T \cite{Aprile:2019dbj}, and the projected sensitivity from XENONnT.\smallskip 

\begin{figure}
    \centering
    \includegraphics[width=0.55\linewidth]{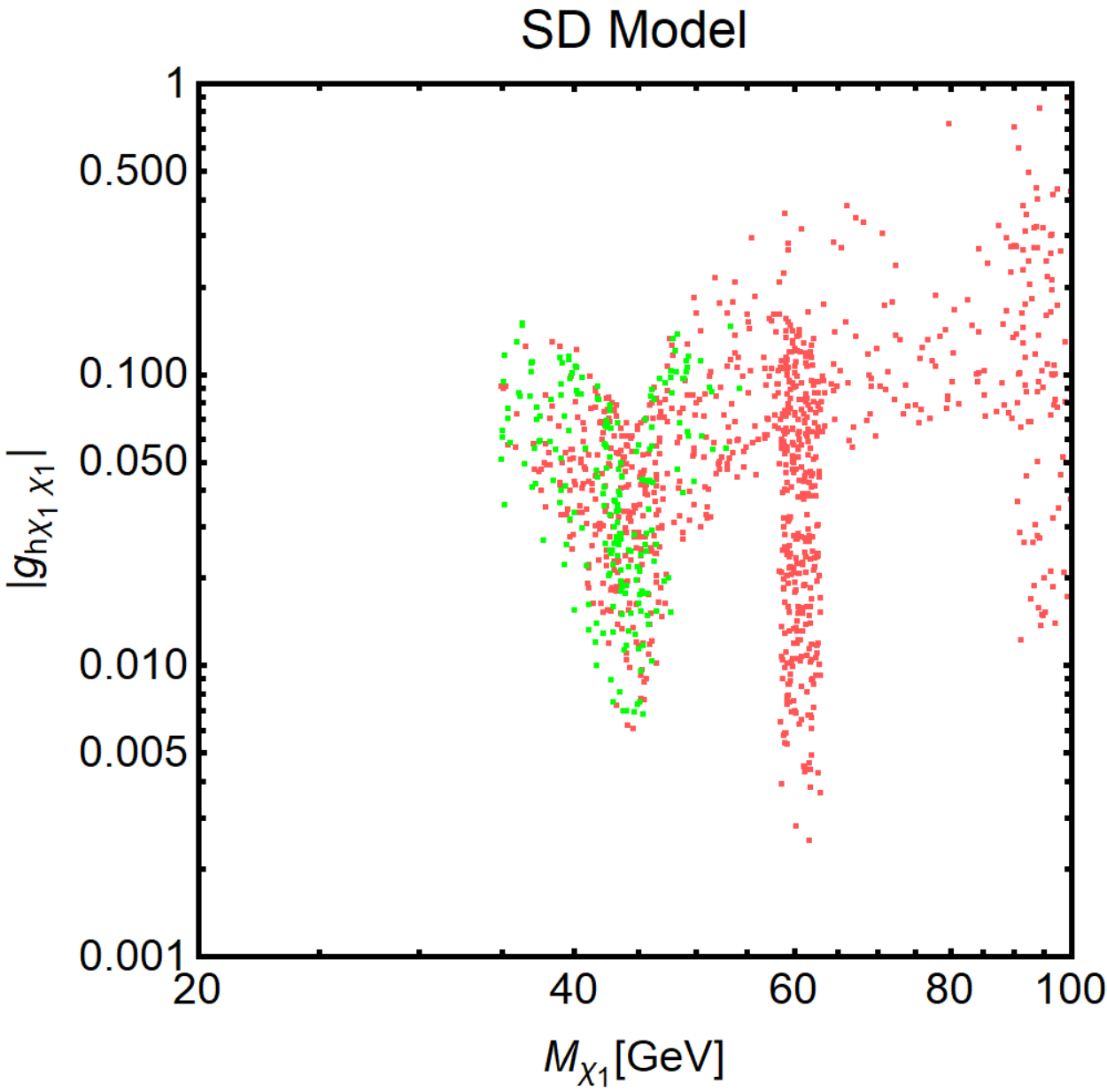}
    \caption{Model points (red), in the $[M_{\chi_1},c_{H \chi_1 \chi_1}]$ plane with an invisible Higgs rate BR$(H \rightarrow \mbox{inv})<11\,\%$ and complying with constraints from DM phenomenology. The model points in green are those which have, in addition,  
BR$(H \rightarrow \mbox{inv})> 2.5\, \%$.}
    \label{fig:pBrSDrelic}
\end{figure}

Fig. \ref{fig:pBrSDrelic} considers in more detail the impact from DM phenomenology. There, we display the model points with an invisible Higgs branching ratio that pass the constraints from LHC Higgs searches, BR$(H \rightarrow \,\mbox{inv})< 11\,\%$ and which, in addition, feature the correct WIMP relic density, as well as DM scattering cross sections, both spin dependent and spin independent, compatible with the current experimental limits. 

Finally, the points with invisible branching fraction above BR$(H \rightarrow \,\mbox{inv})=2.5\,\%$ have been highlighted in green. As can be easily seen, a viable relic density is obtained, within the kinematically favored region for invisible Higgs decays, only for $40\! \lesssim M_{\chi_1}\! \lesssim \!60\,\mbox{GeV}$, where the DM annihilation cross section can be enhanced around the $Z$ and $H$ poles.   

\subsection{Vector DM from higher dark gauge groups}

\subsubsection{The dark SU(2) option} 

UV-complete realizations of the Higgs-portal scenario with a vector DM can be more elaborate than the minimal dark U(1) model introduced and discussed in the previous section. Indeed, the DM state can belong to a more involved hidden sector that is subject to larger gauge symmetries. A next-to-minimal option would consist of a dark SU(2) sector (see e.g. \cite{Hambye:2009fg,Bernal:2015ova,Gross:2015cwa}). In such case, the vector DM sector is represented by three mass degenerate states $V^{1,2,3}$ whose masses originate from the vev of a dark Higgs doublet field. Similarly to the dark U(1) model of the previous section, one can define a mixing angle $\theta$ such that the relevant physical Lagrangian will be given by 
\begin{align}
    & \mathcal{L}=\frac{\tilde{g} M_{V}}{2}\left(-\sin\theta H_1 +\cos\theta H_2\right)\sum_{a=1,3}V_{\mu}^a V^{\mu\,a}\nonumber\\
    & +\tilde{g}\epsilon_{abc}\partial_\mu V_\nu^a V^{b\,\mu}V^{c\,\nu}-\frac{\tilde{g}^2}{4}\left[ {\left(V_\mu^a V^{a\,\mu}\right)}^2-\left(V_\mu^a V_\nu^a V^{b\,\mu} V^{b\,\nu}\right)\right]\, , 
\end{align}    
where $\tilde g$ is the new gauge coupling constant and $M_V$ is the mass of the vector states.\smallskip

\begin{figure}
    \centering
    \subfloat{\includegraphics[width=0.48\linewidth]{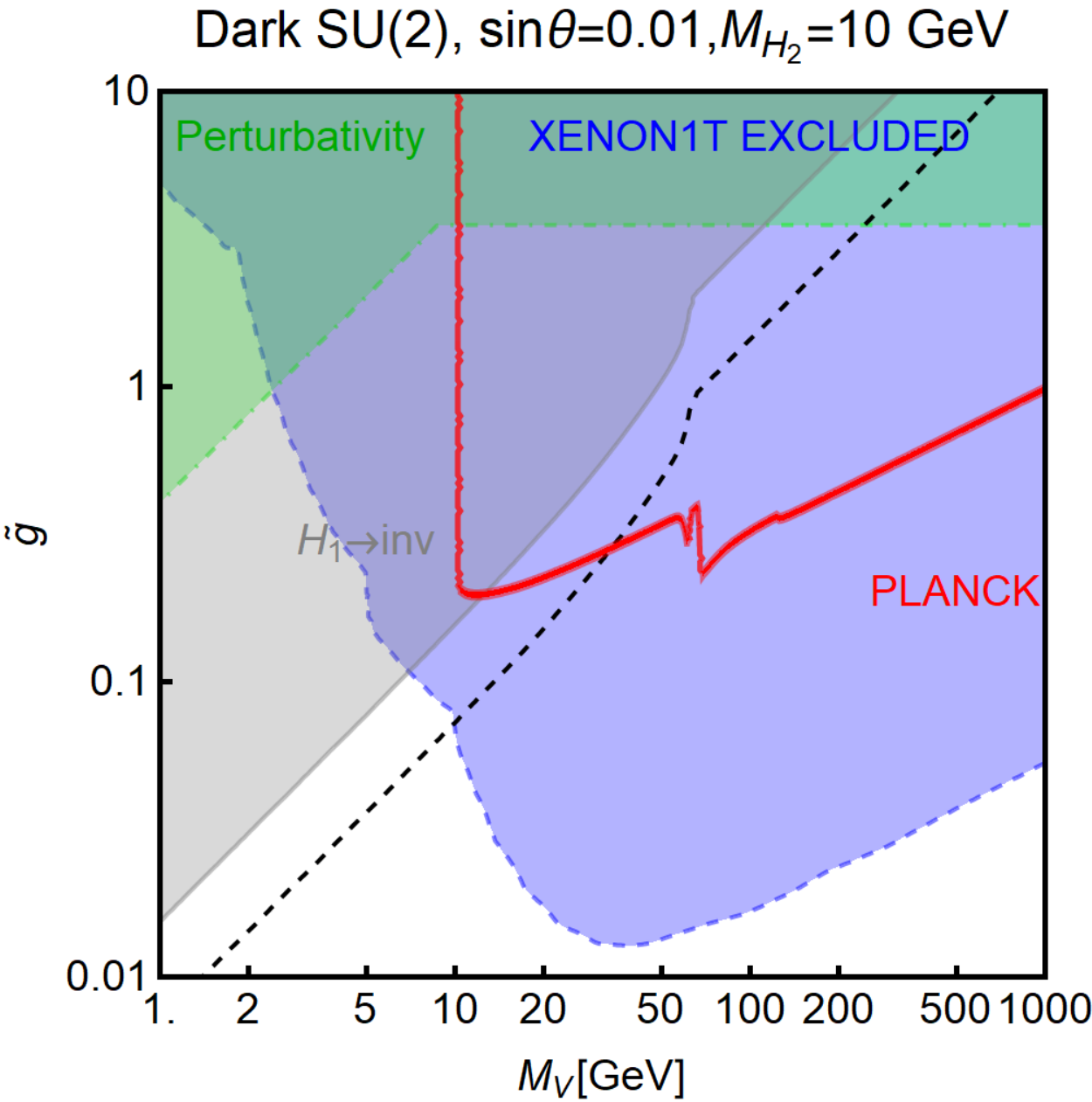}}
    \subfloat{\includegraphics[width=0.48\linewidth]{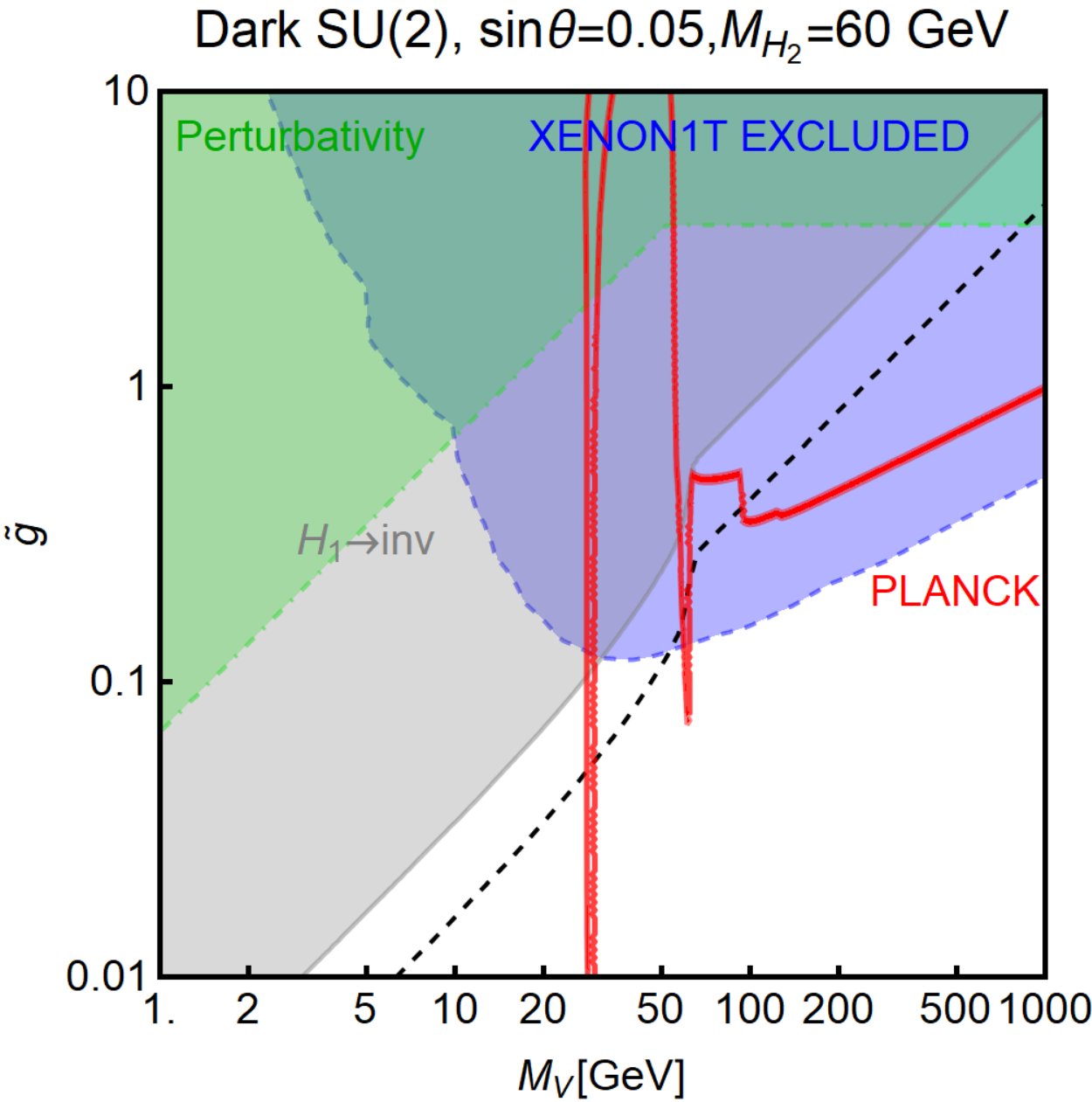}}\\
    \subfloat{\includegraphics[width=0.48\linewidth]{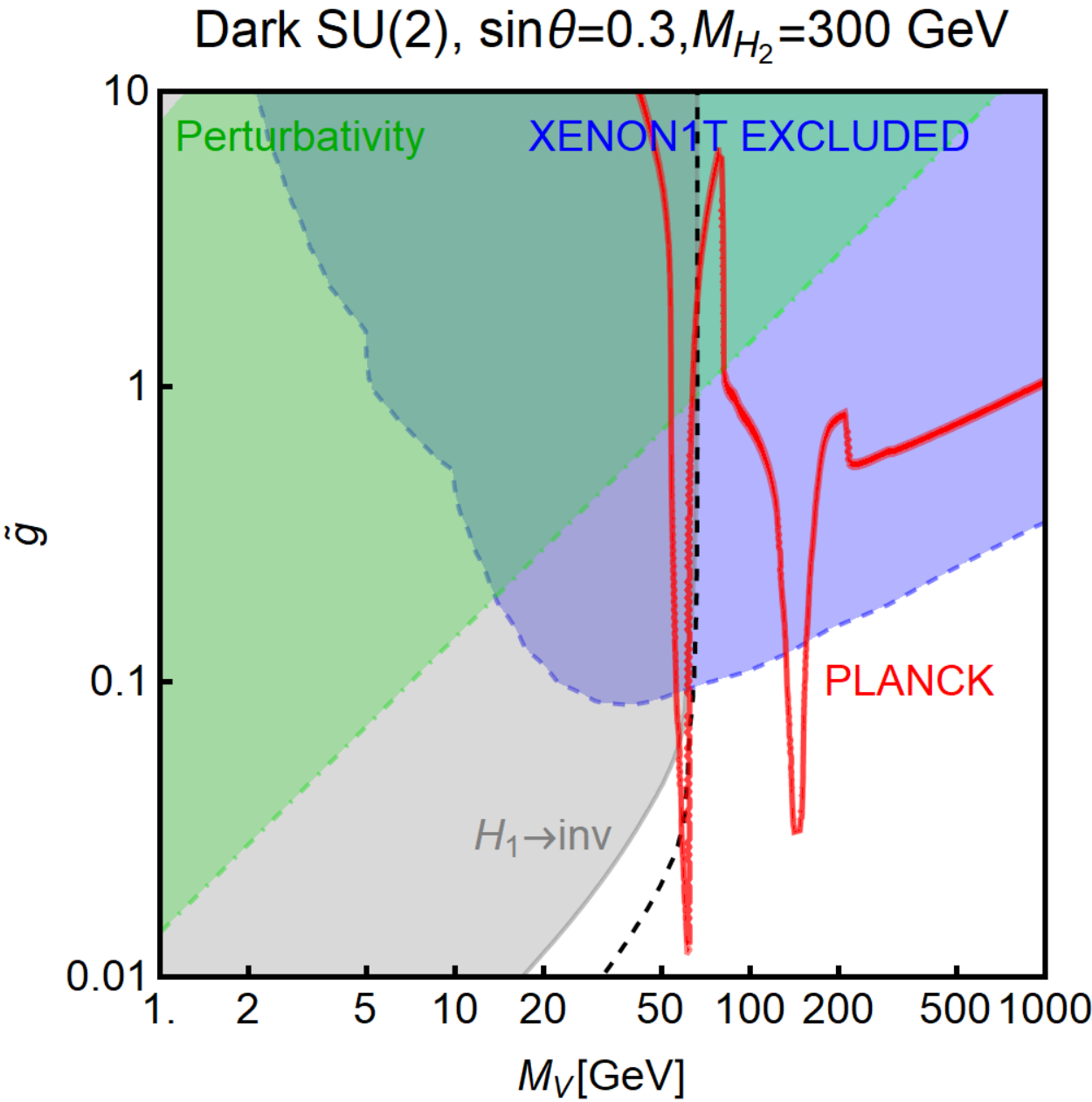}}
    \subfloat{\includegraphics[width=0.48\linewidth]{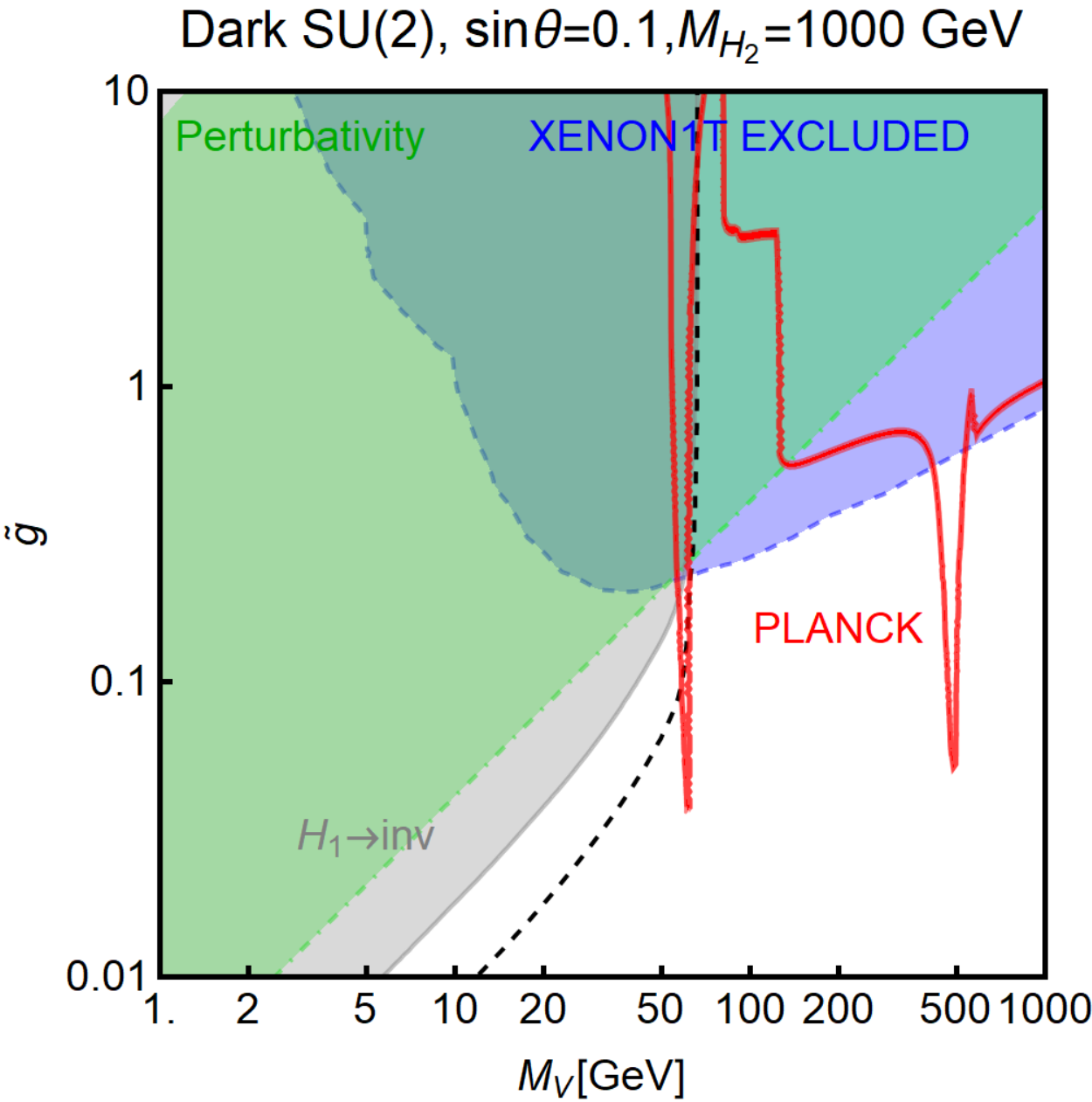}}
    \caption{\footnotesize{The same as Fig.~\ref{fig:U1heavy} but for the dark SU(2) model.}}
    \label{fig:dark_SU(2)}
\end{figure}

The three SU(2) vectors are exactly mass degenerate and have the same couplings to the $H_1$ and $H_2$ fields \footnote{Such mass degeneracy can be removed by considering a more complicated dark Higgs sector. See e.g. \cite{Nomura:2020zlm} for a recent example.}. They therefore behave essentially like a single DM particle. The phenomenology, hence, strongly resembles that of the dark U(1) model, with a redefinition of the gauge coupling $\tilde{g}$. For this reason, we will not analyze the model in detail and simply summarize in Fig.~\ref{fig:dark_SU(2)} the outcome of the various constraints in the same benchmarks used for the dark U(1) model given in Fig.~\ref{fig:U1heavy}. For simplicity, we have omitted the subdominant indirect bounds  for these benchmarks. As can be seen, the two sets of figures are very similar and, hence,  this dark SU(2) option can be ignored in the rest of our discussion. \smallskip

\subsubsection{The Dark SU(3) extension: fields and interactions}

A significantly different phenomenology is obtained when considering a dark SU(3) symmetry. As was already shown in Ref.~\cite{Gross:2015cwa}, the minimal option to break the SU(3) symmetry is through two Higgs fields $\phi_1$ and $\phi_2$ that belong to the fundamental representation of the dark gauge symmetry group. In the unitary gauge, the two fields decompose as follows
\begin{equation}
    \phi_1= \frac{1}{\sqrt{2}}\left(
    \begin{array}{c}
         0  \\
         0  \\
         v_1+h_1
    \end{array}
    \right),\,\,\,\,\phi_2= \frac{1}{\sqrt{2}}\left(
    \begin{array}{c}
         0  \\
         v_2+h_2  \\
         v_3+h_3+i \left(v_4+ h_4\right)
    \end{array}
    \right) \, . 
\end{equation}
The Lagrangian of the dark SU(3) model, including the Higgs part, can then be written as 
\begin{align}
    & \mathcal{L}_{\rm Higgs}=-\frac{\lambda_H}{2}|\phi|^4-m_H^2 |\phi|^2 \, , \nonumber\\
    & \mathcal{L}_{\rm portal}=-\lambda_{H11}|\phi|^2 \phi_1^2-\lambda_{H22}|\phi|^2 \phi_2^2+\left(|\phi|^2 \phi_1^{\dagger}\phi_2+\mbox{h.c}\right) \, ,  \nonumber\\
    & \mathcal{L}_{\rm hidden}=-\frac{1}{2}\mbox{Tr}\left\{V_{\mu \nu}V^{\mu \nu}\right\}+|D_\mu \phi_1|^2+|D_\mu \phi_2|^2-V_{\rm hidden} \, , \nonumber\\
    & V_{\rm hidden}=m_{11}^2 |\phi_1|^2+m_{22}^2 |\phi_2|^2-m_{12}^2\left(\phi_1^{\dagger}\phi_2+\mbox{h.c.}\right) \, ,  \nonumber\\
    & +\frac{\lambda_1}{2}|\phi_1|^4+\frac{\lambda_2}{2}|\phi_2|^4+\lambda_3 |\phi_1|^2 |\phi_2|^2+\lambda_4 |\phi_1^{\dagger}\phi_2|^2 \, ,  \nonumber\\
    & +\left[\frac{\lambda_5}{2}\left(\phi_1^{\dagger}\phi_2\right)^2+\lambda_6 |\phi_1|^2\left(\phi_1^{\dagger}\phi_2\right)+\lambda_7 |\phi_2|^2\left(\phi_1^{\dagger}\phi_2\right)+\mbox{h.c.}\right] \, , 
\end{align}
where we use the usual notation $V_{\mu \nu}=\partial_\mu V_\nu-\partial_\nu V_\mu+i \tilde{g} \left[V_\mu,V_\nu\right]$, $D_\mu \phi_i=\left(\partial_\mu+i \tilde{g}V_\mu \right)\phi_i$.\smallskip

The model introduced above features several free parameters and new states. As was shown in Refs.~\cite{Gross:2015cwa,Arcadi:2016kmk,Arcadi:2016qoz}, the model can be reduced, with a rather modest loss of generality, to a vector DM Higgs-portal model. By setting to zero the $m_{12}^2, \lambda_{H12}, \lambda_6, \lambda_7$ parameters, and assuming the hierarchy $v_{3,4} \ll v_{1,2}$, the mass terms of the Higgs sector will read
\begin{equation}
    \mathcal{L}=-\frac{1}{2}\Phi^T \mathcal{M}_{\rm CP-even}^2 \Phi-\frac{1}{4}\left(\lambda_4-\lambda_5\right) (v_1^2+v_2^2) \psi^2 \, , 
\end{equation}
where $\psi \equiv h_4$ is, in the CP conserving limit with $v_4 =0$, a CP-odd state while $\Phi={\left(H,h_1,h_2,h_3\right)}^T$ are, instead, CP-even states. The mass matrix of the latter is given by
\begin{equation}
    \mathcal{M}_{\rm CP-even}^2=\left(\begin{array}{cccc}
    \lambda_H v^2  &  \lambda_{H11}v v_1  &  \lambda_{H22}v v_2     &  0  \\
     \lambda_{H11}v v_1  &  \lambda_1 v_1^2   &  \lambda_3 v_1 v_3  & 0 \\
     \lambda_{H22}v v_2 &  \lambda_3 v_1 v_3  & \lambda_2 v_2^2   & 0 \\
     0 & 0 & 0 & \frac{1}{2}\left(\lambda_4+\lambda_5\right)(v_1^2+v_2^2)
    \end{array}
    \right) \, .
\end{equation}
Following again Refs.~\cite{Gross:2015cwa,Arcadi:2016kmk,Arcadi:2016qoz},  the choice $\lambda_{H11}=\lambda_3 \ll 1$ is adopted such that Higgs eigenstates are given by 
\begin{align}
     H_1 \simeq \cos\theta H- \sin\theta h_2 \, , \
     H_2 \simeq \sin \theta H + \cos\theta h_2 \, , \
     H_3  \simeq h_3 \, , \
     H_4 \simeq h_1 \, , 
\end{align}
with masses
\begin{align}
    & M_{H_1,H_2}^2\simeq \frac{1}{2}\left(\lambda_2 v_2^2+\lambda_H v^2\right)\mp \frac{\lambda_2 v_2^2-\lambda_H v^2}{2 \cos 2 \theta} \, , \nonumber\\
    & M_{H_3}^2=\frac{1}{2}\left(\lambda_4+\lambda_5\right)(v_1^2+v_2^2)\, , \
    M_{H_4}^2=\lambda_1 v_1^2 \, , 
\end{align}
with $H_1$ being the 125 GeV SM-like Higgs boson. Under our assumptions, there is a sizable mixing of the SM Higgs state only with the dark state $h_2$ with a mixing angle $\theta$ defined by
\begin{equation}
    \tan 2\theta \simeq \frac{2 \lambda_{H22}v v_2}{\lambda_2 v_2^2-\lambda_H v^2} \, . 
\end{equation}
The other two physical eigenstates $h_3$ and $h_4$ will have negligible, but not exactly zero couplings with the SM particles.\smallskip

Under the above assumptions, the vector sector will be comprised of three pure vector states forming degenerate pairs with masses given by
\begin{equation}
    M_{V_1}^2=M_{V_2}^2=\frac{1}{4}\tilde{g}^2v_2^2,\,\,\,\, M_{V_4}^2=M_{V_5}^2=\frac{1}{4}\tilde{g}^2v_1^2,\,\,\,\,
    M_{V_6}^2=M_{V_7}^2=\frac{1}{4}\tilde{g}^2(v_1^2+v_2^2),\,\,\,\,
\end{equation}
and two mixed states
\begin{align}
    V_3^{'}=V_3 \cos \alpha +V_8 \sin \alpha \, , \ \
    V_8^{'}=-V_3 \sin \alpha +V_8 \cos \alpha \, , 
\end{align}
with masses
\begin{equation}
\label{eq:primed_vectors}
    M_{V^{'}_3}^2= \frac{\tilde{g}^2 v_2^2}{4}\left(1-\frac{\tan\alpha}{\sqrt{3}}\right),\,\,\,\,\, M_{V^{'}_8}^2= \frac{\tilde{g}^2 v_1^2}{4}\frac{1}{\left(1-{\tan\alpha}/{\sqrt{3}}\right)} \, , 
\end{equation}
where, in the case $2 v_1^2 > v_2^2$, one would have (for the more general expression, we refer to Ref.~\cite{Gross:2015cwa,Arcadi:2016kmk}) 
\begin{equation}
    \alpha= \frac{1}{2} \arctan\left(\frac{\sqrt{3}v_2^2}{2 v_1^2+v_2^2}\right) \, . 
\end{equation}

The breaking of the dark SU(3) gauge symmetry leaves a global $\mathbb{Z}_2 \times {\rm U(1)}$ symmetry unbroken. The latter has, in turn, a discrete $\mathbb{Z}_2 \times \mathbb{Z}_2^{'}$ subgroup. The fields of the dark sector have different charges under this discrete symmetry and they are summarized in Tab.~\ref{tab:dmSU3}. It should be noticed that $\mathbb{Z}_2 \times \mathbb{Z}_2^{'}$ is a symmetry of the dark sector only if CP is preserved, i.e. when $v_4=0$. Given the charges displayed in Tab.~\ref{tab:dmSU3}, there is one degenerate pair $V_{1,2}$ ($V_{4,5}$) that is cosmologically stable if $v_2 < v_1$ ($v_2 > v_1$). The lightest between the states $V_3^{'}$ and $\psi$ is stable as well. Their mass ratio will be given by
\begin{equation}
\label{eq:A3_ratio}
\frac{m_{\psi}^2 }{m^2_{V'^3}}=2 \frac{\lambda_4-\lambda_5}{\tilde g^2} f(v_2^2/v_1^2)
\quad
\textrm{with}
\quad
f(r)=\frac32 \frac{r+1}{r+1-\sqrt{1+r(r-1)}} \,.
\end{equation}
For $r \equiv v_2^2/v_1^2$ not much smaller than unity, one has $f(r) = 3 + \mathcal{O}((1-r)^2)$, while for $r \ll 1$ one has $f(r) \simeq 1/r+ \mathcal{O}(1)$.

\begin{table}[!h]
\begin{center}
\renewcommand{\arraystretch}{1.6}
\begin{tabular}{ccc}\hline
gauge eigenstates & mass eigenstates &
 $\mathbb{Z}_2\times\mathbb{Z}_2^{\prime}$
\\\hline\hline
$h,h_1,h_2,h_3,A_\mu^7$ &
$H_1, H_2, H_3, H_4,\tilde{V}_\mu^7$ & $(+,+)$\\
$V_\mu^1,V_\mu^4$ &
$V_\mu^1,V_\mu^4$ & $(-,-)$\\
$V_\mu^2,V_\mu^5$ &
$V_\mu^2,V_\mu^5$ & $(-,+)$\\
$h_4,V_\mu^3,V_\mu^6,V_\mu^8$ &
$\psi,V_\mu^{\prime3},V_\mu^6,V_\mu^{\prime8}$ & $(+,-)$\\\hline
\end{tabular}
\caption{$\mathbb{Z}_2\times\mathbb{Z}_2^{\prime}$ assignments of the various fields of the SU(3) dark model.}
\label{tab:dmSU3}
\end{center}
\end{table}

The different possibilities, concerning the cosmologically stable states, are summarized in Table \ref{tab:dm-patterns}.

\begin{table}[!h]
\renewcommand{\arraystretch}{1.6}
\begin{center}
\begin{tabular}{ccccc}\hline
& Case I & Case II & Case III & Case IV\\\hline\hline
dark matter 
 & ($V_{\mu}^1$,$V_{\mu}^2$,$\psi$)
 & ($V_{\mu}^4$,$V_{\mu}^5$,$\psi$)
 & ($V_{\mu}^1$,$V_{\mu}^2$,$V_{\mu}^{\prime3}$)
 & ($V_{\mu}^4$,$V_{\mu}^5$,$V_{\mu}^{\prime3}$)\\\hline
 parameter & $v_2/v_1<1$ & $v_2/v_1>1$ & $v_2/v_1<1$ & $v_2/v_1>1$\\
 choice & $\lambda_4-\lambda_5\ll1$ & $\lambda_4-\lambda_5\ll1$ & $\lambda_4-\lambda_5=\mathcal{O}(1)$ & $\lambda_4-\lambda_5=\mathcal{O}(1)$\\\hline
\end{tabular}
\caption{Natural parameter choices for different multi-component dark matter scenarios.}
\label{tab:dm-patterns}
\end{center}
\end{table}

Assuming $v_1 > v_2$, the relevant Lagrangian for DM phenomenology then reduces to \begin{align}
\label{eq:SU3_lagrangian}
     \mathcal{L}& =\frac{\tilde{g} M_{V}}{2}\left(-\sin\theta H_1 +\cos\theta H_2\right)\left(\sum_{a=1,2}V_{\mu}^a V^{\mu\,a}+{\left(\cos\alpha-\frac{\sin\alpha}{\sqrt{3}}\right)}^2V_\mu^3 V^{\mu\,3}\right)\nonumber\\
    & +\tilde{g}\cos\alpha \sum_{a,b,c}\epsilon_{abc}\partial_\mu V_\nu V_\nu^a V^{b\,\mu}V^{c\,\nu}-\frac{\tilde{g}^2}{2}\cos^2 \alpha \sum_{a=1,2}\left(V_\mu^a V^{a\,\mu}V_\nu^3 V^{3\,\nu}-{\left(V_\mu^a V^{a\,\mu}\right)}^2\right)\nonumber\\
    & -\frac{1}{2}m_\psi^2 \psi^2 +\left[\frac{\tilde{g}}{2 M_V}\left(-\sin\theta H_1 +\cos\theta H_2\right)-\frac{1}{4}\left(\lambda_{\psi \psi 11}H_1^2+2 \lambda_{\psi \psi 12}H_1 H_2+\lambda_{\psi \psi 22}H_2^2\right)\right]\psi^2\nonumber\\
    & -\frac{k_{111}}{2}v H_1^3-\frac{k_{112}}{2}H_1^2 H_2 v \sin\theta-\frac{\kappa_{221}}{2}H_1 H_2^2 v \cos\theta-\frac{\kappa_{222}}{2}H_2^3 v \nonumber\\
     & +\frac{H_1 \cos\theta+H_2 \sin\theta}{v}\left(2 M_W^2 W_\mu^{+}W^{\mu -}+M_Z^2 Z_\mu Z^\mu -m_f \bar f f\right) \, , 
\end{align}
where the trilinear couplings $\kappa$ are the same as in eq.~(\ref{eq:trilinear}) while the new ones are given by  
\begin{align}
    & \lambda_{\psi \psi 11}=\frac{\tilde{g}}{2 M_V v}\sin\theta\left(\cos^3 \theta \left(M_{H_2}^2-M_{H_1}^2\right)+\frac{\tilde{g}}{2 M_V v}\sin\theta\left(\sin^2 \theta M_{H_1}^2+\cos^2 \theta M_{H_2}^2\right)\right) ,  \nonumber \\
    & \lambda_{\psi \psi 12}=\frac{\tilde{g}}{2 M_V v}\sin\theta \cos\theta\left(\sin\theta \cos\theta \left(M_{H_2}^2\!-\!M_{H_1}^2\right)\!-\!\frac{\tilde{g}}{2 M_V v}\sin\theta\left(\sin^2 \theta M_{H_1}^2\!+\!\cos^2 \theta M_{H_2}^2\right)\right) , \nonumber\\
    & \lambda_{\psi \psi 22}=\frac{\tilde{g}}{2 M_V v}\cos\theta \left(\sin^3 \theta \left(M_{H_2}^2-M_{H_1}^2\right)+\frac{\tilde{g}}{2 M_V v}\cos\theta\left(\sin^2 \theta M_{H_1}^2+\cos^2 \theta M_{H_2}^2\right)\right) \, . 
\end{align}

The Lagrangian eq.~(\ref{eq:SU3_lagrangian}) resembles that of the effective Higgs-portal model with vector DM. Two additional light states, the spin--1 boson $V^3$ and the spin--0 boson $\psi$, are nevertheless present and might be important for phenomenology.
Two different scenarios can then be considered.\vspace*{-2mm}

\begin{itemize}
    \item CP is exactly conserved, and $\mathbb{Z}_2 \times {\rm U(1)}$ is then also a symmetry of the scalar sector. The lightest of the  $V_3$ and $\psi$ states is accounted for as an additional DM component. The two possibilities have been studied in detail in Refs.~\cite{Gross:2015cwa} and \cite{Arcadi:2016kmk} respectively.\vspace*{-2mm}

    \item CP is slightly violated, and both $\psi$ and $V_3$ can decay with a lifetime shorter than the age of the universe. They nevertheless strongly impact the DM relic density since they represent additional annihilation final states for the DM. In particular, this is true for $V_3$ as it is always lighter than the DM, as can be seen from eq.~(\ref{eq:primed_vectors}).\vspace*{-2mm} 
\end{itemize}

The phenomenology of these two options will be discussed in the next subsections. 

\subsubsection{Dark SU(3) model with V/V$_3$ DM}

We first consider the scenario in which two spin--1 states compose the lightest dark sector, the case of a spin--1/spin--0 component will be, instead, considered in the next subsection. Both scenarios will encode two possibilities, depending on whether CP is exact or slightly violated in the dark Higgs sector: a single DM component and a lighter (meta-stable) state or a two-component DM particle. \smallskip

The case of only vector DM components is potentially the closest to the dark U(1) scenario and, consequently, to the EFT vector Higgs-portal. Even for a modest hierarchy between the two vev's $v_1$ and $v_2$ we have that $\sin\alpha \ll 1$. As can be easily seen from eq.~\ref{eq:SU3_lagrangian}, this implies that the mass of the $V_{3}$ state becomes very close (with a relative mass difference smaller than $10\,\%$) to the one of the $V_{1,2}$ and its coupling with the $H_{1,2}$ states substantially coincides with the ones of $V_{1,2}$. In other words, we have basically BR$(H_1 \rightarrow \mbox{inv})|_{\rm SU(3)} \simeq 3 {\rm BR}(H_1 \rightarrow \mbox{inv})|_{\rm U(1)}$ in most cases. \smallskip

\begin{figure}
    \centering
    \subfloat{\includegraphics[width=0.48\linewidth]{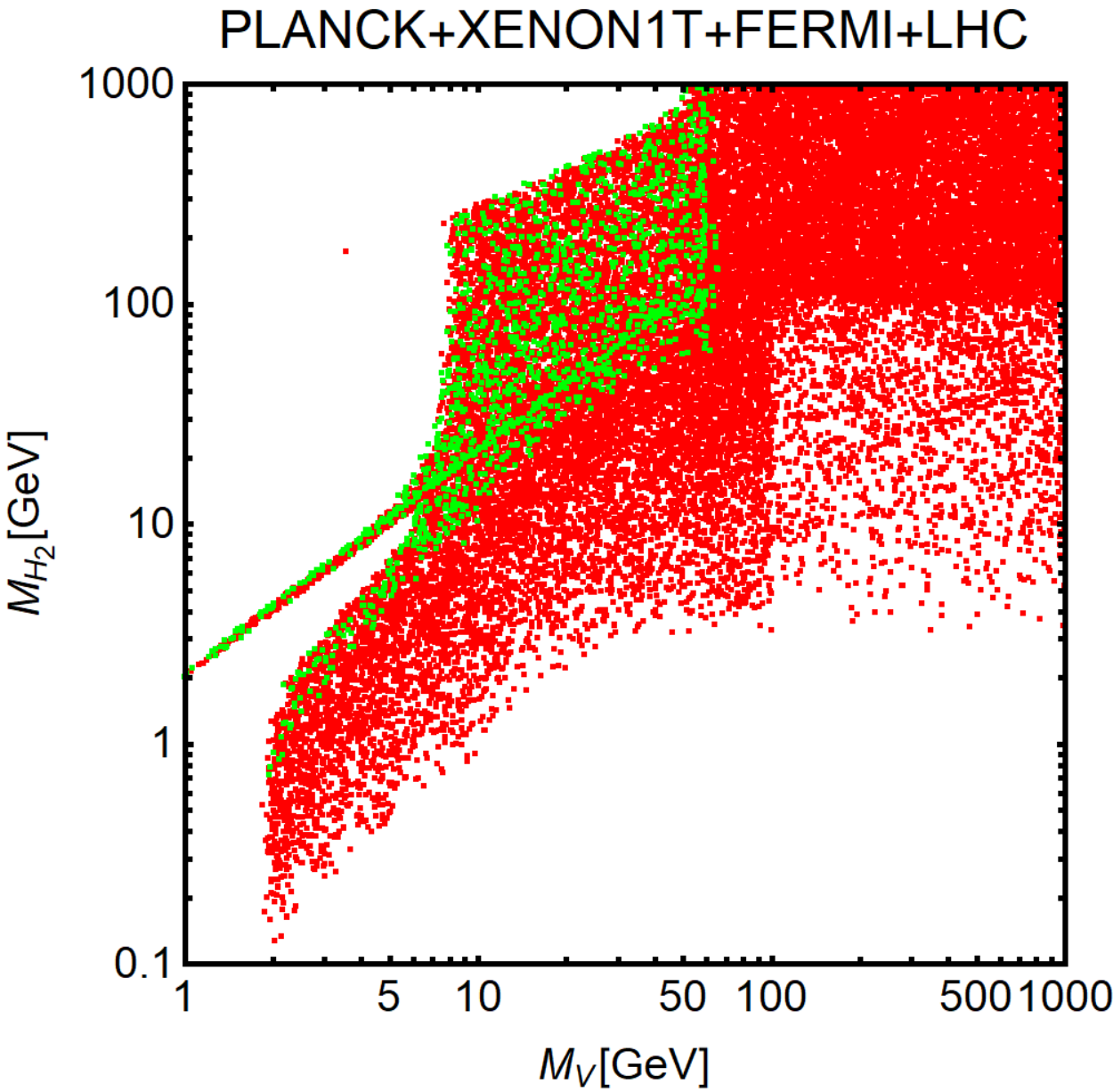}}
    \subfloat{\includegraphics[width=0.48\linewidth]{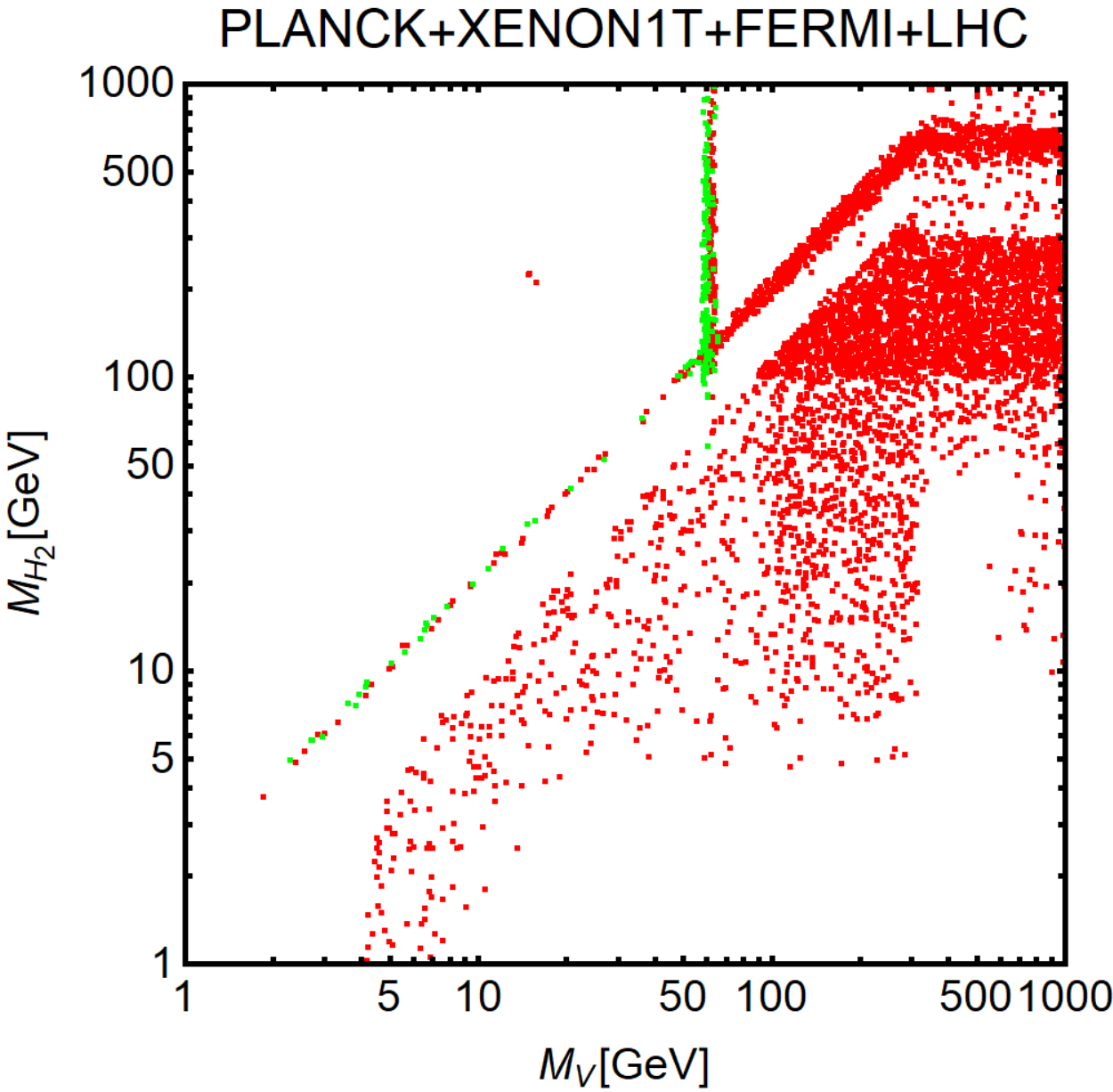}}
\vspace*{-1mm}
    \caption{The same as Fig.~\ref{fig:scan_U1} but for the dark SU(3) model with a single component DM  $V_{1,2}$ (left panel) and multicomponent $V_{1,2,3}$ DM (right panel).}
    \label{fig:scan_SU3}
\end{figure}

A general survey of the model parameters $M_V=M_{V_{1,2}}$, $\tilde{g}$, $M_{H_2}$ and $\sin\theta$ is carried out. The other parameters of the model have been set to  $\sin\alpha=0.08$ and $M_\psi=300\,\mbox{GeV}$. The results are shown in Fig.~\ref{fig:scan_SU3} in the $[M_V,M_{H_2}]$ plane. The left and right panels represent, respectively, the case of single component $V_{1,2}$ and multi-component $V_{1,2,3}$ DM.\smallskip
 
The comparison of the two panels of Fig.~\ref{fig:scan_SU3} shows strong differences among the two scenarios. In the case of single component DM, we see that the viable model points cover a broad region of the $[M_V,M_{H_2}]$ plane. In particular, the appropriate  relic density is achieved with invisible decays of the $H_1$ state that are within the expected future experimental sensitivity, $2.5\% \leq {\rm BR}(H_1 \to {\rm inv}) \leq 
11\%$ with at the same time, $M_{H_2}$ larger than $M_{H_1}$. This is a consequence of the presence of the additional $V V \rightarrow V_3 V_3$ channel, which is always kinematically accessible and which is efficient enough to guarantee an annihilation cross section compatible with the thermally favored value, while evading the experimental constraints, especially the ones from direct detection. The same does not occur, in turn, in the scenario with a two component DM. In such a case, most of the cosmological relic density is achieved by the lightest DM component~\cite{Gross:2015cwa}, i.e. the $V_3$ state.\smallskip

\begin{figure}
\centerline{    
    \subfloat{\includegraphics[width=0.44\linewidth]{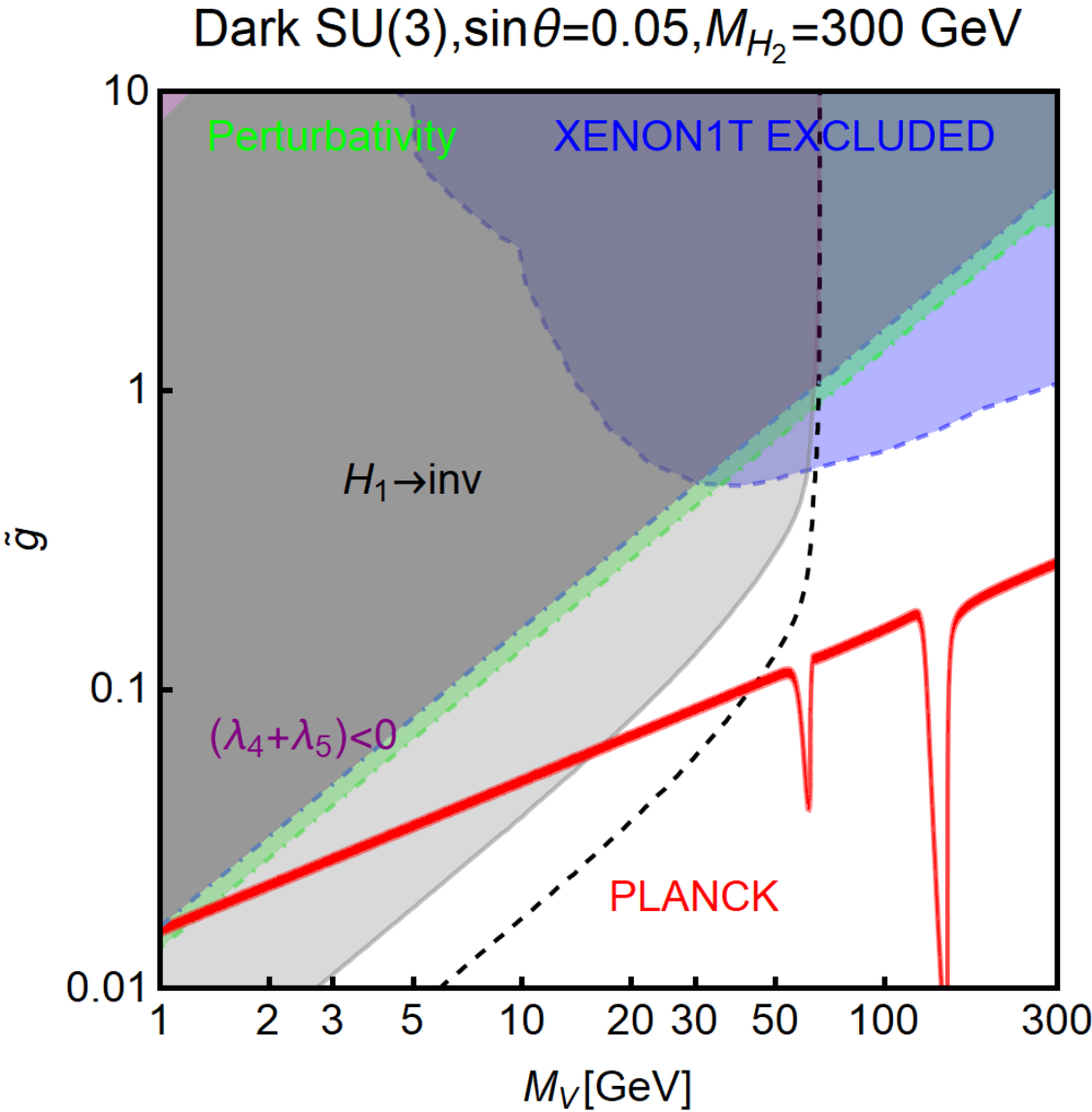}}~~
    \subfloat{\includegraphics[width=0.44\linewidth]{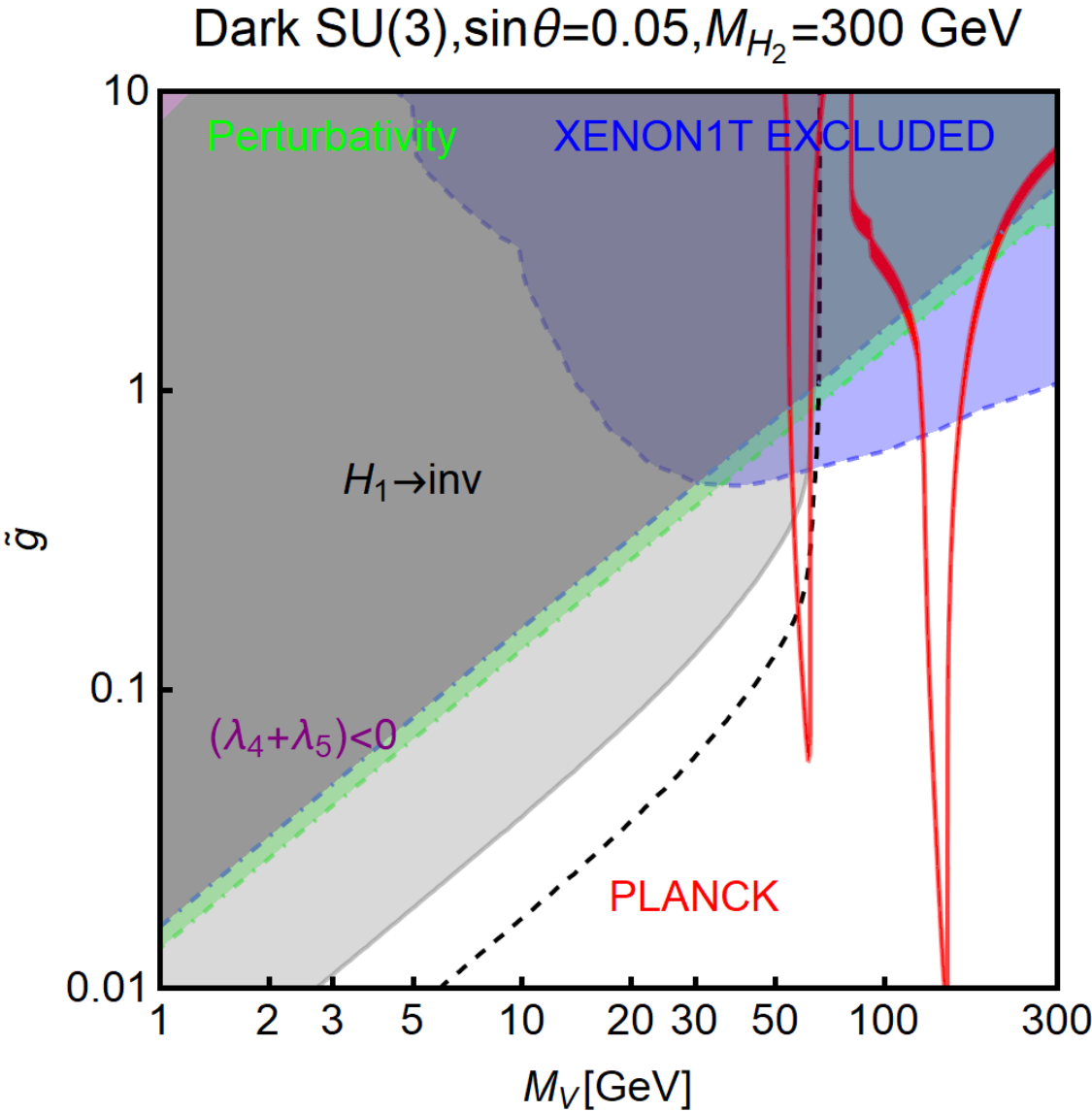}} }
\centerline{
     \subfloat{\includegraphics[width=0.44\linewidth]{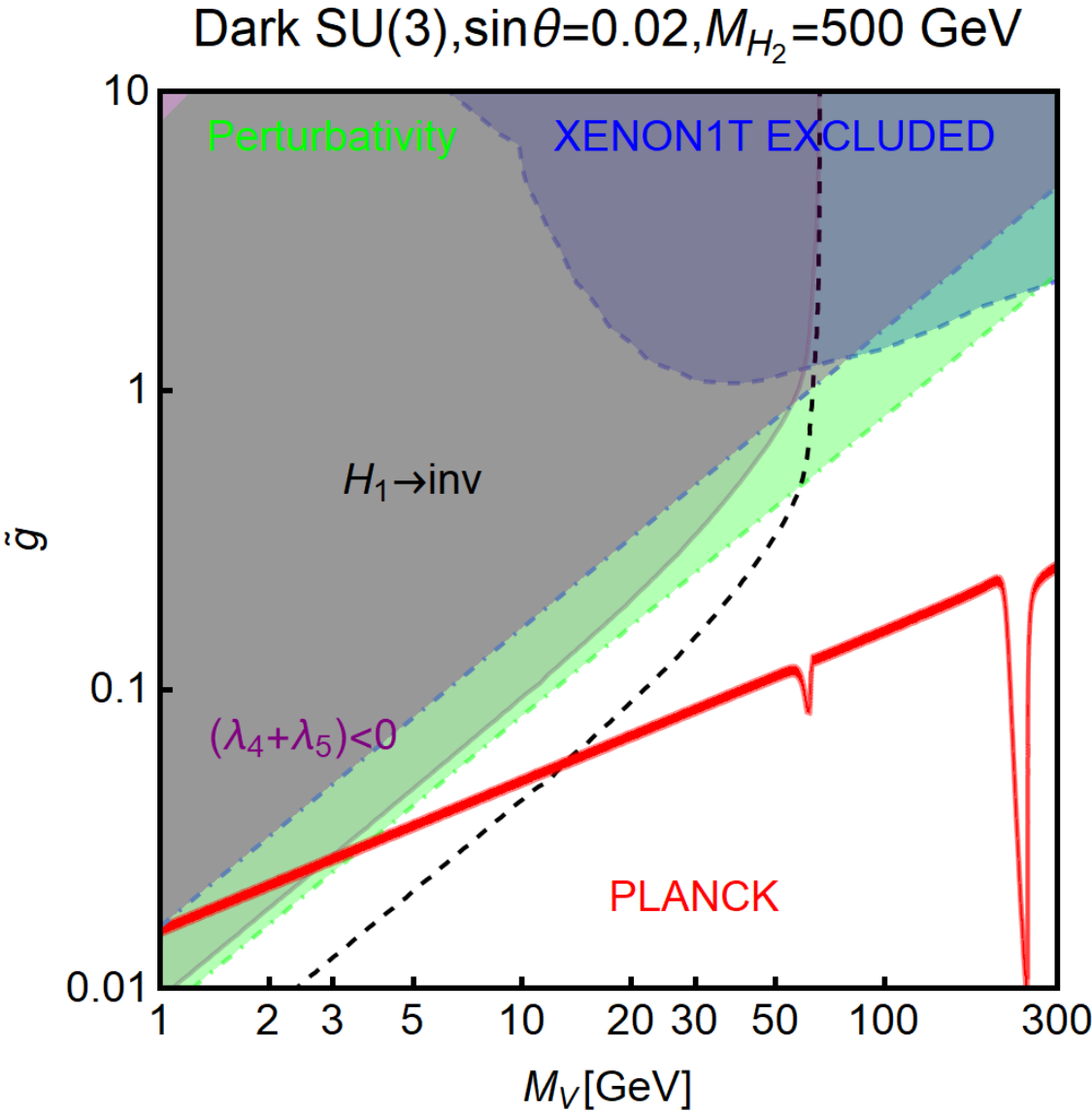}}~~
     \subfloat{\includegraphics[width=0.44\linewidth]{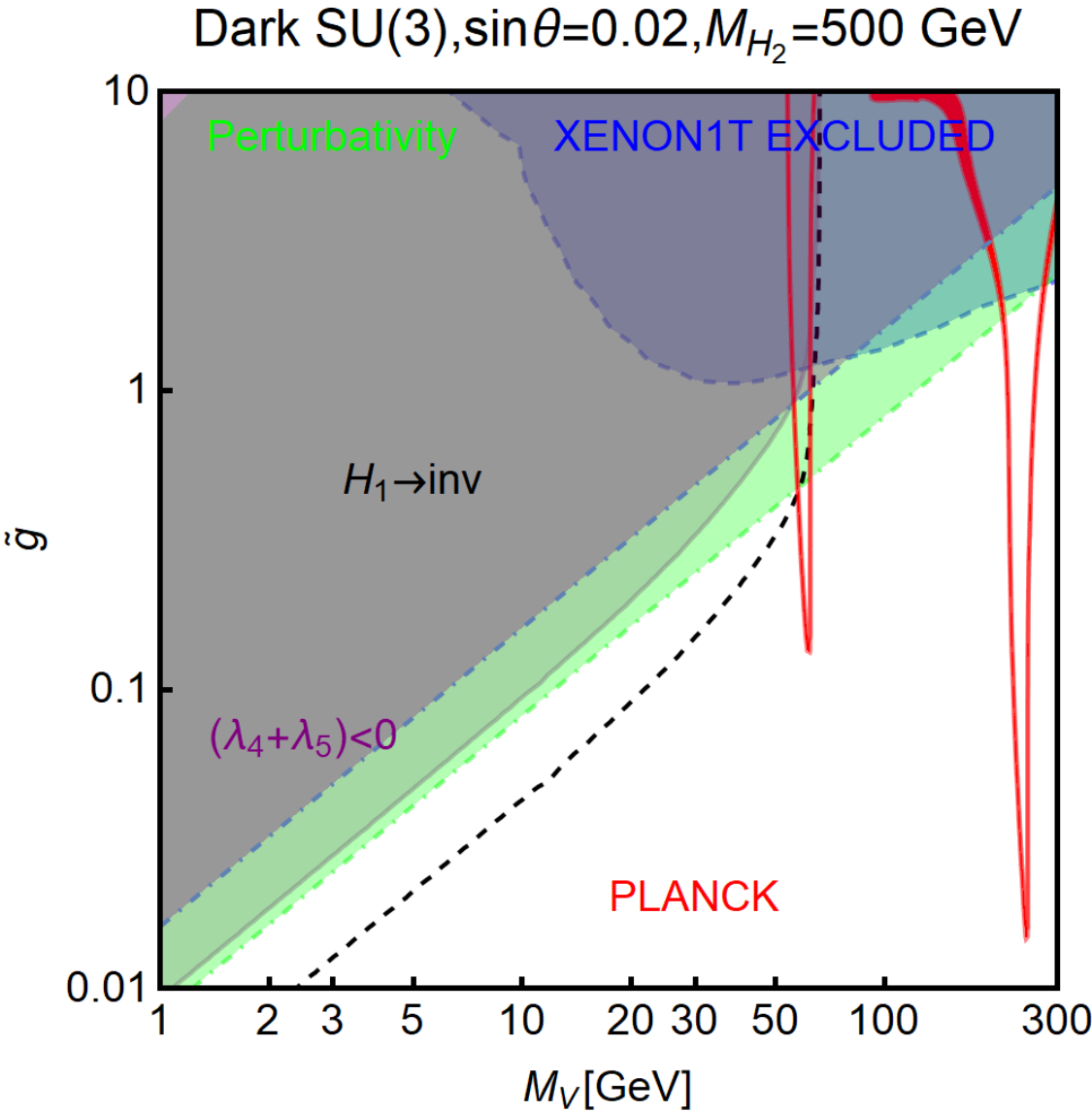}} }
\centerline{ 
    \subfloat{\includegraphics[width=0.44\linewidth]{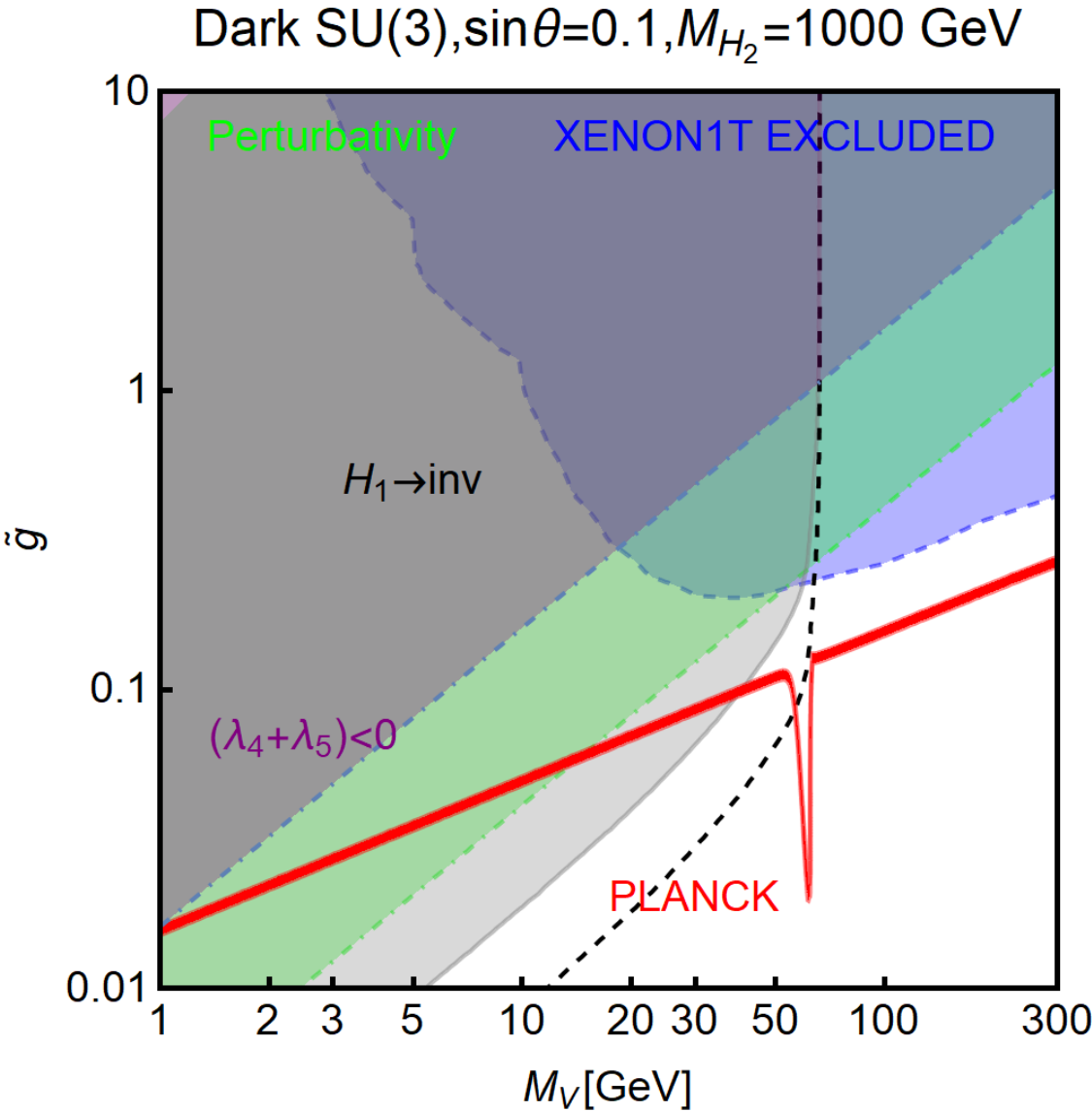}}~~
    \subfloat{\includegraphics[width=0.44\linewidth]{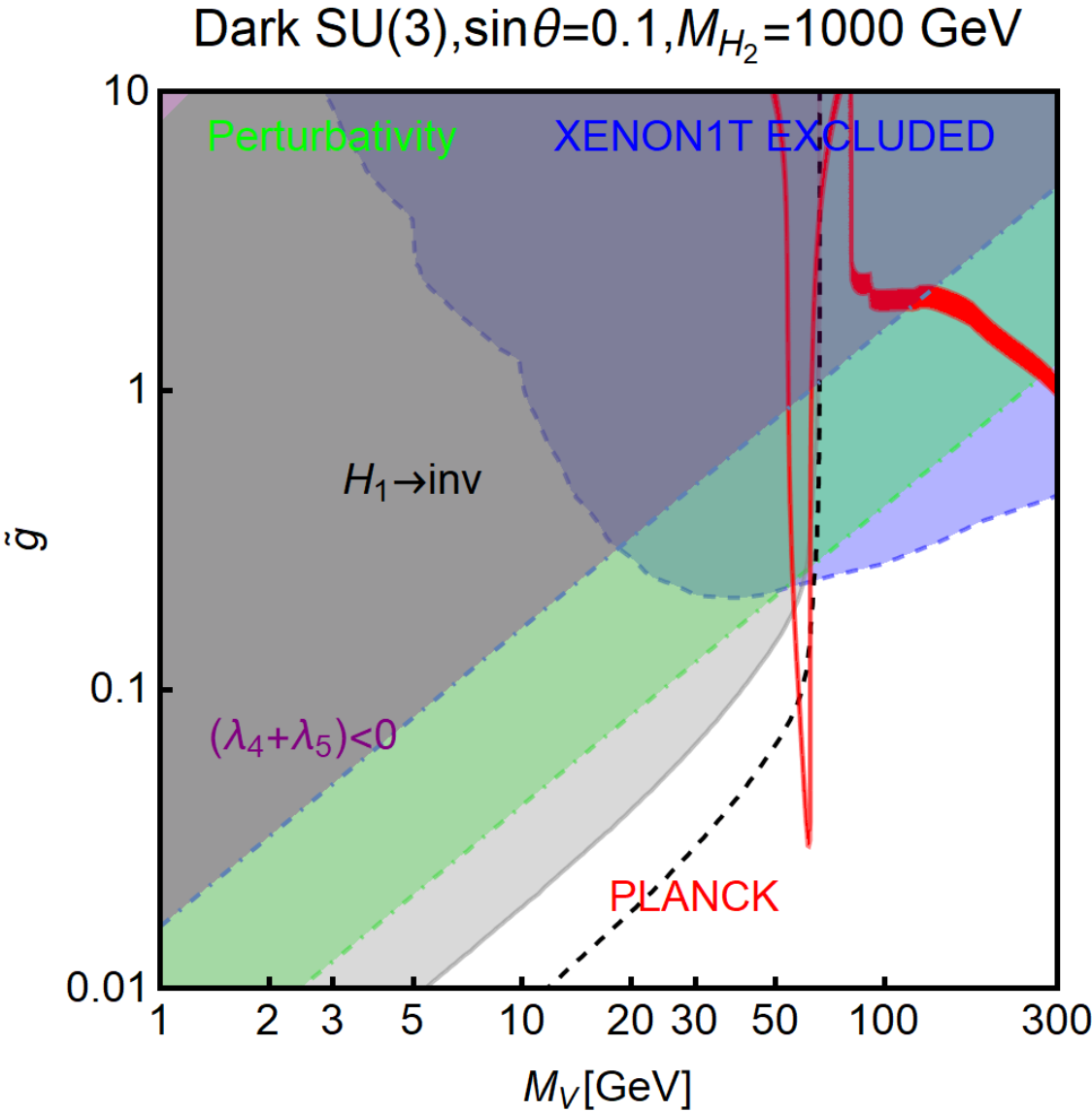}} }
\vspace*{-1mm}
    \caption{Summary of constraints for three benchmark scenarios, within the dark SU(3) realization with single DM component (left panel) and multi-component (right panel) DM. The color code is the same as Fig.~\ref{fig:U1heavy}.}
    \label{fig:SU3heavy}
\end{figure}

The outcome of this analysis is similar to that of the dark U(1) case with viable solutions possible only in the secluded regime, i.e. with $M_V > M_{H_2}$ and in correspondence of the $M_V \simeq \frac12 M_{H_1}$ and $\frac12 M_{H_2}$ poles. We also performed a dedicated study in three benchmark scenarios, with different $M_{H_2}$ and $\sin\theta$ parameter values, considering both the cases of single- and multi-component DM. We have investigated only the case in which $M_{H_2}> M_{H_1}$, where the most significant differences with the previously considered model arise. Our results are displayed in Fig. \ref{fig:SU3heavy} for three different values of the $(M_{H_2}, \sin\theta)$ pairs which are explicitly reported on top of each  panel.\smallskip

As shown in Fig.~\ref{fig:SU3heavy}, the bounds from the invisible Higgs branching ratio are more competitive than the ones from direct detection which are evaded due to the enhancement of the DM annihilation cross section through the $VV \rightarrow V^3 V^3$ process. Comparing the different panels, one can note that the impact of the perturbative unitarity bounds as $M_{H_2}$ increases. In particular, for $M_{H_2}=1\,\mbox{TeV}$, there are no scenarios with $M_V \lesssim \frac12 M_{H_1}$. \smallskip

\begin{figure}
    \centering
    \includegraphics[width=0.55\linewidth]{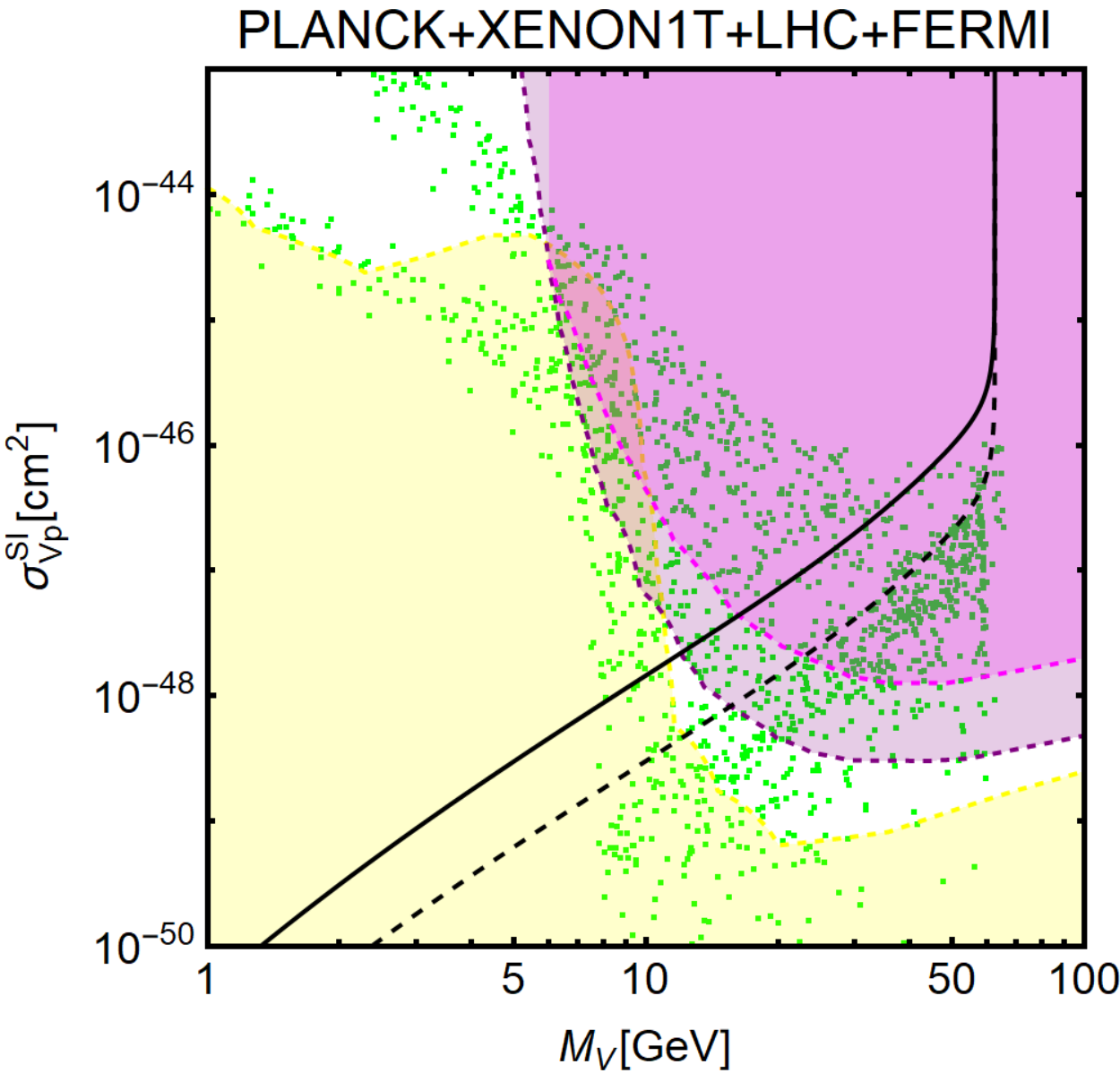}
    \caption{Model points of the dark SU(3) model, in the $[M_V,\sigma_{Vp}^{\rm SI}]$ 
plane, complying with DM constraints and featuring $2.5\,\% \leq {\rm BR}(H_1 \rightarrow\, \mbox{inv}) \leq 11\,\%$. The solid (dashed) isocontours represent the predicted cross sections of the effective vector Higgs-portal corresponding  
to BR$(H_1 \rightarrow\, \mbox{inv})= 11\,\%\, (2.5\,\%)$. The magenta/purple regions represent future sensitivities from XENONnT/DARWIN and the yellow regions the neutrino floor.}
    \label{fig:scan_SU3_bis}
\vspace*{-3mm}
\end{figure}

Fig.~\ref{fig:scan_SU3_bis} shows, in the usual $[M_V,\sigma_{\chi p}^{\rm SI}]$ plane, the model points of our scan that have a viable DM relic density, evade the direct and indirect detection constraints and feature an invisible branching ratio of $2.5\,\% \leq {\rm BR}(H_1 \rightarrow \mbox{inv}) \leq 11\,\%$. For most of the model points the effective vector Higgs-portal limit is valid. While  experimental constraints can be compatible with the mass hierarchy $M_{H_2}> M_{H_1}$, perturbative unitarity constraints forbid a complete decoupling of the $H_2$ state which can subsequently affect the DM scattering cross section. Furthermore, the invisible Higgs branching fraction is three times higher, for the same assignments of the parameters $\tilde{g}$ and $\sin\theta$, than for the case of the dark U(1) model. This does not occur for the DM scattering cross section which, instead, coincides for the two models. Having considered in our parameter scan the case $M_{H_2}< M_{H_1}$ as well, Fig.~\ref{fig:scan_SU3_bis}  also shows a distribution of points with  cross sections well above the effective Higgs-portal limit. 

\subsubsection{Dark SU(3) model with V/$\psi$ DM}

A more peculiar realization of the dark SU(3) scenario is the case with a mass hierarchy  $M_\psi < M_{V^3},M_V$ which features two DM components with different spins, if CP--symmetry is preserved by the dark Higgs sector. The (pseudo)scalar DM component of the dark SU(3) model has a very interesting property. Substituting the analytical expression of the couplings of $\Psi$ with the the $H_{1,2}$ states, \begin{eqnarray}
     g_{H_{1,2}\psi \psi}&=&\lambda_2 v_2 \left(-\sin\theta H_1+\cos\theta H_2\right)+\lambda_{H22}\left(\cos\theta H_1+\sin\theta H_2\right) \nonumber\\
    &=& \frac{\tilde{g}}{2 M_V}\sin\theta \cos\theta \left(-H_1 M_{H_1}^2\sin\theta+H_2 M_{H_2}^2 \cos\theta\right)  
\end{eqnarray}  
into the conventional expression, 
\begin{equation}
    \sigma_{\Psi p}^{\rm SI}\propto \left \vert  \left({g_{H_1 \Psi \Psi}}/{M_{H_1}^2}\cos\theta+{g_{H_2 \Psi \Psi}}/{M_{H_2}^2}\sin\theta\right) \right \vert^2\ , 
\end{equation}
of the scattering cross section of $\Psi$ into protons, it is immediate to notice an automatic exact cancellation, independent on the values of the $(M_{H_2},\sin\theta)$ pair \cite{Arcadi:2016kmk}, among the contributions associated to, respectively, the $t$-channel $H_1$ and $H_2$ exchange processes. In other words, the scattering cross section of the $\Psi$ component of the DM identically vanishes.\smallskip

What actually vanishes is the leading order contribution as,  an additional coupling responsible for spin-independent interactions, is generated by the $V^{6\,\mu}\partial_\mu h_4$ operator, leading to a cross section of the form
\begin{equation}
    \sigma_{\Psi p}=\frac{\mu_{\Psi p}^2}{4 \pi}\frac{M_\Psi^2 M_V^2}{M_{V^6}^4}\sin^2 \theta \cos^2 \theta \left(\frac{1}{M_{H_1}^2}-\frac{1}{M_{H_2}^2}\right)\frac{[Z f_p+(A-Z)f_n]^2}{A^2}, 
\end{equation}
which becomes negligible when the limit ${M_{V}^2}/{M_{V^6}^2}\sim {v_2^2}/{v_1^2}\ll 1$ is approached. This is a very peculiar feature of the UV-complete model which cannot be reproduced by an effective Higgs-portal scenario\footnote{This type of cancellation mechanism has been explored, from a more general perspective, in Ref.~\cite{Gross:2017dan}.}.\smallskip

\begin{figure}
\vspace*{-1mm}
    \centering
    \subfloat{\includegraphics[width=0.55\linewidth]{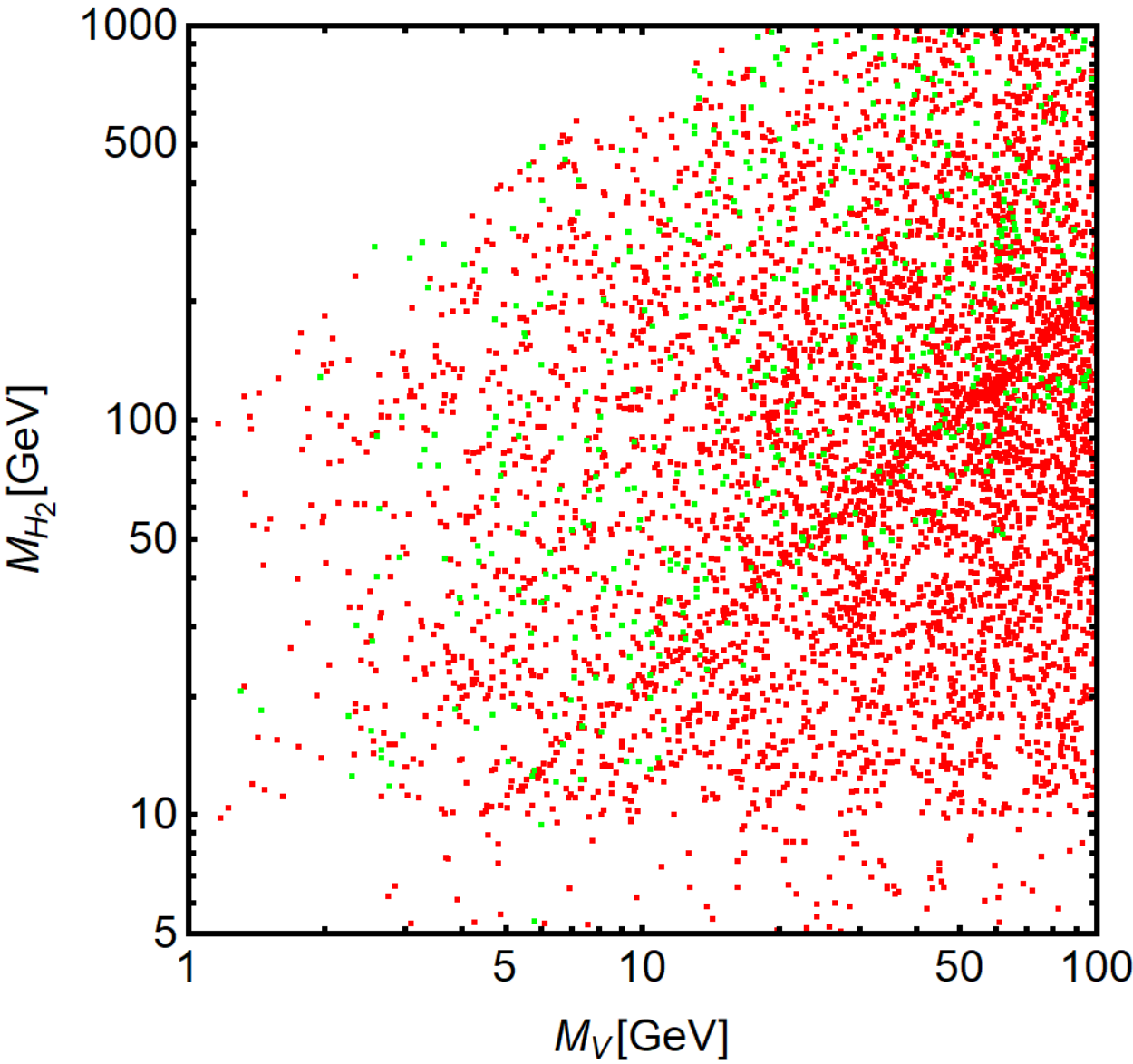}}
    \subfloat{\includegraphics[width=0.55\linewidth]{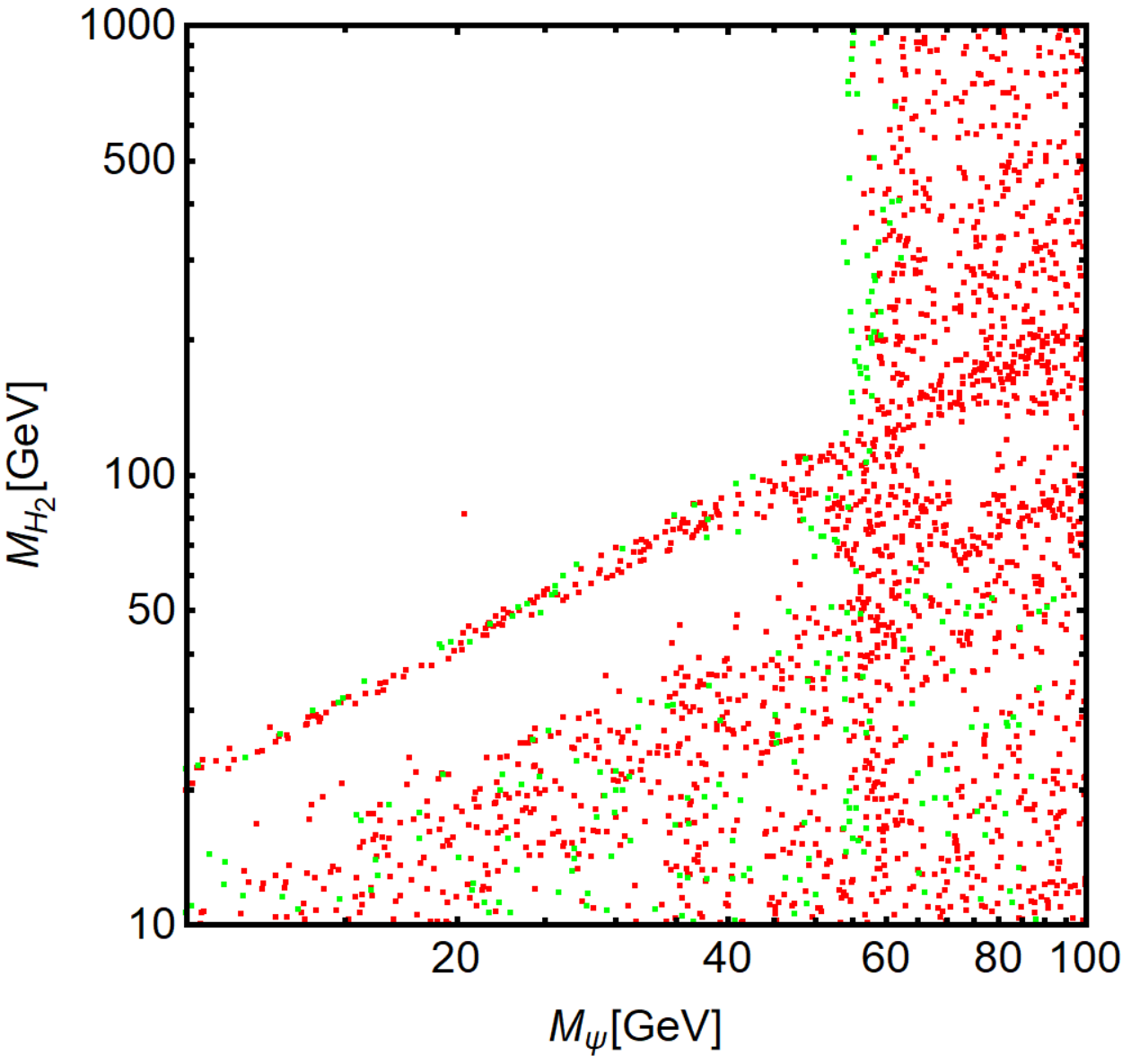}}
\vspace*{-1mm}
    \caption{ The same as Fig.~\ref{fig:scan_U1} but for the dark SU(3) model with 
a spin--1/spin--0 DM with a single component (left) and two component DM (right). In the latter case, the $x$-axis is for the mass $M_\Psi$ of the spin--0 component.}
    \label{fig:SU3Vchi_2CDM_scan}
\vspace*{-1mm}
\end{figure}

Searches for invisible decays of the SM-like Higgs boson hence represent an indispensable probe for this scenario, at least in the case of a light DM state. We show in Fig.~\ref{fig:SU3Vchi_2CDM_scan} the results of a scan over the model parameters, similarly to the approach that we have adopted before. The $(M_V,M_{H_2},\tilde{g},\sin\theta)$ input parameters are varied in the same ranges as the ones considered in the previous section while for the mass $M_\Psi$, we have considered the relatively low range $M_\Psi \in [10,100]\,\mbox{GeV}$. \smallskip

The outcome of the parameter survey is shown in Fig.~\ref{fig:SU3Vchi_2CDM_scan} where,  similarly to Fig.~\ref{fig:scan_SU3}, the results are illustrated in the $[M_{\rm DM},M_{H_2}]$ plane distinguishing the cases of single and double component DM. The single component DM scenario with a lighter metastable partner shown in the left panel of the figure, features a wider region of allowed parameter space since the $VV \rightarrow \Psi \Psi$ annihilation process ensures a viable relic density without conflicting with experimental bounds. For the scalar DM case, as discussed in Ref.~\cite{Arcadi:2016kmk}, the $\Psi$ DM component retains most of the total relic density. The right panel of Fig.~\ref{fig:SU3Vchi_2CDM_scan} shows the results in the $[M_\Psi,M_{H_2}]$  plane. 
For $M_\Psi \lesssim M_W$, the DM annihilation into SM states is suppressed unless an enhancement around the $\frac12 M_{H_2}$ pole occurs. Alternatively, $\Psi \Psi \rightarrow H_2 H_2$ annihilation would be required, which implies $M_\Psi \geq M_{H_2}$. For heavier DM, instead, the annihilation rate into $WW$ and $ZZ$ becomes progressively allowed, so we can have viable points also for a heavy $H_2$ state.

\begin{figure}
\vspace*{-1mm}
    \centering
    \subfloat{\includegraphics[width=0.48\linewidth]{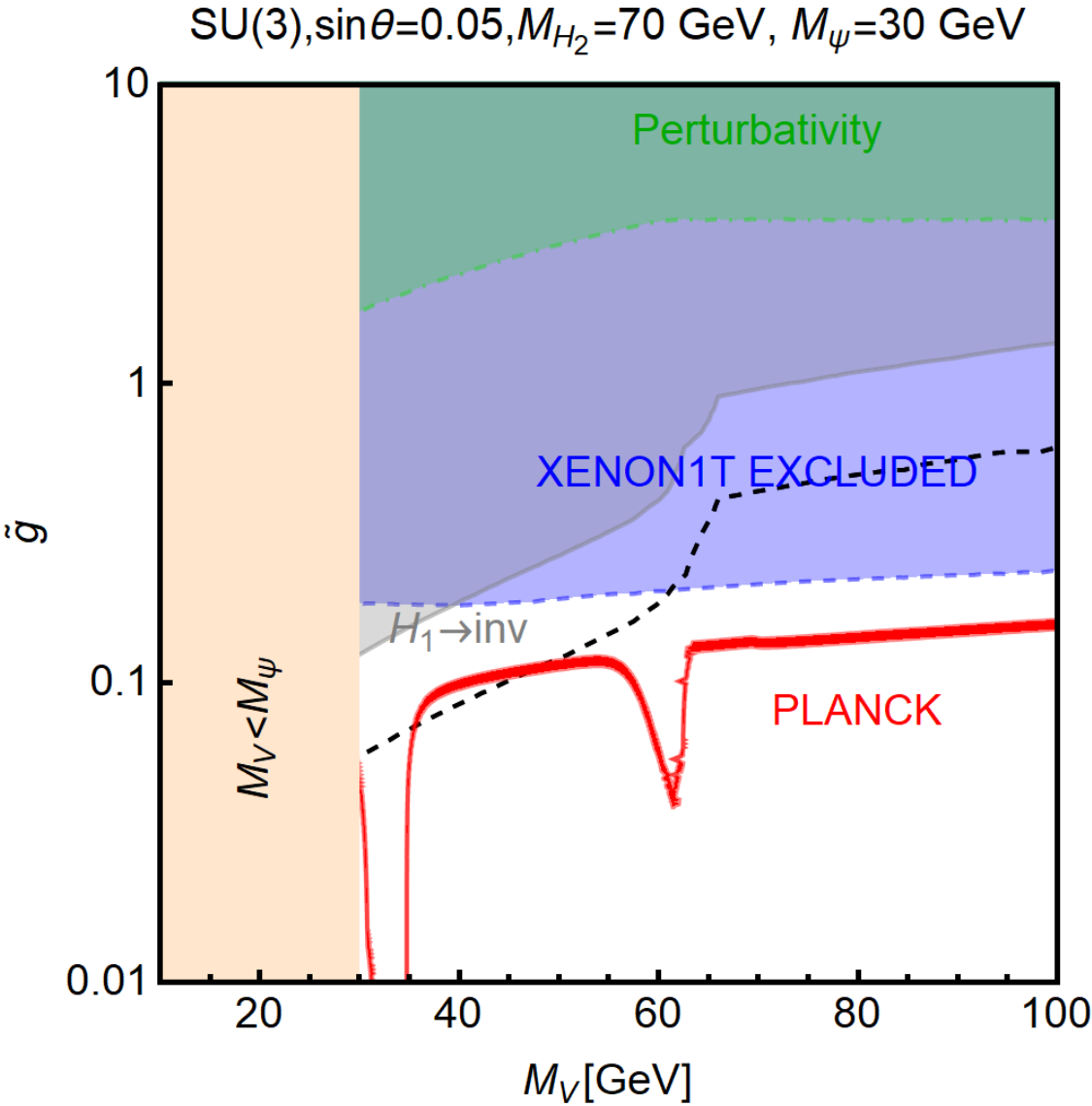}}
     \subfloat{\includegraphics[width=0.48\linewidth]{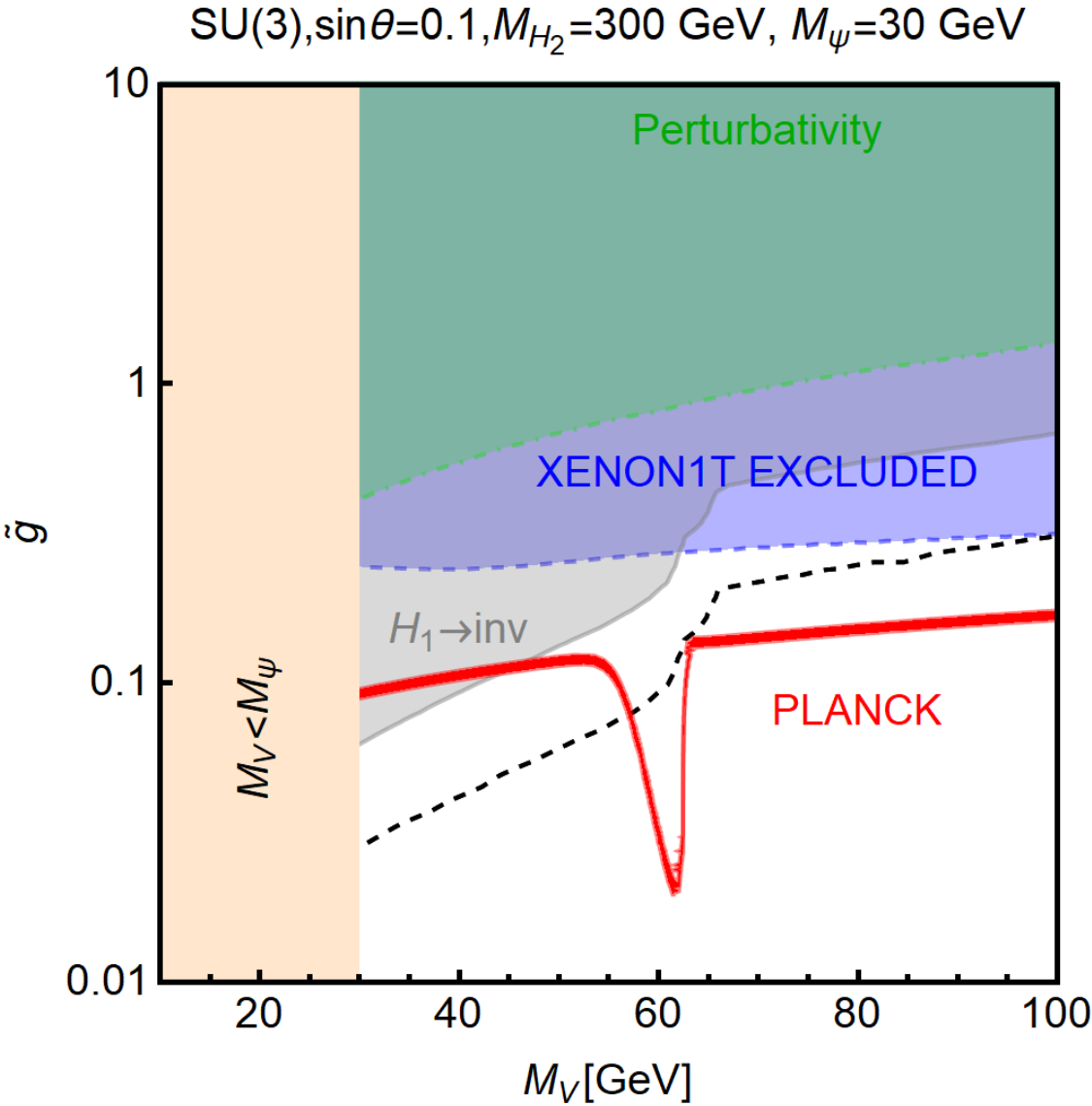}}\\
     \subfloat{\includegraphics[width=0.48\linewidth]{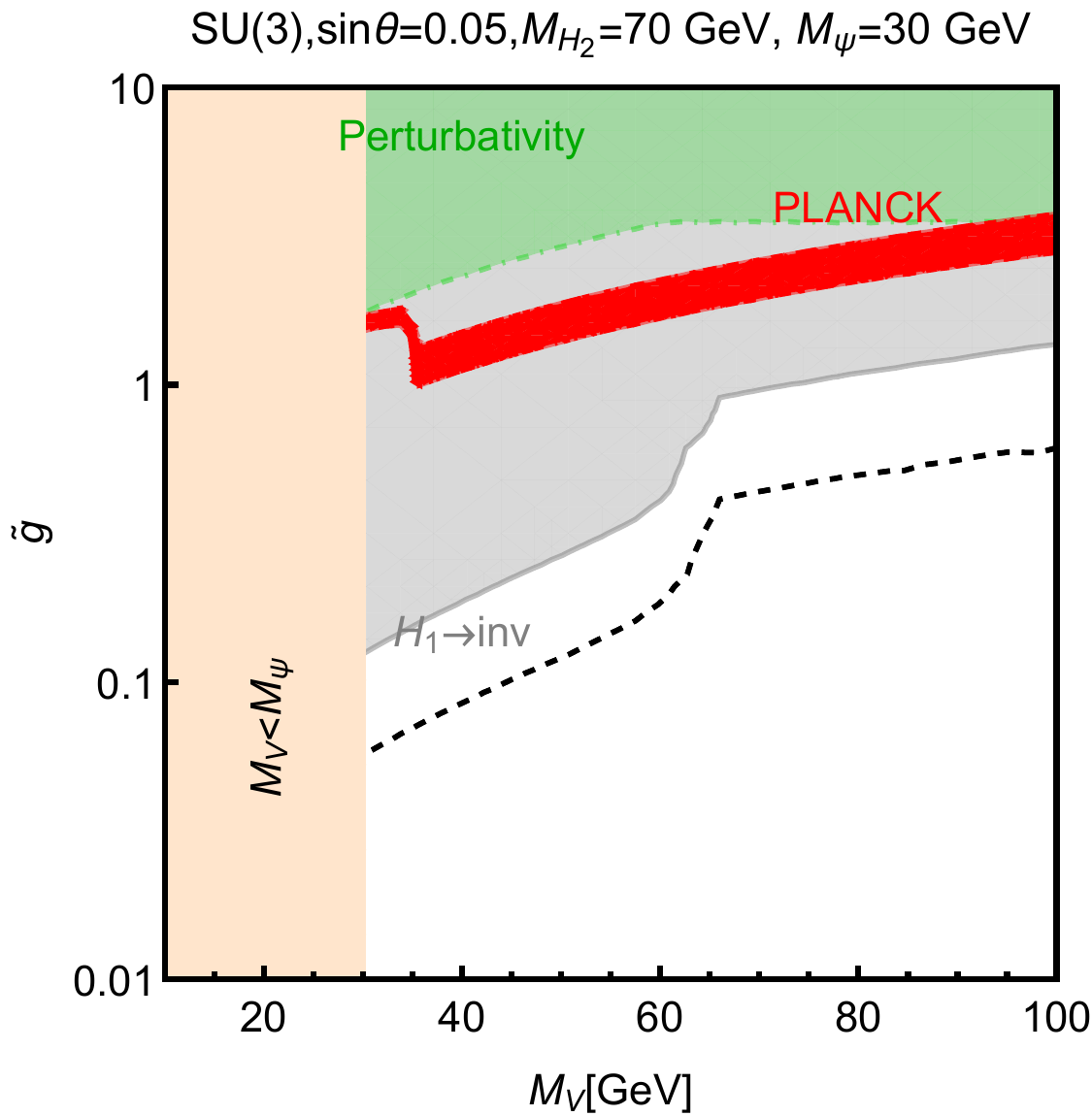}}
     \subfloat{\includegraphics[width=0.48\linewidth]{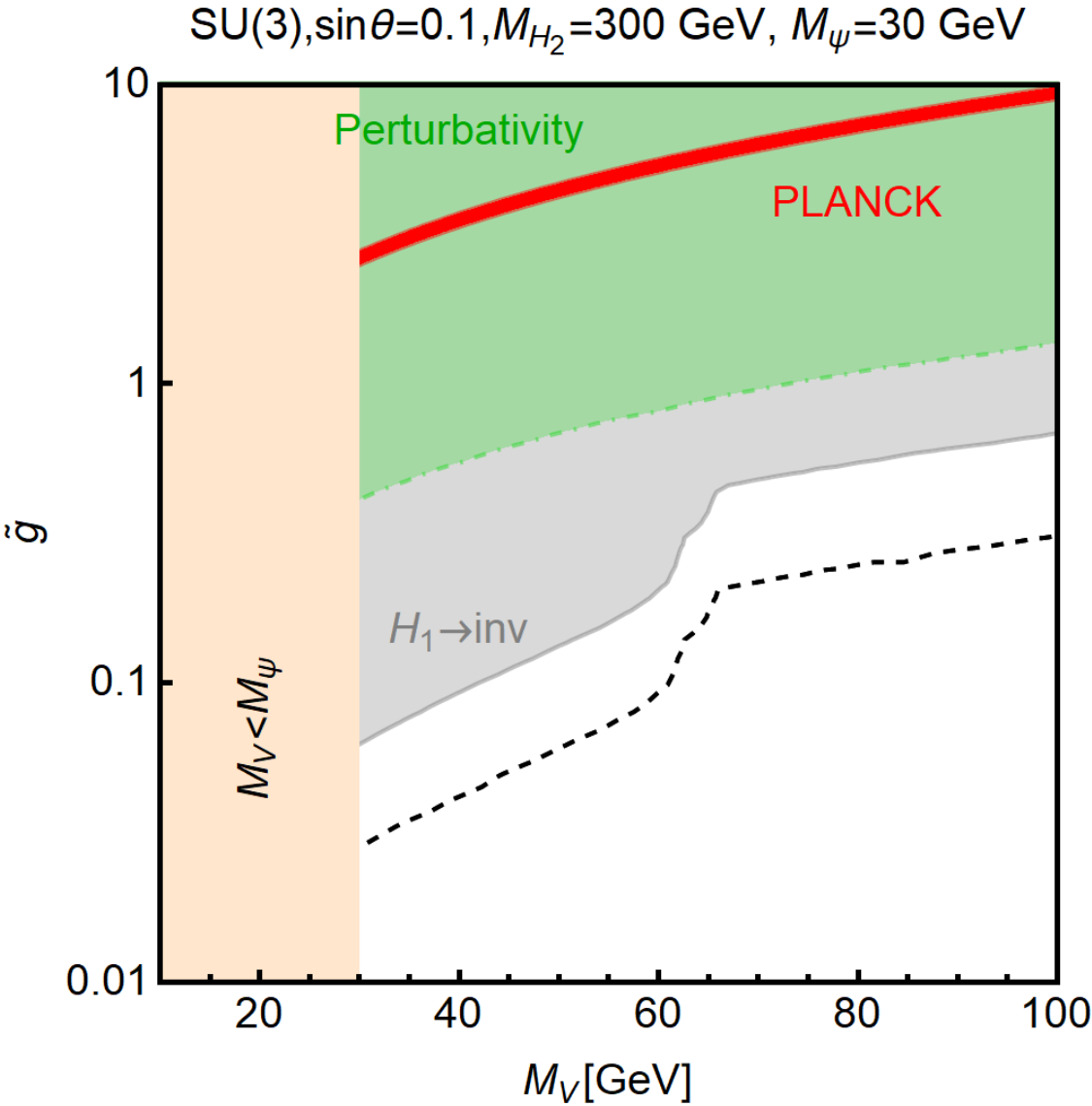}}
\vspace*{-1mm}
    \caption{Constraints for two benchmarks points of the dark SU(3) model with light $\Psi$ for which the mass is fixed to 30 GeV. In the upper row panels the DM has only the $V$ component while $\Psi$ can decay on cosmological scales thanks to a small amount of CP-violation. The bottom row panels represent the same two benchmark assignments of the model parameters but with $V/\Psi$ double component DM.}
    \label{fig:SU3Vchi}
\vspace*{-1mm}
\end{figure}

Fig. \ref{fig:SU3Vchi} shows in the $[M_V,\tilde g]$ plane, the combined constraints on two benchmark assignments of the pair $(M_{H_2},\sin\theta)$ of parameters, namely $(70\,\mbox{GeV},0.05)$ and $(300\,\mbox{GeV},0.1)$, keeping $M_\Psi= 30$ GeV. The upper and lower rows of the plots refer, respectively, to the case in which CP is weakly violated in the new scalar sector, so that $V$ is the unique DM state, and when it is exactly conserved, so that $\Psi$ is an additional DM component.\smallskip

These two scenarios lead to very different outcomes. In the case of a single DM component, we see that the exclusion bound from XENON1T can be easily evaded with the enhancement of the DM annihilation cross section due to the $\Psi \Psi$ final state. The two considered benchmarks are, at least partially, in the reach of future collider experiments that will search for invisible decays of the $H_1$ state. If the $\Psi$ state is cosmologically stable, however, the situation will be significantly different. As a result of a more suppressed annihilation cross section, the spin--0 DM component retains most of the relic density and values of the coupling $\tilde{g}$ have to be chosen above unity in order to comply with the cosmological constraints. As a consequence, the first benchmark is excluded by searches of invisible Higgs decays while the second is in conflict with perturbative unitarity constraints.

\begin{figure}
\vspace*{-1mm}
    \centering
    \subfloat{\includegraphics[width=0.48\linewidth]{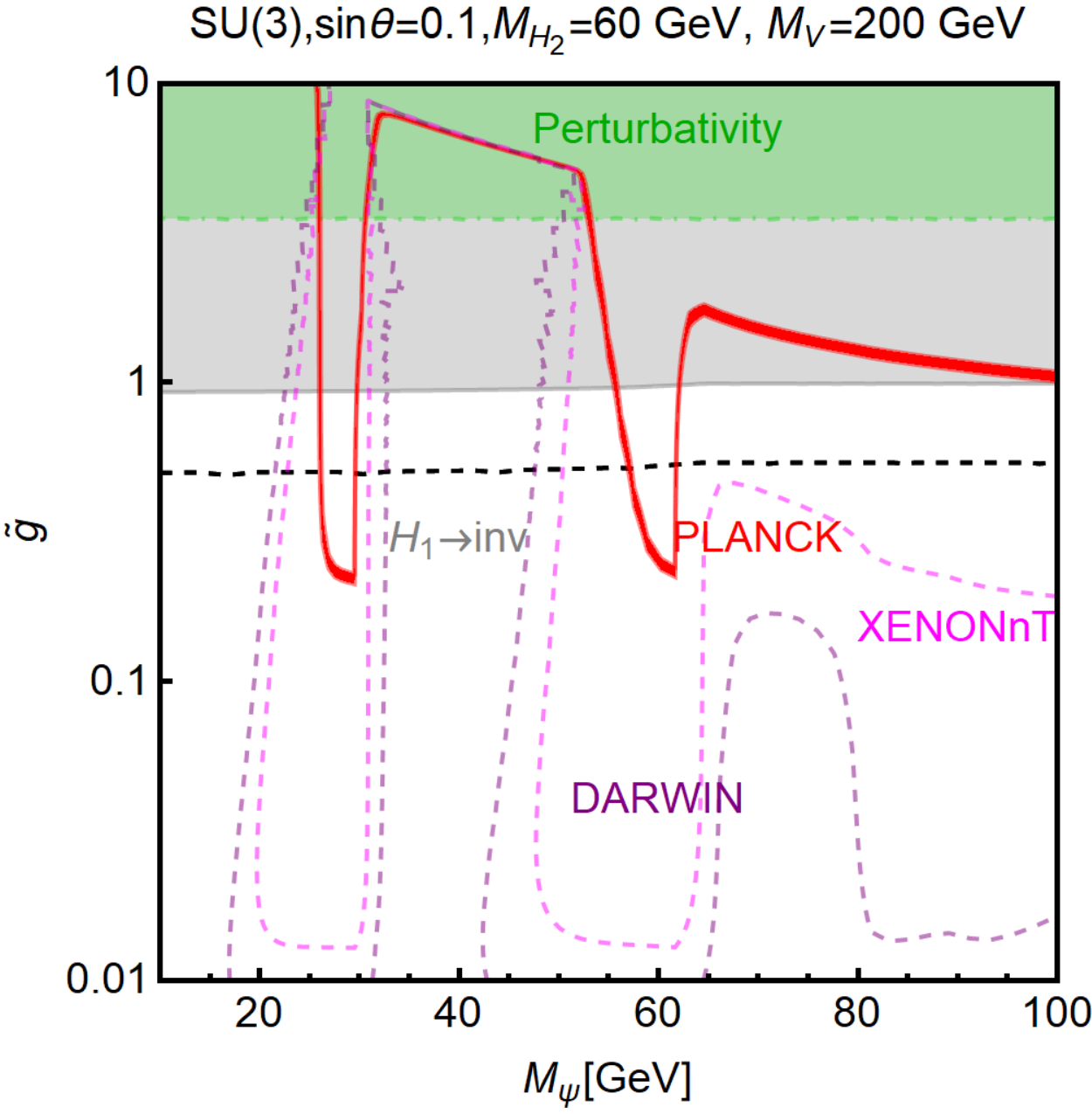}}
    \subfloat{\includegraphics[width=0.48\linewidth]{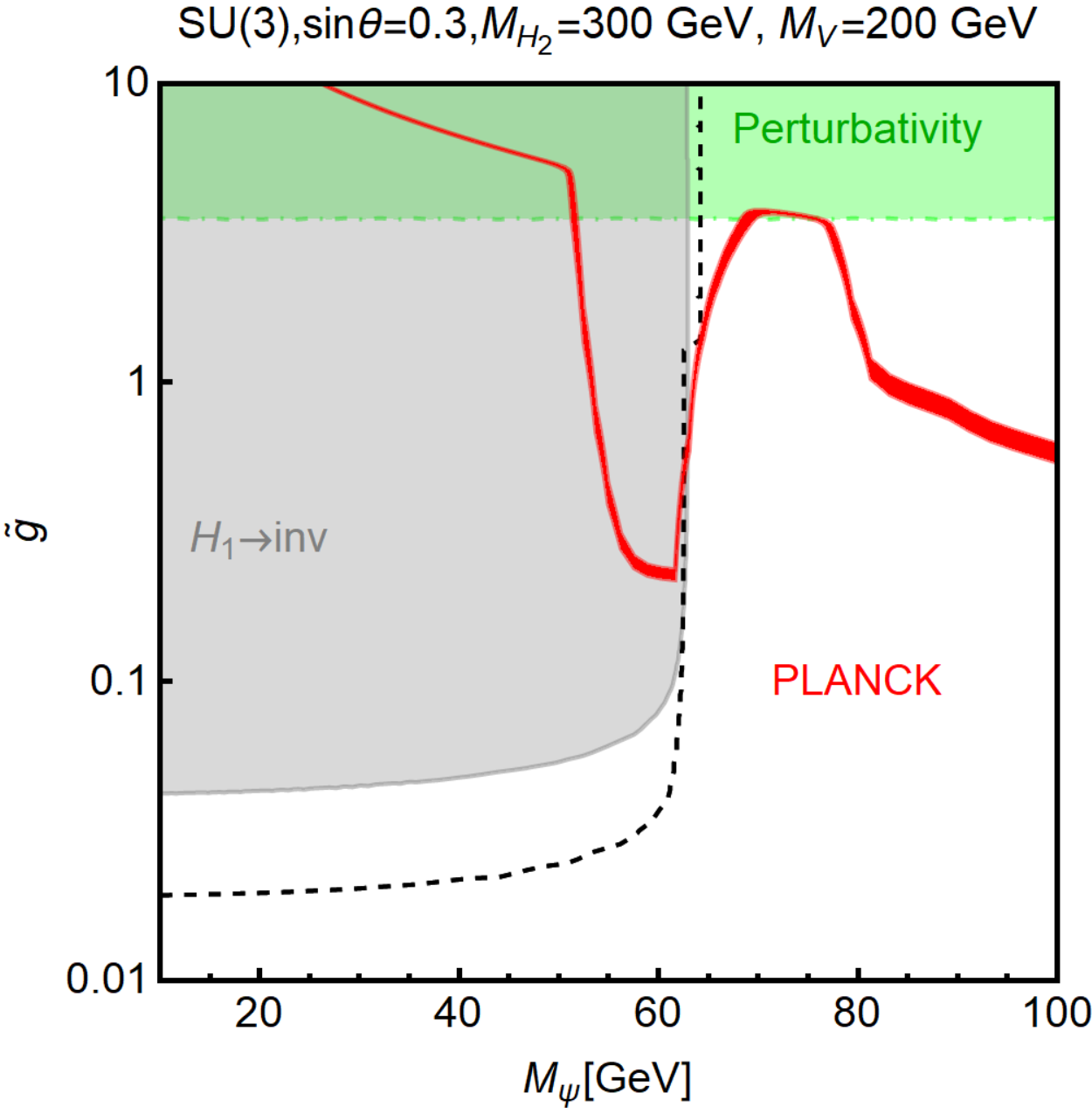}}
\vspace*{-1mm}
    \caption{Summary of DM constraints for two benchmark models of the SU(3) realization with $V/\Psi$ two-component DM  in the $(M_\Psi,\tilde{g})$  plane, where $M_V$ is fixed to 200 GeV. The red contours represent the correct DM relic density and the gray region (black dashed curve) is the current exclusion (projected sensitivity) from searches of the invisible branching fraction of the SM-like Higgs boson.}
    \label{fig:SU3Vchi_2CDM_chivar}
\vspace*{-1mm}
\end{figure}

Fig.~\ref{fig:SU3Vchi_2CDM_chivar} also considers a two component DM scenario but, in this case, the mass $M_\Psi$ is varied while the mass $M_V$ is fixed to 200~GeV. The assignments of the parameters $M_{H_2}$ and $\sin\theta$ have been modified as well to $(60\,\mbox{GeV},0.1)$ and $(300\,\mbox{GeV},0.3)$, respectively. The XENON1T constraints are ineffective since most of the DM relic density is dominated by the $\Psi$ component which has a suppressed spin-independent scattering cross section. Searches for invisible Higgs decays are instead effective in constraining the parameter space. Upcoming direct detection experiments like XENONnT and DARWIN could nevertheless provide competitive constraints as they will be sensitive to the scattering of the DM vector component even if its scattering rate is suppressed by its subdominant contribution to the DM relic density. \smallskip

\begin{figure}
\vspace*{1mm}
    \centering
    \includegraphics[width=0.55\linewidth]{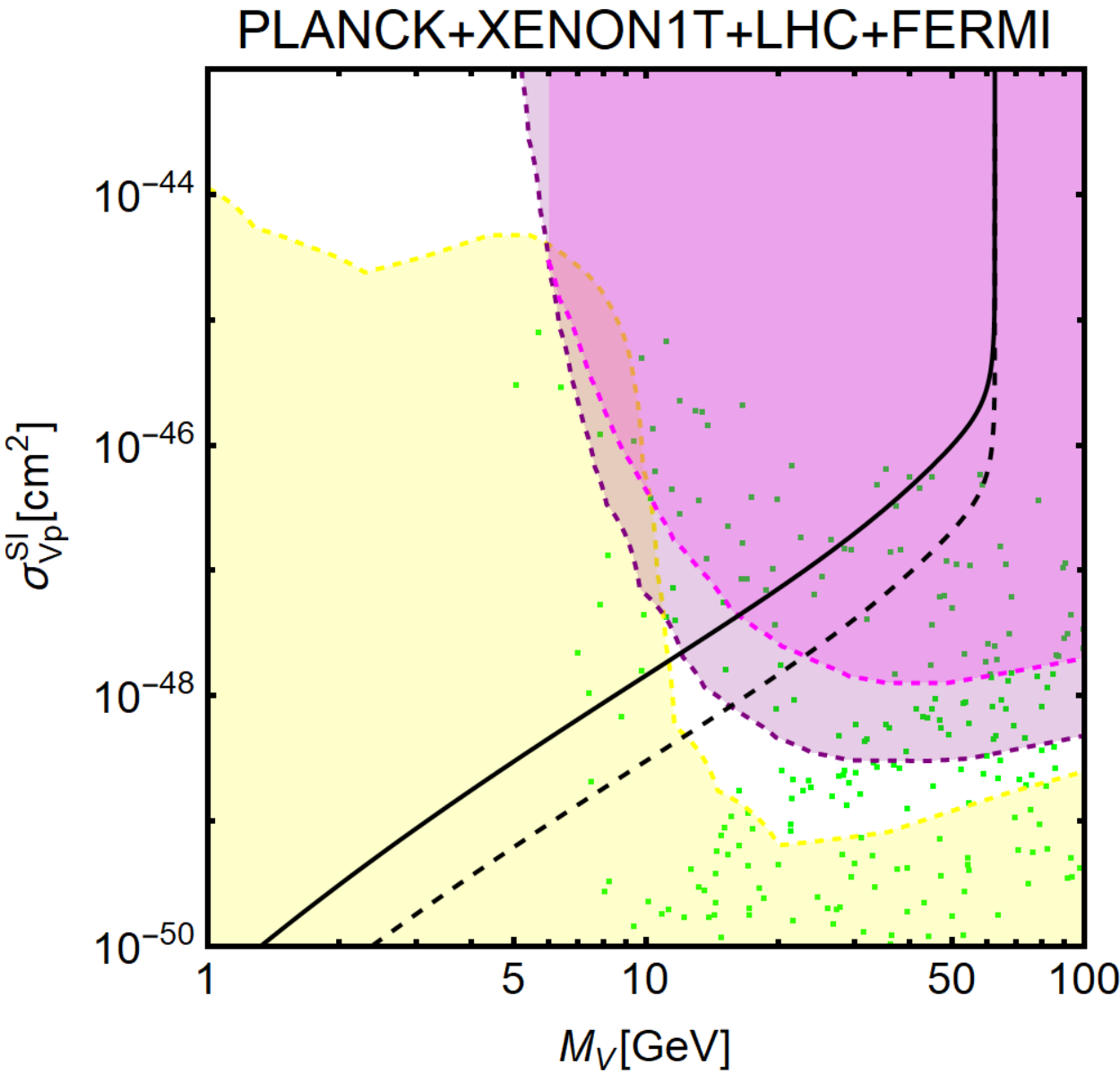}
\vspace*{-1mm}
    \caption{Exclusion limits for the dark SU(3) model with vector DM plus light metastable $\Psi$
    complying with DM constraints and featuring $2.5\,\% \leq {\rm BR}(H_1 \rightarrow\, \mbox{inv}) \leq 11\,\%$. The solid (dashed) contours represent the predicted cross sections of the effective vector Higgs-portal corresponding to BR$(H_1 \rightarrow\, \mbox{inv})= 11\,\%\, (2.5\,\%)$. The magenta/purple regions represent future sensitivities from XENONnT/DARWIN and the yellow ones the neutrino floor.}
    \label{fig:SU3chi_corr}
\vspace*{-2mm}
\end{figure}

In the case of two component DM, the $\Psi$ component retains almost the entire density fraction over most of the parameter space. The scalar Higgs portal is not, however, a viable limit because of the cancellation mechanism in the DM scattering cross section. The case in which the only DM component is the vector $V$ is investigated in Fig. \ref{fig:SU3chi_corr}.\smallskip

Also for this scenario, the majority of allowed model points lie well within the effective vector Higgs portal model limit. A few points, nevertheless, are above the exclusion limit. For these cases both the DM $V$ and the $\Psi$ particle contribute to the invisible width of the Higgs boson and the two states are not degenerate. Therefore due to their different spins, the decay rates of the $H_1 \rightarrow VV$ and $H_1 \rightarrow \Psi \Psi$ processes have different functional dependence on the model parameters.\smallskip 

Hence, contrary to the $V/V^3$ DM scenario, the LHC correlation plot cannot properly capture the phenomenological features of the model.

\section{Conclusions}

The effective field theory approach to the Higgs-portal scenario for weakly interacting massive particles is a powerful tool which allows to study in a simple and model-independent manner the phenomenology of the dark matter particles at both collider and in astroparticle physics experiments. In particular, it allows to straightforwardly exploit the complementarity between two different search strategies: the search for invisible decays of the standard Higgs boson at high--energy colliders such as the LHC and the direct detection of the DM particles in astroparticle physics experiments such as XENON.\smallskip

The objective of the work presented herein is to assess the validity of the effective approach in interpreting the exclusion limits from DM searches at collider  experiments. We have compared the simple effective Higgs-portal models with spin--0, spin--$\frac12$ and spin--$1$ DM states, which in general are subject to theoretical problems such as non-renormalizability and unitarity violation,  with a series of more realistic and ultraviolet complete models and analyzed the conditions under which the interpretation holds and can  be used. \smallskip

We have first considered the case in which the DM interactions with the Higgs sector of the theory occur as a result of the presence of an additional scalar singlet field that mixes with the SM-like Higgs field. This new field  dynamically generates the masses of the fermionic and vector DM particles, thus solving  the renormalizability issue of the effective Higgs-portal models in these two specific cases. This scenario introduces a new degree of freedom in addition to the DM particle, which could significantly alter the phenomenology and possibly spoil the validity of the effective theory interpretation. We have shown, however, that in the conventional Higgs-portal models for spin--$\frac12$, 1 and for completeness also spin--0 DM particles, this happens only when the additional scalar boson is light. Assuming a sufficiently heavy additional scalar, with a mass in the TeV range, fully resolves this inconsistency and is consistent with the spirit of an effective limit. \smallskip

We have also shown that if one insists to enforce the WIMP paradigm, the DM state cannot reproduce the measured cosmological relic density in most parts of the parameter space, except for narrow areas such as the Higgs boson ``pole" and the light DM regions, which are within the kinematical reach for invisible decays of the SM-like Higgs boson, unless the extra scalar degree of freedom is relatively light. In this case, the simplest effective field theory approach does not apply.  \smallskip

We have then analyzed the possibility of more complicated dark sectors for the three different spin assignments of the DM particle.  For the spin--0 and spin--$\frac12$ DM possibilities, we have considered, respectively,  the inert Higgs doublet model and the fermionic singlet-doublet model, in which the DM particle  has an SU(2) component and comes with new neutral and charged partners. In the light DM regime, these models are analogous to the scalar and fermionic Higgs-portal scenario and inherit the difficulty of achieving the correct relic density outside the Higgs pole region. \smallskip

In the case of vector DM, we have considered more complex DM dark sectors by increasing the rank of the dark gauge group. While the SU(2) case does not change the picture drastically, the case of the SU(3) group revealed itself to be quite interesting. In particular, it features a vector DM candidate accompanied by a lighter metastable state. This allows to obtain the correct DM relic density without relying on the presence of additional scalar degrees of freedom. In addition, the impact on the interpretation of invisible Higgs searches at the LHC is very modest since the DM and the additional vector states are sufficiently close in mass to be treated as a single particle in collider searches. \smallskip

The effective Higgs-portal approaches provide robust and consistent limits for a very wide range of more complete and realistic concrete models. These effective models, therefore, provide excellent benchmarks in all three DM particles hypotheses of spin--0, $\frac12$ and spin--1, for interpretation in terms of  DM-nucleon spin independent cross sections. While the measured cosmological DM relic density can be obtained only in narrow regions of parameter space, this strong assumption can be relinquished in favor of alternatives to the conventional thermal paradigm, such as  non-thermal production mechanisms and/or modified cosmologies. 

\bigskip

\noindent {\bf Acknowledgements:}\smallskip

We warmly thank Oleg Lebedev for reading the manuscript and for his valuable comments. We thank Abdesslam Arhrib, Ketevi Assamagan, Martti Raidal and Rui Santos for discussions.  This work is supported in part by the Estonian Mobilitas Plus grant MOBTT86 and by the Junta de Andalucia through the Talentia Senior program.

\bibliography{MajDM}
\bibliographystyle{JHEPfixed}

\end{document}